\documentclass[sigconf,nonacm]{acmart}

\AtBeginDocument{%
  \providecommand\BibTeX{{%
    \normalfont B\kern-0.5em{\scshape i\kern-0.25em b}\kern-0.8em\TeX}}}
    
\usepackage{graphicx}
\usepackage{balance}  
\usepackage{hyperref}
\usepackage{mathtools}
\usepackage{tikz}

\usepackage[noend]{algpseudocode}
\algrenewcommand\textproc{\textit}
\usepackage{algorithm}
\usepackage{amsmath}
\usepackage{etoolbox}

\usepackage{dcolumn}

\usepackage{graphicx}
\usepackage{listings}
\usepackage{color}
\usepackage{subfig}

\usepackage{amsfonts}
\usepackage{hyperref,xcolor}
\usepackage{pgfplots}
\usetikzlibrary{patterns}
\usepackage{blindtext}

\newcommand{\system}{SPF}
\newcommand{\systemlong}{Star Pattern Fragments}
\newcommand{\Tau}{\mathcal{T}}

\newcolumntype{d}[1]{D{.}{.}{#1}}

\makeatletter
\newcommand*{\algrule}[1][\algorithmicindent]{%
  \makebox[#1][l]{%
    \hspace*{.2em}
    \vrule height .75\baselineskip depth .25\baselineskip
  }
}

\newcount\ALG@printindent@tempcnta
\def\ALG@printindent{%
    \ifnum \theALG@nested>0
    \ifx\ALG@text\ALG@x@notext
    \else
    \unskip
    \ALG@printindent@tempcnta=1
    \loop
    \algrule[\csname ALG@ind@\the\ALG@printindent@tempcnta\endcsname]%
    \advance \ALG@printindent@tempcnta 1
    \ifnum \ALG@printindent@tempcnta<\numexpr\theALG@nested+1\relax
    \repeat
    \fi
    \fi
}
\patchcmd{\ALG@doentity}{\noindent\hskip\ALG@tlm}{\ALG@printindent}{}{\errmessage{failed to patch}}
\patchcmd{\ALG@doentity}{\item[]\nointerlineskip}{}{}{} 
\makeatother

\newtheorem{definition}{Definition}

\usepackage{setspace}

\newcommand*{\blankpage}{%
\vspace*{\fill}
{\centering\it This page was intentionally left blank.\par}}
\makeatletter
\renewcommand*{\cleardoublepage}{\clearpage\if@twoside \ifodd\c@page\else
\blankpage
\thispagestyle{empty}
\newpage
\if@twocolumn\hbox{}\newpage\fi\fi\fi}
\makeatother

\begin{document}

\pagestyle{plain}


\title{Star Pattern Fragments: Accessing Knowledge Graphs through Star Patterns}

\author{Christian Aebeloe}
\email{caebel@cs.aau.dk}
\affiliation{%
  \institution{Aalborg University, Denmark}
}

\author{Ilkcan Keles}
\email{ilkcan@cs.aau.dk}
\affiliation{%
  \institution{Aalborg University, Denmark}
}
\email{ilkcan.keles@turkcell.com.tr}
\affiliation{%
  \institution{Turkcell, Turkey}
}

\author{Gabriela Montoya}
\email{gmontoya@cs.aau.dk}
\affiliation{%
  \institution{Aalborg University, Denmark}
}

\author{Katja Hose}
\email{khose@cs.aau.dk}
\affiliation{%
  \institution{Aalborg University, Denmark}
}

\renewcommand{\shortauthors}{Aebeloe et al.}

\begin{abstract}
SPARQL endpoints offer access to a vast amount of interlinked information. 
While they offer a well-defined interface for efficiently retrieving results for complex SPARQL queries, complex query loads can easily overload or crash endpoints as all the computational load of answering the queries resides entirely with the server hosting the endpoint. 
Recently proposed interfaces, such as Triple Pattern Fragments, have therefore shifted some of the query processing load from the server to the client at the expense of increased network traffic in the case of non-selective triple patterns. 
%
This paper therefore proposes Star Pattern Fragments (SPF), an RDF interface enabling a better load balancing between server and client by decomposing SPARQL queries into star-shaped subqueries, evaluating them on the server side. 
%
Experiments using synthetic data (WatDiv), as well as real data (DBpedia), show that SPF does not only significantly reduce network traffic, it is also up to two orders of magnitude faster than the state-of-the-art interfaces under high query load.
\end{abstract}


\maketitle

\section{Introduction}\label{sec:introduction}

Over the past decade, the Semantic Web community has seen a rapid increase in the volume of data available as Linked Open Data (LOD)~\cite{DBLP:reference/bdt/HartigHS19,DBLP:books/crc/linked2014}.
Multiple LOD datasets have been released spanning a broad range of different topics, such as geography (e.g., LinkedGeoData~\cite{DBLP:journals/semweb/StadlerLHA12}), life sciences (e.g., Bio2RDF~\cite{DBLP:conf/semweb/DumontierCCAEBD14}), government data (e.g., US Government LOD~\cite{DBLP:journals/expert/HendlerHMT12}), and general knowledge (e.g., DBpedia~\cite{dbpedia}).
Today, such open datasets can have several billions of triples, for example DBpedia~\cite{dbpedia} where the English language dataset alone has over a billion triples, Wikidata~\cite{DBLP:journals/cacm/VrandecicK14} with around 12 billion triples, and Bio2RDF~\cite{DBLP:conf/semweb/DumontierCCAEBD14} with over 10 billion triples.
Such datasets are often made available through public endpoints, dereferenceable URIs, or downloadable data dumps.
However, this kind of access relies totally on the individual data providers to provide access to their data.

As multiple previous studies have highlighted~\cite{DBLP:journals/ws/VerborghSHHVMHC16,DBLP:conf/esws/AebeloeMH19}, this presents a huge burden for the data providers and, in situations with limited resources on the server, often results in the performance of such public endpoints deteriorating quickly as the load increases~\cite{DBLP:journals/ws/VerborghSHHVMHC16}, and in worst case this leads to unavailability~\cite{DBLP:conf/semweb/ArandaHUV13,DBLP:journals/semweb/VandenbusscheUM17}.

Despite recent efforts to speed up SPARQL query processing under high query load~\cite{DBLP:journals/ws/VerborghSHHVMHC16,DBLP:journals/corr/HartigA16,DBLP:conf/semweb/MontoyaAH18,sage,smartkg}, answering SPARQL queries remains an expensive task.
In fact, deciding whether a set of bindings is an answer to a query has been shown to be at least NP-complete~\cite{DBLP:journals/tods/PerezAG09}.
Still, Triple Pattern Fragments (TPF)~\cite{DBLP:journals/ws/VerborghSHHVMHC16} have provided interesting insights into the problem and a novel way to approach it.
TPF limits the load on the server by sharing the computational load between the server and the client. 
While the server evaluates individual triple patterns, the client handles the remaining query processing tasks.
This increases the availability of the server and ensures more efficient query processing during periods with high load.

\begin{figure}[htb]
\begin{lstlisting}[captionpos=b, caption=Find Germans and Norwegians that have won the same award and their birth dates, label=lst:q1,floatplacement=tbp,basicstyle=\footnotesize\ttfamily,mathescape=true]
PREFIX dbo: <http://dbpedia.org/ontology/>
PREFIX dbr: <http://dbpedia.org/resource/>
select distinct * where {
 ?p1 dbo:country dbr:Germany. # tp1: 18,174 matches
 ?p1 dbo:award ?a .           # tp2: 90,933 matches
 ?p1 dbo:birthDate ?bd1 .     # tp3: 1,740,614 matches
 ?p2 dbo:country dbr:Norway . # tp4: 5,520 matches
 ?p2 dbo:award ?a .           # tp5: 90,933 matches
 ?p2 dbo:birthDate ?bd2       # tp6: 1,740,614 matches
}
\end{lstlisting}
\end{figure}

Nevertheless, there are cases where TPF is significantly less efficient than  SPARQL endpoints.
Consider, for example, the SPARQL query shown in Listing~\ref{lst:q1} over DBpedia version 2016-04~\cite{dbpedia}.
Executing this query using TPF requires transferring a huge number of intermediate results.
In addition, the TPF client sends a server request for each binding obtained from the previously evaluated triple patterns.
This results in a large number of intermediate results being transferred to the client, as well as in a large number of calls to the server.
This creates a significant overhead when processing the query, decreasing the overall performance.

Linked Data Fragments (LDF) interfaces, such as Bindings-Restricted Triple Pattern Fragments (brTPF)~\cite{DBLP:journals/corr/HartigA16} and hybridSE~\cite{DBLP:conf/semweb/MontoyaAH18}
present different ways of addressing this issue.
brTPF uses block nested loop-like joins, where a triple pattern is evaluated once per group of \textit{N} bindings obtained from the previously evaluated triple patterns (\textit{5 $\leq$ N $\leq$ 50} in~\cite{DBLP:journals/corr/HartigA16}). 
While this results in significantly fewer calls to the server, it still incurs a relatively high network traffic (364 calls to process \texttt{tp2} given the bindings found for \texttt{tp1} in Listing~\ref{lst:q1}).

What all these approaches ignore though is the potential of evaluating conjunctive subqueries.
Such subqueries can
(i) be computed relatively efficiently on the server~\cite{DBLP:journals/tods/PerezAG09} and 
(ii) reduce the network traffic since fewer intermediate results are transferred. 
Subqueries, such as subqueries \texttt{\{tp1$\,.\,$tp2$\,.\,$tp3\}} and \texttt{\{tp4$\,.\,$tp5$\,.\,$tp6\}} in Listing~\ref{lst:q1}, do not require full SPARQL expressiveness.
While there could potentially be several ways to decompose SPARQL queries (e.g., based on shared variables between triple patterns~\cite{DBLP:journals/ws/VerborghSHHVMHC16}), the specific decomposition strategy is not the focus of this paper.
Nevertheless, decomposition into star-shaped subqueries is a widely used decomposition strategy~\cite{smartkg,DBLP:journals/pvldb/AbdelazizMOAK17} that is used in this paper.

This paper investigates the limitations due to large numbers of intermediate results that most LDF interfaces suffer from, and the effects of evaluating conjunctive subqueries on the server while still processing queries on the client, on the network usage and server load under high query load.
This paper introduces a novel interface called \systemlong{} (\system) that improves the overall query processing performance, while also ensuring high availability by combining a lower network load with a comparatively low server load through decomposing SPARQL queries into star-shaped subqueries\footnote{Code is available on the SPF website \url{http://relweb.cs.aau.dk/spf}}.
By doing so, SPF is able to reduce the amount of intermediate results transferred to the client compared to other LDF interfaces.

In summary, this paper makes the following contributions:

\begin{itemize}
	\item Definition of \systemlong{} (\system), an LDF interface that reduces network usage while keeping the server load comparatively low.
    \item Formalization and implementation of an \system{} server.
    \item Client-side query processing strategies to efficiently compute answers to SPARQL queries using an SPF server to process star-shaped subqueries and process queries with any SPARQL operator.
\end{itemize}

To assess the effects of processing star-shaped subqueries on the server while executing the queries on the client, a thorough evaluation of SPF using three different sized WatDiv~\cite{watdiv} datasets with up to 10 billion triples, using large query loads for stress testing, is provided.
Moreover, SPF is evaluated against DBpedia~\cite{dbpedia} using queries posed by real users~\cite{DBLP:conf/semweb/SaleemAHMN15} to evaluate how the approach performs in real-world scenarios.

This paper is organized as follows.
Section~\ref{sec:relatedWork} discusses related work, 
Section~\ref{sec:background} introduces the terminology used in this paper, 
Section~\ref{sec:characteristics} presents a formal characterization of the Star Pattern Fragments interface, 
Section~\ref{sec:queryprocessing} describes the SPF server and client details,
Section~\ref{sec:evaluation} discusses experimental results, and 
Section~\ref{sec:conclusion} concludes the paper and provides a perspective on future work. 

\section{Related Work}\label{sec:relatedWork}
One of the most popular interfaces for querying RDF data is SPARQL endpoints.
SPARQL endpoints are Web services that implement the SPARQL protocol and usually provide an HTTP interface that accepts SPARQL queries.
However, several studies~\cite{DBLP:conf/semweb/ArandaHUV13,DBLP:journals/semweb/VandenbusscheUM17} have previously highlighted the fact that such endpoints are often unavailable, meaning that accessing data can sometimes be impossible.

Decentralization of the data storage and distribution of query processing between clients and servers is often referred to when discussing solutions to dataset availability~\cite{DBLP:journals/ws/VerborghSHHVMHC16,DBLP:conf/esws/AebeloeMH19,solid,DBLP:conf/semweb/PolleresKFTM18,DBLP:conf/semco/Marx0LN18}.
For example, the Solid project~\cite{solid} uses decentralized Personal Online Datastores (PODs) to separate personal information from applications.
Users can decide for themselves where their POD is stored, and which application have access to it.
Thus when loading a Solid application, it must query data from multiple sources located on the Web.
However, Solid focuses mostly on the security of personal datasets, whereas this paper focuses on efficient query processing during high server loads.

Previous work increased the data availability by decentralizing the data storage, using federated query processing, or decentralizing the query processing effort.
The remainder of this section contains an analysis of each approach and an explanation of the pitfalls of such approaches.

\subsection{Decentralized Architectures}
Decentralized architectures have previously been shown to increase the availability of the data.
For instance, Peer-to-Peer (P2P) architectures~\cite{DBLP:conf/esws/AebeloeMH19,ppbfs,DBLP:conf/www/CaiF04,DBLP:journals/ws/KaoudiKKMMP10,DBLP:conf/icde/KarnstedtSRMHSJ07} remove the central server altogether and instead let clients also act like servers with a limited local datastore; by replicating each data fragment across several such nodes, P2P systems are able to ensure that the data is available even if the original node fails.
However, P2P systems are typically either vulnerable to churn (when nodes frequently leave or join the network)~\cite{DBLP:conf/www/CaiF04,DBLP:journals/ws/KaoudiKKMMP10,DBLP:conf/esws/GrallFMSMSV17} or cause high network traffic for queries with a large number of intermediate results~\cite{DBLP:conf/esws/AebeloeMH19,ppbfs}.
Hence, several approaches~\cite{grall:hal-01805154,DBLP:conf/semweb/MolliS17} focus on sharing query processing tasks across  networks of web browsers based on the functionality offered by the browsers and caching of recently used datasets.
While this lowers the load on each individual node, Web browsers are usually quite limited in processing power and storage capabilities.
However, Star Pattern Fragments (SPF) are orthogonal to the aforementioned decentralized architectures.

\subsection{Federated Systems}
Federated query engines~\cite{DBLP:conf/esws/SchwarteHHSS11,DBLP:conf/semweb/AcostaVLCR11,DBLP:conf/esws/SaleemN14,DBLP:conf/i-semantics/CharalambidisTK15,DBLP:conf/esws/IbragimovHPZ15,DBLP:conf/semweb/GorlitzS11,DBLP:conf/semweb/MontoyaSH17} divide SPARQL query processing over multiple SPARQL endpoints. 
Nonetheless, they sometimes fail to generate optimal query plans that transfer the minimum amount of data from endpoints to the engine and therefore increase the load on SPARQL endpoints~\cite{DBLP:conf/esws/JakobsenMH19}.
This means that they sometimes still suffer from relatively high server load.
Query optimization techniques for federated engines, such as~\cite{DBLP:conf/semweb/MontoyaVA12}, consider decomposing  
SPARQL queries into star-shaped subqueries that can be evaluated by a single SPARQL endpoint. 
Star-shaped query decomposition has also been used
in~\cite{DBLP:conf/esws/VidalRLMSP10} to improve the query execution time. 
While these approaches use a similar query decomposition scheme as SPF, they mainly target situations where the server side is made stronger by endpoint federations.
As mentioned earlier, such endpoints suffer from unavailability~\cite{DBLP:conf/semweb/ArandaHUV13,DBLP:journals/semweb/VandenbusscheUM17}.
Instead, other optimization techniques for federated engines~\cite{DBLP:conf/semweb/MontoyaSH17,DBLP:conf/i-semantics/0002PSHN18} focus on estimating the selectivity of joins to produce better query execution plans.
These approaches could be combined with SPF and provide the benefits highlighted in this paper to federated systems as well.

\subsection{Client-Server Architectures}
Linked Data Fragments (LDF) interfaces, such as Triple Pattern Fragments (TPF)~\cite{DBLP:journals/ws/VerborghSHHVMHC16}, were proposed to improve the server availability under load. 
TPF servers only process individual triple patterns and therefore have a lower processing burden than SPARQL endpoints. 
TPF clients rely on either a greedy algorithm~\cite{DBLP:journals/ws/VerborghSHHVMHC16}, a metadata based strategy~\cite{DBLP:conf/esws/HerwegenVMW15}, or adaptive query processing techniques and star-shaped decomposition~\cite{DBLP:conf/semweb/AcostaV15} to determine the execution order of the triple patterns. 
While TPF reduces the load on the server in general, it puts much more load on the client and incurs more network traffic.
Furthermore, the performance of TPF is heavily affected by aspects such as the triple pattern type~\cite{DBLP:conf/semweb/HelingAMS18} (defined with respect to the position of variables in a triple pattern) and the query shape~\cite{DBLP:conf/semweb/MontoyaKH19,DBLP:journals/corr/abs-1912-08010}.
Bindings-Restricted TPF (brTPF)~\cite{DBLP:journals/corr/HartigA16} was proposed to reduce network traffic by coupling triple patterns and bindings obtained from previously evaluated triple patterns.
%
Despite improving the availability of RDF data, all these approaches cause a large number of calls to the server during query processing. 
hybridSE~\cite{DBLP:conf/semweb/MontoyaAH18} combines SPARQL endpoints and brTPF servers to process queries more efficiently than the TPF-based interfaces; SPARQL subqueries with a large number of intermediate results are evaluated using SPARQL endpoints to overcome limitations of TPF clients. 
%
However, since hybridSE may send complex subqueries to the endpoint, and endpoints have downtime~\cite{DBLP:conf/semweb/ArandaHUV13}, which leaves the approach vulnerable to downtime.
The LDF interfaces mentioned above process individual triple patterns on the server.
This causes a large overhead on the network usage since many intermediate results have to be transferred.
Instead, SPF processes conjunctive subqueries on the server, decreasing the amount of intermediate results that have to be transferred over the network and improving performance overall.

Other client-server architectures use different techniques to address some of the issues posed by TPF.
SaGe~\cite{sage}, for example, uses a preemptive model that suspends queries after a fixed time quantum, as to not starve simpler queries of system resources, after which they can be resumed upon client request.
While the time quantum ensures that long-running queries will not starve system resources, these long-running queries tend to cause a high number of requests to the server since they have to be resumed several times. 
Attempting to decrease the number of requests by increasing the time quantum may result in the server resources being exhausted, lowering performance overall.
Smart-KG~\cite{smartkg} ships star-shaped partitions to the client during query processing.
This decreases the number of requests issued to the server, since partitions already shipped to the client can be evaluated directly on the client.
However, this can in some cases lead to unnecessary data transfer during query processing, since the entire partition is shipped regardless of object bindings obtained from previously evaluated triple patterns.
SPF is able to both avoid long-running queries exhausting server resources and causing a high number of requests by only processing star-shaped joins on the server.
Such computations do not significantly increase the server load because star-shaped subqueries can be answered in linear complexity~\cite{DBLP:journals/tods/PerezAG09}.
In doing so, SPF also achieves a reduction on the data transfer and the execution time without having a significant impact on availability.
As a result, SPF achieves better query processing performance for complex workloads over large datasets and under high load compared to both SaGe and Smart-KG as shown in Section~\ref{sec:evaluation}.


\section{Preliminaries}
\label{sec:background}
The recommended format for storing semantic data is the Resource Description Framework (RDF)\footnote{\url{https://www.w3.org/TR/rdf11-concepts/}}. 

\begin{definition}[RDF Triple]
Given the infinite and disjoint sets $U$ (set of all URIs), $B$ (set of all blank nodes), and $L$ (set of all literals), an RDF triple is a triple of the form $(s,p,o)\in (U\cup B)\times U \times (U\cup B\cup L)$, where $s$, $p$, $o$ are called \textit{subject}, \textit{predicate}, and \textit{object}.
\end{definition}

A knowledge graph (RDF graph) $\mathcal{G}$ is a finite set of RDF triples.
Today, SPARQL\footnote{\url{https://www.w3.org/TR/sparql11-query/}} is the standard language for querying RDF data.
A SPARQL query contains a set of \textit{triple patterns}, which, given the additional infinite set $V$ (disjoint with $U$, $B$ and $L$) of all variables, are triples of the form $(s,p,o)\in (U\cup B\cup V)\times (U\cup V) \times (U\cup B\cup L\cup V)$. 

In the following, a \textit{star pattern} is defined to be one of two types of star patterns: subject-based star patterns or object-based star patterns.

\begin{definition}[Subject-Based Star Pattern]\label{def:sbsp}
A \textit{subject-based star pattern} is a set of \textit{n} triple patterns, $\{(s_1,p_1,o_1),\dots,(s_n,p_n,o_n)\}$, such that the subjects of all these triple patterns are the same, i.e., $s_i=s_j$ for all $1\leq i,j\leq n$. 
\end{definition}

\begin{definition}[Object-Based Star Pattern]\label{def:obsp}
An \textit{object-based star pattern} is a set of \textit{n} triple patterns, $\{(s_1,p_1,o_1),\dots,(s_n,p_n,o_n)\}$, such that the objects of all these triple patterns are the same, i.e., $o_i=o_j$ for all $1\leq i,j\leq n$. 
\end{definition}

Given the definition of subject-based star patterns and object-based star patterns, a star pattern is defined as follows.

\begin{definition}[Star Pattern]\label{def:sp}
A \textit{star pattern} $S$ is a set of \textit{n} triple patterns, $S=\{(s_1,p_1,o_1),\dots,(s_n,p_n,o_n)\}$, such that $S$ is either a subject-based star pattern or an object-based star pattern. 
\end{definition}

Corollary, an \textit{RDF star} is a set of RDF triples that has the same properties as in Definition~\ref{def:sp}.

%
%

\subsection{Linked Data Fragments}\label{sec:ldfs_prel}
A Linked Data Fragment (LDF) of a knowledge graph $\mathcal{G}$ consists of a subset of $\mathcal{G}$'s triples (a fragment) coupled with metadata about the fragment and controls to retrieve similar LDFs.
The following description of LDFs follows~\cite{DBLP:journals/ws/VerborghSHHVMHC16}.
LDFs consider only blank-node-free RDF triples.
An LDF is defined as follows.

\begin{definition}[Linked Data Fragment~\cite{DBLP:journals/ws/VerborghSHHVMHC16}]
Given a knowledge graph $\mathcal{G}$, a Linked Data Fragment (LDF) consists of the following three elements:

\begin{itemize}
\item \textbf{Data}: A subset of $\mathcal{G}$'s triples
\item \textbf{Metadata}: RDF triples that describe the data
\item \textbf{Controls}: Links and forms to retrieve other LDFs of the same or other knowledge graphs
\end{itemize}
\end{definition}

Any knowledge graph made available on the Web, in any format, can be described as an LDF.
For example, a data dump can be described as a single LDF with the following components~\cite{DBLP:journals/ws/VerborghSHHVMHC16}:

\begin{itemize}
\item \textbf{Data}: All triples in the data dump
\item \textbf{Metadata}: Data about the dump, e.g., version number, author, etc.
\item \textbf{Controls}: No controls, since the entire data dump is given in the LDF.
It could, however, contain controls to other versions of the data dump.
\end{itemize}

\noindent
Given a knowledge graph $\mathcal{G}$, each LDF of $\mathcal{G}$ contains triples that somehow belong together.
To obtain triples from $\mathcal{G}$ to form a fragment, a \textit{selector function} is used, and defined as follows.

\begin{definition}[Selector Function~\cite{DBLP:journals/ws/VerborghSHHVMHC16}]
Given $\Tau^*=U\times U \times (U\cup L)$, the set of all blank-node-free RDF triples, a selector function $s$ is a function such that $s:2^{\Tau^*}\rightarrow 2^{\Tau^*}$.
\end{definition}

That is, a selector function takes as input a set of blank-node-free RDF triples, and outputs a set of blank-node-free RDF triples.
Note that the output could in principle contain triples that are not in the input, e.g., \texttt{CONSTRUCT} queries. 
However, in most cases, the output corresponds to a subset of the input.

\begin{definition}[Hypermedia Controls~\cite{DBLP:journals/ws/VerborghSHHVMHC16}]\label{def:hypermediacontrols}
A hypermedia control is a function that maps from some set to $U$.
\end{definition}

A URI is a zero-argument hypermedia control, i.e., a constant function, and a form is a multi-argument hypermedia control.
In the case of LDF, the domain of a hypermedia control is a set of selector functions, encoded as URLs.

\begin{definition}[Linked Data Fragment~\cite{DBLP:journals/ws/VerborghSHHVMHC16}]
Given a knowledge graph $\mathcal{G}$, a Linked Data Fragment (LDF) of $\mathcal{G}$ is a $5$-tuple $f=\langle u,s,\Gamma,M,C\rangle$, with 
\begin{itemize}
\item a source URI $u$,
\item a selector function $s$,
\item the result of applying $s$ to $\mathcal{G}$, $s(\mathcal{G})=\Gamma$,
\item a set of additional triples $M$ that describes metadata, and
\item a finite set of hypermedia controls $C$.
\end{itemize}
\end{definition}

\noindent An LDF server should divide each fragment $f=\langle u,s,\Gamma,M,C\rangle$ into reasonably sized \textit{LDF pages} $\phi=\langle u',u_f,s_f,\Gamma',M',C'\rangle$, containing 
(i) the URI $u'$ from which $\phi$ could be obtained and $u'\neq u$,
(ii) $u_f=u$,
(iii) $s_f=s$
(iv) $\Gamma'\subseteq \Gamma$,
(v) $M'\supseteq M$, and
(vi) $C'\supseteq C$.
$M'$ and $C'$ are supersets of $M$ and $C$, since they also contain additional metadata and controls that are specific to the LDF page.
Having additional metadata and controls makes it possible for clients to avoid downloading very large chunks of data accidentally~\cite{DBLP:journals/ws/VerborghSHHVMHC16}.

\section{\systemlong}
\label{sec:characteristics}
In between SPARQL endpoints, which handle all the query processing load on the server, and TPF, which processes only triple patterns on the server and handles the rest of query processing load on the client, there is a lot of potential for other interfaces that provide a better way of sharing query processing load between server and client.
For instance, processing conjunctive subqueries (e.g., star patterns) on the server can result in less network traffic while it does not impose a high additional server load, which is evident from the experiments in Section~\ref{sec:evaluation}.

\begin{figure*}[htb]
\centering
\tikzset{every picture/.style={line width=0.75pt}} 

\begin{tikzpicture}[x=0.7pt,y=0.65pt,yscale=-1,xscale=1]

\draw [line width=1.5]    (98,153) -- (560,153) ;
\draw [shift={(563,153)}, rotate = 180] [fill={rgb, 255:red, 0; green, 0; blue, 0 }  ][line width=1.5]  [draw opacity=0] (13.4,-6.43) -- (0,0) -- (13.4,6.44) -- (8.9,0) -- cycle    ;
\draw [shift={(95,153)}, rotate = 0] [fill={rgb, 255:red, 0; green, 0; blue, 0 }  ][line width=1.5]  [draw opacity=0] (13.4,-6.43) -- (0,0) -- (13.4,6.44) -- (8.9,0) -- cycle    ;
\draw [line width=1.5]    (116.71,140.71) -- (116.71,152.71) ;

\draw [line width=1.5]    (180.71,140.71) -- (180.71,152.71) ;

\draw [line width=1.5]    (258.71,140.71) -- (258.71,152.71) ;

\draw [line width=1.5]    (537.71,140.71) -- (537.71,152.71) ;

\draw [line width=1.5]    (272.71,152.71) -- (272.71,164.71) ;

\draw [line width=1.5]    (350.71,140.71) -- (350.71,152.71) ;






\draw (140,179) node [scale=0.8] [align=left] {\textit{{\scriptsize generic requests}}\\\textit{{\scriptsize high client effort}}\\\textit{{\scriptsize high server availability}}};
\draw (526,167) node [scale=0.8] [align=left] {\textit{{\scriptsize specific requests}}};
\draw (522,180) node [scale=0.8] [align=left] {\textit{{\scriptsize high server effort}}};
\draw (512,193) node [scale=0.8] [align=left] {\textit{{\scriptsize low server availability}}};
\draw (119,123) node [scale=0.8] [align=left] {\textbf{{\small data}}\\\textbf{{\small dump}}};
\draw (183,122) node [scale=0.8] [align=left] {\textbf{{\small Linked Data}}\\\textbf{{\small document}}};
\draw (261,123) node [scale=0.8] [align=left] {{\small \textbf{triple pattern}}\\{\small \textbf{fragments}}};
\draw (540,122) node [scale=0.8] [align=left] {\textbf{{\small SPARQL}}\\\textbf{{\small result}}};
\draw (280,180) node [scale=0.8] [align=left] {{\small \textbf{bindings-restricted}}\\\textbf{{\small triple pattern fragments}}};
\draw (353,122) node [scale=0.8] [align=left] {\textbf{{\small \textit{star pattern}}}\\\textbf{{\small \textit{fragments}}}};

\end{tikzpicture}
\caption{HTTP interfaces for RDF data (adapted from~\cite{DBLP:journals/corr/HartigA16,DBLP:journals/ws/VerborghSHHVMHC16}).}\label{fig:interfaces}
\end{figure*}
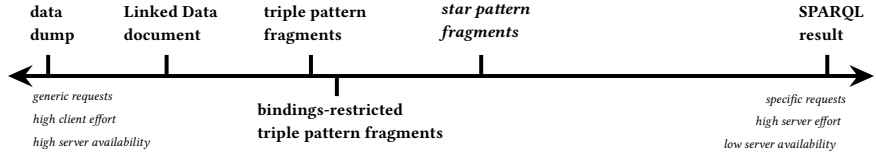

This section contains a formal definition of Star Pattern Fragments (SPF), as an extension of brTPF~\cite{DBLP:journals/corr/HartigA16}, that exposes an HTTP interface for processing star pattern queries in addition to processing individual triple pattern queries.
This increases the server load slightly; however, for queries with large intermediate results (such as Listing~\ref{lst:q1}), this is preferable to ensure fewer requests to the server, which results in lower network traffic and faster query processing.
The relative position of SPF between different RDF interfaces is shown in Figure~\ref{fig:interfaces}.

Logically, an SPF over a given knowledge graph $\mathcal{G}$ has the following properties:

\begin{itemize}
\item \textbf{Data}: All RDF stars in $\mathcal{G}$ that match a given star pattern
\item \textbf{Metadata}: An estimate of the number of stars that match the given star pattern
\item \textbf{Controls}: A hypermedia form that allows the client to retrieve any SPF of the same knowledge graph
\end{itemize}

\noindent
As with TPF~\cite{DBLP:journals/ws/VerborghSHHVMHC16}, SPF is able to prevent long-running queries to exhaust the server resources by dividing the results into reasonably sized pages. SPF pages contain a bound number of stars, but the number of triples varies with the number of triple patterns in the star pattern. Moreover, 
since conjunctive subqueries can be answered efficiently by the server~\cite{DBLP:journals/tods/PerezAG09}, each request can be answered relatively quickly.

The remainder of this section formalizes SPFs by adapting the general formalizations of TPF~\cite{DBLP:journals/ws/VerborghSHHVMHC16} and brTPF~\cite{DBLP:journals/corr/HartigA16}, and provides a response format for an SPF request in the form of a Hydra formalization~\cite{DBLP:conf/www/LanthalerG13a}.

\subsection{Formal Definition}\label{sec:fd}

Let $[[S]]_\mathcal{G}$ be the answer to a star pattern $S$ over a knowledge graph $\mathcal{G}$.
$[[S]]_\mathcal{G}$ is a set of \emph{solution mappings}, i.e., partial mappings $\mu:{V}\mapsto({U}\cup{L})$.
A set of blank-node-free RDF triples $T$ is said to be \textit{matching triples} for a star pattern $S$, denoted $T[S]$, if there exists a solution mapping $\mu$ in $[[S]]_\mathcal{G}$ such that $T=\mu[S]$ where $\mu[S]$ denotes the triples (or triple patterns) obtained by replacing the variables in $S$ with values according to $\mu$.

Similar to how brTPF~\cite{DBLP:journals/corr/HartigA16} couples bindings and triple patterns, SPF couples bindings obtained from previously evaluated star patterns with subsequent star patterns to decrease the network traffic.

\begin{definition}[Star Pattern-Based Selector Function]\label{def:selector}
Given a star pattern $S$ and a finite sequence of solution mappings $\Omega$, the star pattern-based selector function for $S$ and $\Omega$, $s_{(S,\Omega)}$, is the selector function that, for every knowledge graph $\mathcal{G}$, is defined as follows.

$$
	 s_{(S,\Omega)}(\mathcal{G}) = \\
	\begin{cases}
		\{t\in T\mid T\subseteq\mathcal{G} \land T[S] & \text{if } \Omega=\emptyset  \\
		\{t\in T\mid T\subseteq\mathcal{G} \land T[S] \, \land \\ \;\;\;\;\;\exists\mu\in[[S]]_\mathcal{G},\mu'\in\Omega: \\\;\;\;\;\;\;\; \mu[S]=T\wedge\mu'\subseteq\mu\} & \text{ otherwise.}
	\end{cases} $$


\end{definition}

The simplest star pattern consists of a single triple pattern. 
For this reason, SPF is backwards compatible with both TPF~\cite{DBLP:journals/ws/VerborghSHHVMHC16} and brTPF~\cite{DBLP:journals/corr/HartigA16}, as a star pattern request with a single triple pattern corresponds to a single triple pattern request for TPF and brTPF.
As such, applying the star pattern-based selector function in this case would be equivalent to applying either the triple pattern-based selector function or the bindings-restricted triple pattern-based selector function.

Consider, for example, the star pattern $S$ and the knowledge graph $\mathcal{G}$ given in Figure~\ref{fig:example}. The star pattern-based selector function $s_{(S,\emptyset)}(\mathcal{G})$ retrieves
the three triples from $\mathcal{G}$ that include \textsf{dbr:Jens\_Bratlie} as subject, as shown in Figure~\ref{fig:queryD}.

\newbox\mybox
\begin{lrbox}{\mybox}
	\begin{lstlisting}[basicstyle=\tiny\sffamily,language=sparql,mathescape=true]
S = {(?p2, dbo:country, dbr:Norway),  
      (?p2, dbo:award, ?a),            
      (?p2, dbo:birthDate, ?bd2)}
      
$\mu$(?p2)=dbr:Jens_Bratlie
$\mu$(?db2)=1856-1-17
$\mu$(?a)=dbr:Order_of_St._Olav

$s_{(S,\emptyset)}(\mathcal{G})$={(dbr:Jens_Bratlie, dbo:country, dbr:Norway),  
     (dbr:Jens_Bratlie, dbo:award, dbr:Order_of_St._Olav),            
     (dbr:Jens_Bratlie, dbo:birthDate, 1856-1-17)}
	\end{lstlisting}
\end{lrbox}

\begin{figure*}[htb]
	\raisebox{3cm}{\subfloat[$S$ and $s_{(S,\emptyset)}(\mathcal{G})$]{\label{fig:queryD}\usebox\mybox}}
	\hspace*{-0.2cm}\subfloat[RDF Graph]{\label{fig:triples}\includegraphics[width=0.5\textwidth]{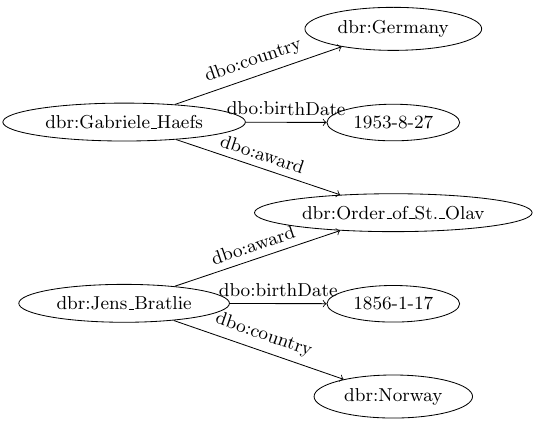}}
	\caption{Star Pattern, Star Pattern-Based Selector Function, and RDF Graph} 
	\label{fig:example}
\end{figure*}

Formally, SPF adapts the general definition of LDF given in \cite{DBLP:journals/ws/VerborghSHHVMHC16}.
%
%
Given a maximum number of distinct solution mappings that can be sent to the server $\mathtt{maxMpR}$, an SPF is defined as follows:

\begin{definition}[Star Pattern Fragment] \label{def:spf} Given a control c, a c-specific LDF collection F is called a Star Pattern Fragment collection if, for every possible star pattern $S$ and any finite sequence $\Omega$ of at most $\mathtt{maxMpR}$ distinct solution mappings, there exists one LDF $\langle u,s,\Gamma,M,C\rangle\in F$, called a Star Pattern Fragment, that has the following properties:
\begin{enumerate}
  \item $s$ is the star pattern-based selector function for $S$ and $\Omega$.
  \item There exists a triple \textsf{$<$u, void:triples, cnt$>$} $\in$ \textit{M} with \textit{cnt} representing an estimate of the cardinality of $\Gamma$, that is, \textit{cnt} is an integer that has the following two properties:
  \begin{enumerate}
  	\item If $\Gamma=\emptyset$, then \textit{cnt} = 0.
  	\item If $\Gamma\neq\emptyset$, then \textit{cnt} $>$ 0 and $abs(|\Gamma|-cnt)\leq\epsilon$ for some F-specific threshold $\epsilon$.
  \end{enumerate}	
  \item $c\in C$.
\end{enumerate}		
\end{definition}

Notice that SPF, like TPF and brTPF, is hypermedia and therefore contains hypermedia controls (Definition~\ref{def:hypermediacontrols}).
An SPF can be obtained by forming a request from a star pattern and including already bound values (e.g., object values).
Furthermore, by obtaining an arbitrary SPF from the server, it is possible to directly reach all other SPFs spanning all triples in the knowledge graph and all possible star patterns.
To account for the pagination of the results, the remainder of the paper uses the notation for LDF pages introduced in Section~\ref{sec:ldfs_prel} when describing and using SPF pages.

The query semantics of a BGP over an SPF collection follows logically from the query semantics of TPF and brTPF.
Given that the answer to a BGP $B$ over a knowledge graph $\mathcal{G}$ is denoted $[[B]]_{\mathcal{G}}$, the answer to $B$ over an SPF collection $F$ over $\mathcal{G}$ is determined by the following query semantics.

\begin{definition}[Query Semantics~\citep{DBLP:journals/ws/VerborghSHHVMHC16}]
Given a knowledge graph $\mathcal{G}$ and some SPF collection $F$ over $\mathcal{G}$, the evaluation of a BGP $B$ over $F$, denoted by $[[B]]_{F}$, is $[[B]]_{F}=[[B]]_{\mathcal{G}}$.
\end{definition}

The definition of SPF, and its hypermedia controls, allows for both subject-based and object-based star patterns to be evaluated on the server.
This allows the client to employ a complex decomposition strategy that can utilize both types of star patterns.
However, in order to investigate the applicability of the model independently of possibly complex query decomposition strategies that would be necessary on the client if both types of star patterns are considered, and since subject-based star patterns are much more common in real query loads~\cite{10.1145/3308558.3313556}, the rest of the paper will focus on subject-based star patterns only.

\subsection{Hypermedia Controls}
As previously mentioned, SPF is hypermedia, and a response to a star pattern request must thus contain controls to access other SPFs of the same collection.
The response to an SPF request consists of three fields: data, metadata and controls.

\textbf{Data.}
The data field of an SPF response is $\Gamma$, i.e., the result of applying $s_{(sp,\Omega)}(\mathcal{G})$ to $\mathcal{G}$, however, it should be paged according to Section~\ref{sec:fd}.
The metadata should thus contain pointers to other pages within the same SPF collection (i.e., next and previous pages).
The data field in an SPF request consists of triples that is part of an answer to the star pattern.
Triples that answer the star pattern can be grouped into resulting stars in the response do allow for faster interpretation on the client.

\textbf{Metadata.}
The metadata field contains a set of RDF triples that are not part of the data field of the response.
Given that an SPF $f$ is obtained by the URI $u$, the estimated total number of stars in the entire fragment is represented, in each page, as the triple \textsf{$<$u, void:triples, cnt$>$} where $cnt$ is the cardinality estimation of the star pattern and has the type \texttt{xsd:integer}.

\begin{figure*}[htb]
\begin{lstlisting}[captionpos=b, caption=Example of the controls of an SPF request and an example SPF request applying the template provided in the controls to $S$ in Figure~\ref{fig:queryD}, label=lst:controls,floatplacement=tbp,basicstyle=\footnotesize\ttfamily,mathescape=true]
<http://example.org/dbpedia#dataset>
        a                   void:Dataset , hydra:Collection ;
        void:subset         <http://example.org/dbpedia> ;
        hydra:search        [ hydra:mapping   [ hydra:property  rdf:subject ; hydra:variable  "s" ] ;
                              hydra:mapping   [ hydra:property  xsd:integer ; hydra:variable  "triples" ] ;
                              hydra:mapping   [ hydra:property  xsd:string ; hydra:variable  "star" ] ;
                              hydra:mapping   [ hydra:property  xsd:string ; hydra:variable  "values" ] ;
                              hydra:template  "http://example.org/dbpedia{?s,triples,star,values}" ] .
                              
http://example.org/dbpedia?triples=3&star=[p1,dbo:country;o1,dbr:norway;p2,dbo:award;p3,dbo:birthDate]
\end{lstlisting}
\end{figure*}

\textbf{Controls.}
The controls of an SPF is described with the Hydra Core Vocabulary~\cite{DBLP:conf/www/LanthalerG13a} as templated URIs~\cite{DBLP:books/sp/Wilde11}.
An example of such controls, as well as an example of an SPF request applying the template obtained from the controls to the star pattern $S$ in Figure~\ref{fig:queryD}, can be seen in Listing~\ref{lst:controls}.
The template for an SPF request has the following fields:
\begin{itemize}
\item \textbf{subject} (line 4): Since the paper focuses on subject-based star patterns, the subject of each triple pattern is the same, and thus just has one field in the template.
Accommodating for object-based star patterns can easily be done by renaming this field to \textit{vertex} and adding a field describing whether the vertex is a subject or object.
\item \textbf{triples} (line 5): The number of triple patterns in the star pattern.
\item \textbf{star} (line 6): Grouped predicate/object values.
In the case of object-based star patterns, this would instead be subject/predicate values.
\item \textbf{values} (line 7): In the case of already bound variables, this field can be set with the same syntax as the \texttt{VALUES} field in a SPARQL query.
\end{itemize}

While this section contained the most important aspects of an SPF response, the full SPF specification with examples can be found on the SPF website\footnote{\url{http://relweb.cs.aau.dk/spf}}.

\section{Query Processing}
\label{sec:queryprocessing}
The SPF interface processes queries using resources from both the server and the client.  
The server provides fragments as answers to requests whereas the client 
processes all other SPARQL operators. 
Differently from RDF interfaces such as TPF and brTPF, SPF does not define fragments based on triple patterns but rather based on star patterns. 

Query processing using SPF relies on a server and a client, each managing different tasks.
The general outline of how query processing works for a given SPARQL query $Q$ is as follows:

\begin{enumerate}
\item For each BGP $B\in Q$, decompose $B$ into star-shaped subqueries and determine the join order.
\item Find the first page of the SPF for each of $B$'s subqueries and select the subquery with the lowest cardinality estimation.
\item Compute the $BGP$ result by, for each star pattern $S$ in $B$, incrementally updating the set of bindings by processing $S$ on the server, and using these bindings for subsequent star patterns.
\item Compute the query result by processing all the remaining SPARQL operators in $Q$ on the client.
\end{enumerate}

\subsection{Client-Side Query Processing}
\label{sec:client_side}
To process a SPARQL query, an SPF client first decomposes the query into star-shaped subqueries.
This decomposition is necessary to process more complex SPARQL queries than star-shaped queries using an SPF server.
The rest of this section focuses on Basic Graph Pattern (BGP)\footnote{A BGP is a set of triple patterns, \url{https://www.w3.org/TR/rdf-sparql-query/\#BasicGraphPatterns}} queries. 
Nevertheless, SPF can be used for full SPARQL specification including queries with one or more BGPs combined using operators such as \texttt{OPTIONAL} and \texttt{UNION} and queries with \texttt{FILTER} constraints (experiments in Section~\ref{subsec:load} includes queries with the \texttt{OPTIONAL}, \texttt{UNION} and \texttt{FILTER} operators).
However, this section will not go into detail on such queries since BGPs are the focus of SPF.




\begin{definition}[Subject-Based Star Decomposition]\label{def:stardecomposition}
Given a BGP $B=\{tp_1,\dots,tp_n\}$ with subjects $B_S=\{s_1,\dots,s_m\}$, a subject-based star decomposition of $B$, $\mathcal{S}(B)=\{S_s\mid s\in B_S\}$, is a set of subject-based star patterns $S_s$ for each $s\in B_S$ such that $B=\cup_{s\in B_S}S_s$ where $S_s=\{tp\in B\mid\exists p,o:tp=(s,p,o)\}$.~\cite{smartkg}
\end{definition}

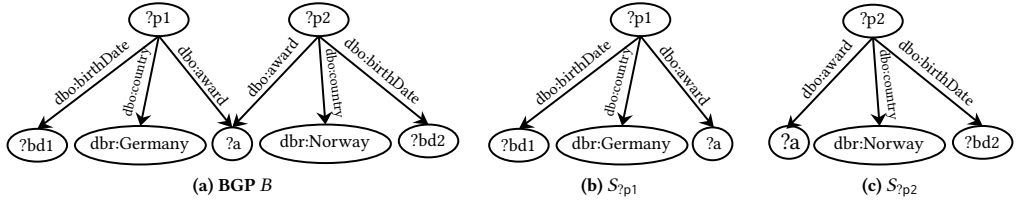
\begin{figure*}[htb]
    \centering
    \subfloat[BGP $B$]{
        \tikzset{every picture/.style={line width=0.75pt}} 

\begin{tikzpicture}[x=0.5pt,y=0.6pt,yscale=-.9,xscale=1]

\draw   (110,138.5) .. controls (110,132.15) and (120.07,127) .. (132.5,127) .. controls (144.93,127) and (155,132.15) .. (155,138.5) .. controls (155,144.85) and (144.93,150) .. (132.5,150) .. controls (120.07,150) and (110,144.85) .. (110,138.5) -- cycle ;
\draw   (231,138.5) .. controls (231,132.15) and (241.07,127) .. (253.5,127) .. controls (265.93,127) and (276,132.15) .. (276,138.5) .. controls (276,144.85) and (265.93,150) .. (253.5,150) .. controls (241.07,150) and (231,144.85) .. (231,138.5) -- cycle ;
\draw   (173,225.5) .. controls (173,219.15) and (179.72,214) .. (188,214) .. controls (196.28,214) and (203,219.15) .. (203,225.5) .. controls (203,231.85) and (196.28,237) .. (188,237) .. controls (179.72,237) and (173,231.85) .. (173,225.5) -- cycle ;
\draw   (69,226.5) .. controls (69,219.04) and (91.16,213) .. (118.5,213) .. controls (145.84,213) and (168,219.04) .. (168,226.5) .. controls (168,233.96) and (145.84,240) .. (118.5,240) .. controls (91.16,240) and (69,233.96) .. (69,226.5) -- cycle ;
\draw   (18,226.5) .. controls (18,220.15) and (28.07,215) .. (40.5,215) .. controls (52.93,215) and (63,220.15) .. (63,226.5) .. controls (63,232.85) and (52.93,238) .. (40.5,238) .. controls (28.07,238) and (18,232.85) .. (18,226.5) -- cycle ;
\draw   (209,225.5) .. controls (209,218.04) and (231.16,212) .. (258.5,212) .. controls (285.84,212) and (308,218.04) .. (308,225.5) .. controls (308,232.96) and (285.84,239) .. (258.5,239) .. controls (231.16,239) and (209,232.96) .. (209,225.5) -- cycle ;
\draw   (314,223.5) .. controls (314,217.15) and (324.07,212) .. (336.5,212) .. controls (348.93,212) and (359,217.15) .. (359,223.5) .. controls (359,229.85) and (348.93,235) .. (336.5,235) .. controls (324.07,235) and (314,229.85) .. (314,223.5) -- cycle ;
\draw    (132.5,150) -- (42.13,213.85) ;
\draw [shift={(40.5,215)}, rotate = 324.76] [fill={rgb, 255:red, 0; green, 0; blue, 0 }  ][line width=0.75]  [draw opacity=0] (10.72,-5.15) -- (0,0) -- (10.72,5.15) -- (7.12,0) -- cycle    ;

\draw    (132.5,150) -- (118.93,211.05) ;
\draw [shift={(118.5,213)}, rotate = 282.53] [fill={rgb, 255:red, 0; green, 0; blue, 0 }  ][line width=0.75]  [draw opacity=0] (10.72,-5.15) -- (0,0) -- (10.72,5.15) -- (7.12,0) -- cycle    ;

\draw    (132.5,150) -- (186.69,212.49) ;
\draw [shift={(188,214)}, rotate = 229.07] [fill={rgb, 255:red, 0; green, 0; blue, 0 }  ][line width=0.75]  [draw opacity=0] (10.72,-5.15) -- (0,0) -- (10.72,5.15) -- (7.12,0) -- cycle    ;

\draw    (253.5,150) -- (189.43,212.6) ;
\draw [shift={(188,214)}, rotate = 315.65999999999997] [fill={rgb, 255:red, 0; green, 0; blue, 0 }  ][line width=0.75]  [draw opacity=0] (10.72,-5.15) -- (0,0) -- (10.72,5.15) -- (7.12,0) -- cycle    ;

\draw    (253.5,150) -- (258.34,210.01) ;
\draw [shift={(258.5,212)}, rotate = 265.39] [fill={rgb, 255:red, 0; green, 0; blue, 0 }  ][line width=0.75]  [draw opacity=0] (10.72,-5.15) -- (0,0) -- (10.72,5.15) -- (7.12,0) -- cycle    ;

\draw    (253.5,150) -- (334.9,210.8) ;
\draw [shift={(336.5,212)}, rotate = 216.76] [fill={rgb, 255:red, 0; green, 0; blue, 0 }  ][line width=0.75]  [draw opacity=0] (10.72,-5.15) -- (0,0) -- (10.72,5.15) -- (7.12,0) -- cycle    ;

\draw (132,138) node  [scale=0.77,align=left] {?p1};
\draw (253,138) node  [scale=0.77,align=left] {?p2};
\draw (189,225) node  [scale=0.77,align=left] {?a};
\draw (118,226) node  [scale=0.77,align=left] {dbr:Germany};
\draw (40,226) node  [scale=0.77,align=left] {?bd1};
\draw (258,225) node  [scale=0.77,align=left] {dbr:Norway};
\draw (336,223) node  [scale=0.77,align=left] {?bd2};
\draw (116,182) node [scale=0.6,rotate=-285] [align=left] {dbo:country};
\draw (81,175) node [scale=0.7,rotate=-320] [align=left] {dbo:birthDate};
\draw (265,183) node [scale=0.6,rotate=-81.87] [align=left] {dbo:country};
\draw (302,174) node [scale=0.7,rotate=-40] [align=left] {dbo:birthDate};
\draw (218,172) node [scale=0.7,rotate=-310] [align=left] {dbo:award};
\draw (166,177) node [scale=0.7,rotate=-52] [align=left] {dbo:award};

\end{tikzpicture}
        \label{fig:decomp1}
    }\hspace*{.5ex}
    \subfloat[$S_\mathtt{?p1}$]{
        \tikzset{every picture/.style={line width=0.75pt}} 

\begin{tikzpicture}[x=0.5pt,y=0.6pt,yscale=-.9,xscale=1]

\draw   (130,158.5) .. controls (130,152.15) and (140.07,147) .. (152.5,147) .. controls (164.93,147) and (175,152.15) .. (175,158.5) .. controls (175,164.85) and (164.93,170) .. (152.5,170) .. controls (140.07,170) and (130,164.85) .. (130,158.5) -- cycle ;
\draw   (192,245.5) .. controls (192,239.15) and (198.72,234) .. (207,234) .. controls (215.28,234) and (222,239.15) .. (222,245.5) .. controls (222,251.85) and (215.28,257) .. (207,257) .. controls (198.72,257) and (192,251.85) .. (192,245.5) -- cycle ;
\draw   (89,246.5) .. controls (89,239.04) and (111.16,233) .. (138.5,233) .. controls (165.84,233) and (188,239.04) .. (188,246.5) .. controls (188,253.96) and (165.84,260) .. (138.5,260) .. controls (111.16,260) and (89,253.96) .. (89,246.5) -- cycle ;
\draw   (38,246.5) .. controls (38,240.15) and (48.07,235) .. (60.5,235) .. controls (72.93,235) and (83,240.15) .. (83,246.5) .. controls (83,252.85) and (72.93,258) .. (60.5,258) .. controls (48.07,258) and (38,252.85) .. (38,246.5) -- cycle ;
\draw    (152.5,170) -- (62.13,233.85) ;
\draw [shift={(60.5,235)}, rotate = 324.76] [fill={rgb, 255:red, 0; green, 0; blue, 0 }  ][line width=0.75]  [draw opacity=0] (10.72,-5.15) -- (0,0) -- (10.72,5.15) -- (7.12,0) -- cycle    ;

\draw    (152.5,170) -- (138.93,231.05) ;
\draw [shift={(138.5,233)}, rotate = 282.53] [fill={rgb, 255:red, 0; green, 0; blue, 0 }  ][line width=0.75]  [draw opacity=0] (10.72,-5.15) -- (0,0) -- (10.72,5.15) -- (7.12,0) -- cycle    ;

\draw    (152.5,170) -- (206.69,232.49) ;
\draw [shift={(208,234)}, rotate = 229.07] [fill={rgb, 255:red, 0; green, 0; blue, 0 }  ][line width=0.75]  [draw opacity=0] (10.72,-5.15) -- (0,0) -- (10.72,5.15) -- (7.12,0) -- cycle    ;

\draw (152,158) node  [scale=0.77,align=left] {?p1};
\draw (208,245) node  [scale=0.77,align=left] {?a};
\draw (138,246) node  [scale=0.77,align=left] {dbr:Germany};
\draw (60,246) node  [scale=0.77,align=left] {?bd1};
\draw (136,202) node [scale=0.6,rotate=-285] [align=left] {dbo:country};
\draw (101,195) node [scale=0.7,rotate=-320] [align=left] {dbo:birthDate};
\draw (186,197) node [scale=0.7,rotate=-52] [align=left] {dbo:award};

\end{tikzpicture}
        \label{fig:decomp2}
    }\hspace*{.5ex}
    \subfloat[$S_\mathtt{?p2}$]{
        \tikzset{every picture/.style={line width=0.75pt}} 

\begin{tikzpicture}[x=0.5pt,y=0.6pt,yscale=-.9,xscale=1]

\draw   (314,178.5) .. controls (314,172.15) and (324.07,167) .. (336.5,167) .. controls (348.93,167) and (359,172.15) .. (359,178.5) .. controls (359,184.85) and (348.93,190) .. (336.5,190) .. controls (324.07,190) and (314,184.85) .. (314,178.5) -- cycle ;
\draw   (292,265.5) .. controls (292,258.04) and (314.16,252) .. (341.5,252) .. controls (368.84,252) and (391,258.04) .. (391,265.5) .. controls (391,272.96) and (368.84,279) .. (341.5,279) .. controls (314.16,279) and (292,272.96) .. (292,265.5) -- cycle ;
\draw   (397,263.5) .. controls (397,257.15) and (407.07,252) .. (419.5,252) .. controls (431.93,252) and (442,257.15) .. (442,263.5) .. controls (442,269.85) and (431.93,275) .. (419.5,275) .. controls (407.07,275) and (397,269.85) .. (397,263.5) -- cycle ;
\draw    (336.5,190) -- (272.43,252.6) ;
\draw [shift={(271,254)}, rotate = 315.65999999999997] [fill={rgb, 255:red, 0; green, 0; blue, 0 }  ][line width=0.75]  [draw opacity=0] (10.72,-5.15) -- (0,0) -- (10.72,5.15) -- (7.12,0) -- cycle    ;

\draw    (336.5,190) -- (341.34,250.01) ;
\draw [shift={(341.5,252)}, rotate = 265.39] [fill={rgb, 255:red, 0; green, 0; blue, 0 }  ][line width=0.75]  [draw opacity=0] (10.72,-5.15) -- (0,0) -- (10.72,5.15) -- (7.12,0) -- cycle    ;

\draw    (336.5,190) -- (417.9,250.8) ;
\draw [shift={(419.5,252)}, rotate = 216.76] [fill={rgb, 255:red, 0; green, 0; blue, 0 }  ][line width=0.75]  [draw opacity=0] (10.72,-5.15) -- (0,0) -- (10.72,5.15) -- (7.12,0) -- cycle    ;

\draw   (258,263.5) .. controls (258,257.15) and (264.72,252) .. (273,252) .. controls (281.28,252) and (288,257.15) .. (288,263.5) .. controls (288,269.85) and (281.28,275) .. (273,275) .. controls (264.72,275) and (258,269.85) .. (258,263.5) -- cycle ;

\draw (336,178) node  [scale=0.77,align=left] {?p2};
\draw (341,265) node  [scale=0.77,align=left] {dbr:Norway};
\draw (419,263) node  [scale=0.77,align=left] {?bd2};
\draw (348,223) node [scale=0.6,rotate=-81.87] [align=left] {dbo:country};
\draw (385,214) node [scale=0.7,rotate=-40] [align=left] {dbo:birthDate};
\draw (301,212) node [scale=0.7,rotate=-310] [align=left] {dbo:award};
\draw (274,263) node  [align=left] {?a};

\end{tikzpicture}
        \label{fig:decomp3}
    }
    \caption{Subject-based star decomposition of $Q$ (Listing~\ref{lst:q1}) into $S_\mathtt{?p1}$ and $S_\mathtt{?p2}$.}\label{fig:decomp}
\end{figure*}

Using Definition~\ref{def:stardecomposition}, a BGP query can be partitioned into a set of star patterns where each star pattern corresponds to a specific value on subject position.
All triple patterns are then part of a specific star pattern with a shared subject.
%
SPF uses a greedy decomposition algorithm that iterates over the triple patterns in the BGP and has linear complexity. 
This algorithm always finds the largest possible star patterns and ensures that the query is decomposed into non-overlapping star patterns. 
SPF thus decomposes singular triple patterns (i.e., triple patterns with unique subjects) into star patterns consisting of just one triple pattern; processing such a singular star pattern is in line with processing the triple pattern individually.

An example of using Definition~\ref{def:stardecomposition} to partition a BGP query $Q$ (Listing~\ref{lst:q1}) is illustrated in Figure~\ref{fig:decomp}.
The star decomposition of $Q$ results in one star pattern per variable on subject position that includes all triple patterns with the corresponding subject value (i.e., the largest possible star patterns).
In this example, variables \texttt{?p1} and \texttt{?p2} are both positioned as the subject of at least one triple pattern, and so the resulting star patterns are rooted in these variables. Figures~\ref{fig:decomp2} and~\ref{fig:decomp3} show the output star patterns $S_\mathtt{?p1}$ and $S_\mathtt{?p2}$, respectively.

\begin{algorithm*}[htb]
\caption{Evaluate a BGP on an SPF client}
\label{algo:general}
\begin{algorithmic}[1]
\Statex \textbf{Input:} A BGP $B=\{tp_1,\dots,tp_n\}$; a control $c$ of a $c$-specific SPF collection $F$; a set of solution mappings $\Omega$
\Statex \textbf{Output:} A set of solution mappings $[[B]]_F$
\Function{evaluateBGP}{$B$,$c$,$\Omega=\emptyset$}
\If{$B=\emptyset$}
  \State \Return $\Omega$;
\EndIf
\State $S_B\leftarrow\mathcal{S}(B)$ such that $\mathcal{S}(B)=\{S_{s_1},\dots,S_{s_k}\}$ is the result of applying Definition~\ref{def:stardecomposition} to $B$;\label{algo:sd}
\ForAll{$S_i\in S_B$}\label{algo:cards}
  \State $\phi^i_1=\langle u^j_1,u_j,s,\Gamma^j_1,M^j_1,C^j_1\rangle\leftarrow \mathtt{GET}\, c(S_i,\Omega)$ resulting in page 1 of the SPF for $S_i$ and $\Omega$;\label{algo:get}
  \State $cnt_j\leftarrow cnt$ where $\langle u_j,\mathtt{void:triples},cnt\rangle\in M^j_1$;
  \If{$cnt_j=0$}
    \State \Return $\Omega$;
  \EndIf
\EndFor\label{algo:carde}
\State $S_\epsilon\leftarrow S_k$ where $S_k\in S_B$ and $cnt_k\leq cnt_j$ for all $S_j\in S_B$;\label{algo:card}
\State $\phi^\epsilon\leftarrow\{\phi^k_1,\phi^\epsilon_2,\dots,\phi^\epsilon_l\}$ through $\mathtt{GET}$ requests for each page $\phi^\epsilon_p$ over $\Omega$ using controls from $\phi^\epsilon_{p-1}$;\label{algo:findall}
\State $\Gamma^\epsilon\leftarrow\bigcup_{\langle u^\epsilon_l,u_\epsilon,s,\Gamma^\epsilon_l,M^\epsilon_l,C^\epsilon_l\rangle\in \phi^\epsilon}\Gamma^\epsilon_l$;
\State $\Omega^\epsilon\leftarrow \Omega\bowtie\{\mu\mid dom(\mu)=vars(S_\epsilon)\text{ and }\mu[S_\epsilon]\in\Gamma^\epsilon\}$;\label{algo:merge}
\State $B'\leftarrow B\setminus S_\epsilon$;
\State \Return $evaluateBGP(B',c,\Omega^\epsilon)$;\label{algo:recurs}
\EndFunction
\end{algorithmic}
\end{algorithm*}

Let $dom(\mu)$ be a function that returns the \textit{domain} of $\mu$ (i.e., the set of variables that are bound in $\mu$) and $vars(S)$ be a function that returns the set of all variables in a star pattern $S$.


Given a control $c$ obtained from an arbitrary fragment on the SPF server and a BGP $B$, the general approach to processing a BGP is shown in Algorithm~\ref{algo:general}.
This algorithm, while similar to the general approach for TPF (Listing 3 in~\cite{DBLP:journals/ws/VerborghSHHVMHC16}), has several key differences to account for due to the nature of star patterns compared to triple patterns as well as coupling bindings with the star patterns sent to the server.
The approach outlined in Algorithm~\ref{algo:general} is an illustration of how to adapt the general approach outlined by TPF to process queries over SPF recursively with a divide-and-conquer strategy.
The \texttt{maxMpR} value (Definition~\ref{def:spf}) is therefore ignored in this algorithm.
A concrete approach using iterators is shown later in this section.

First, applying the subject-based star decomposition (Definition~\ref{def:stardecomposition}) in line~\ref{algo:sd} is similar to splitting the BGP into sub-BGPs as TPF does.
However, since SPF evaluates star patterns on the server, passing each individual triple pattern into sub-BGPs to process them individually is unnecessary.
Instead, the entire set of star patterns is recursively evaluated (line~\ref{algo:recurs}), continuously expanding the set of solution mappings according to the evaluated star pattern (line~\ref{algo:merge}), while sending the incrementally updated set of bindings to the server with the request (line~\ref{algo:get}).
Since the set of obtained bindings can contain bindings for variables not present in the star pattern to be evaluated, and to avoid unnecessary data transfer to the server, $c(S_i,\Omega)$ on line~\ref{algo:get} ensures that only bindings for the variables in $S_i$ are attached to the request.
Second, since SPF couples previously obtained bindings with the star pattern before sending it to the server, Algorithm~\ref{algo:general} takes an additional argument, $\Omega$, being the set of currently obtained bindings.
The result of the algorithm is thus the accumulated set of bindings over each recursive call of the function (one recursive call per star pattern).

The algorithm starts by finding the first page of the corresponding SPFs for each star pattern in the BGP (lines~\ref{algo:cards}-\ref{algo:carde}), and selects the star pattern with the lowest cardinality estimation (line~\ref{algo:card}).
To assess the applicability of the approach regardless of potentially complex join order strategies, SPF uses the same join order strategy as TPF (i.e., based on cardinality estimations provided by the server).
Then, the algorithm finds all relevant bindings for the selected star pattern given the bindings $\Omega$ through consecutive \texttt{GET} requests to the server using controls obtained from each page to find the next page (line~\ref{algo:findall}).
The bindings found for the star pattern are joined with $\Omega$ in order to incrementally update the resulting bindings (line~\ref{algo:merge}).
Last, 
a recursive call is made, giving as argument the remaining BGP (minus the selected star pattern) and the newly obtained bindings (line~\ref{algo:recurs}).

Take, as an example, the BGP $B$ in Listing~\ref{lst:q1}, and assume a control $c$ was obtained from an SPF server giving access to DBpedia version 2016-04~\cite{dbpedia}.
Applying subject-based star decomposition (Definition~\ref{def:stardecomposition}) to $B$ results in $S_\mathtt{?p1}$ and $S_\mathtt{?p2}$ from Figure~\ref{fig:decomp}.
While the cardinalities of each individual triple pattern are large, the cardinality of $S_\mathtt{?p1}$ is 13 and the cardinality of $S_\mathtt{?p2}$ is 71.

When calling $evaluateBGP(B,c)$, the first step is to obtain the first pages of the SPFs for both star patterns and select the star pattern with the lowest cardinality; in this case $S_\mathtt{?p1}$.
The 13 resulting bindings from $S_\mathtt{?p1}$, are then joined with the (currently empty) set $\Omega$.
Then, the function is called recursively with $S_\mathtt{?p1}$ removed from $B$ (i.e., $S_\mathtt{?p2}$). 
The bindings obtained from $S_\mathtt{?p2}$,
are then joined with the ones obtained from $S_\mathtt{?p1}$
and returned as the result to the BGP query.

The following presents a concrete approach to process a BGP with an SPF client that follows the general approach outlined in Algorithm~\ref{algo:general} and uses the iterator pattern presented by Verborgh et al.~\cite{DBLP:journals/ws/VerborghSHHVMHC16}.
A \texttt{RootIterator} returns an empty binding on the first call and \texttt{nil} on subsequent calls.

SPF provides a \texttt{StarPatternIterator} (Algorithm~\ref{algo:spi}), similar to Listing 5 in~\cite{DBLP:journals/ws/VerborghSHHVMHC16}, which, distinctly from the iterator provided in~\cite{DBLP:journals/ws/VerborghSHHVMHC16}, finds a set of solution mappings rather than a single solution mapping.
This is due to the fact that SPF bulks obtained bindings into groups of \texttt{maxMpR} bindings and forwards those to the server along with the next star patterns to obtain.
A \texttt{StarPatternIterator} has two members: $\phi$, the current SPF page, and $\Omega_s$, the most recently read set of \texttt{maxMpR} solution mappings.
If the iterator has already read one or more SPF pages, the next page will be obtained using the controls from the previous page (line~\ref{algo:getpage}).
However, if there is no such control, or the first page has not yet been read, the iterator will retrieve the next set of at most \texttt{maxMpR} solution mappings (this is the case since all iterators are restricted to return sets of at most $\mathtt{maxMpR}$ solution mappings) from the source iterator $I_s$ (line~\ref{algo:getfirstpage}) and use those to obtain the next page (line~\ref{algo:getnextfirstpage}).
After a page has been found, the iterator will attempt to return solution mappings.
Instead of finding one solution mapping, it will iterate through the current page until \texttt{maxMpR} solution mappings have been found, and return those as a set (lines~\ref{algo:bulks}-\ref{algo:bulke}).

Consider, for example, if the star pattern is $S_\mathtt{?p1}$ from Figure~\ref{fig:decomp} with a \texttt{maxMpR} of 50.
In this case, since it is the first evaluated star pattern, the source iterator $I_s$ would be a \texttt{RootIterator} and return an empty set of bindings.
The iterator will therefore request $S_\mathtt{?p1}$ with an empty set of bindings, and thus retrieve the 13 resulting stars.
Since $13<\mathtt{maxMpR}$, all these bindings will be grouped together and returned as a set.

\begin{algorithm*}[htb]
\caption{Star pattern iterator on an SPF client}
\label{algo:spi}
\begin{algorithmic}[1]
\Statex \textbf{Input:} A source iterator $I_s$; a star pattern $S$; a control $c$ of a $c$-specific SPF collection $F$; a maximum amount of distinct solution mapping per request $\mathtt{maxMpR}$
\Statex \textbf{Output:} The next set of solution mappings $\Omega'$ such that $|\Omega'|\leq\mathtt{maxMpR}$, or \texttt{nil} if no such mappings are left
\Function{StarPatternIterator.GetNext()}{}
\If{$\mathtt{self}.\phi$ has not been assigned to previously}
  \State $\mathtt{self}.\phi\leftarrow$ an empty page with no stars or controls;
\EndIf
\While{$\mathtt{self}.\phi$ does not contain unread stars}
  \If{$\mathtt{self}.\phi$ has a control to a next page with URI $u_{\phi'}$}
    \State $\mathtt{self}.\phi\leftarrow \mathtt{GET}\,u_{\phi'}$;\label{algo:getpage}
  \Else
    \State $\mathtt{self}.\Omega_s\leftarrow I_s.\mathtt{GetNext()}$;\label{algo:getfirstpage}
    \State \textbf{if} $\mathtt{self}.\Omega_s=\mathtt{nil}$ \textbf{then} \Return \texttt{nil};
    \State $\mathtt{self}.\phi\leftarrow\mathtt{GET}\, c(S,\mathtt{self}.\Omega_s)$ resulting in page 1 of the SPF for $S$ and $\mathtt{self}.\Omega_s$;\label{algo:getnextfirstpage}
  \EndIf
\EndWhile
\State $\Omega_o\leftarrow\emptyset$;
\While{$|\Omega_o|<\mathtt{maxMpR}$ and $\mathtt{self}.\phi$ contains unread stars}\label{algo:bulks}
  \State $s\leftarrow$ an unread star from $\mathtt{self}.\phi$;
  \State $\mu\leftarrow$ a solution mapping such that $dom(\mu)=vars(S)$ and $\mu[S]=s$;
  \State $\Omega_o\leftarrow\Omega_o\cup\{\mu\}$
\EndWhile
\State \Return $\Omega_o\bowtie\mathtt{self}.\Omega_s$\label{algo:bulke}
\EndFunction
\end{algorithmic}
\end{algorithm*}

\begin{algorithm*}[htb]
\caption{BGP iterator on an SPF client}
\label{algo:bgpi}
\begin{algorithmic}[1]
\Statex \textbf{Input:} A source iterator $I_s$; a BGP $B$ with $|\mathcal{S}(B)|\geq 2$; a control $c$ of a $c$-specific SPF collection $F$;  a maximum amount of distinct solution mapping per request $\mathtt{maxMpR}$
\Statex \textbf{Output:} The next set of solution mappings $\Omega'$ such that $|\Omega'|\leq\mathtt{maxMpR}$, or \texttt{nil} if no such mappings are left
\Function{BasicGraphPatternIterator.GetNext()}{}
\State \textbf{if} $\mathtt{self}.I_p$ has not been assigned to previously \textbf{then} $\mathtt{self}.I_p\leftarrow\mathtt{nil}$;
  \While{$\mathtt{self}.I_p=\mathtt{nil}$}
    \State $\mathtt{self}.\Omega_s\leftarrow I_s.\mathtt{GetNext()}$;
    \State \textbf{if} $\mathtt{self}.\Omega_s=\mathtt{nil}$ \textbf{then} \Return \texttt{nil};
    \ForAll{star patterns $S_j\in\mathcal{S}(B)$}\label{algo:sels}
      \State $\phi^j_1=\langle u^j_1,u_j,s,\Gamma^j_1,M^j_1,C^j_1\rangle\leftarrow \mathtt{GET}\, c(S_j,\mathtt{self}.\Omega_s)$ resulting in page 1 of that SPF;
      \State $cnt_j\leftarrow cnt$ where $\langle u_j,\mathtt{void:triples},cnt\rangle\in M^j_1$;
    \EndFor
    \If{$\forall j:cnt_j>0$}
      \State $\epsilon\leftarrow j$ such that $cnt_j\leq cnt_k\forall S_k\in\mathcal{S}(B)$;\label{algo:sele}
      \State $I_\epsilon\leftarrow\mathtt{StarPatternIterator}(\mathtt{RootIterator}(),S_\epsilon,c,\mathtt{maxMpR})$;\label{algo:createspit}
      \State $\mathtt{self}.I_p\leftarrow\mathtt{BasicGraphPatternIterator}(I_\epsilon,B\setminus S_\epsilon,c,\mathtt{maxMpR})$;\label{algo:createbgpit}
    \EndIf
  \EndWhile
\State \Return $\mathtt{self}.I_p.\mathtt{GetNext}()$;\label{algo:getnextt}
\EndFunction
\end{algorithmic}
\end{algorithm*}

SPF defines a \texttt{BasicGraphPatternIterator} (Algorithm~\ref{algo:bgpi}), similarly to Listing 6 in~\cite{DBLP:journals/ws/VerborghSHHVMHC16}, which in a similar fashion to the \texttt{StarPatternIterator} returns a set of at most \texttt{maxMpR} solution mappings rather than a single solution mapping at a time.
If the BGP contains no subject-based star pattern (i.e., is empty), the \texttt{BasicGraphPatternIterator} constructor creates a \texttt{RootIterator}.
If, instead, the BGP consists of only a single subject-based star pattern, the constructor creates a \texttt{StarPatternIterator}.
Given a BGP $B$, the \texttt{BasicGraphPatternIterator} creates a chained pipeline of iterators, which will incrementally call each other to obtain a set of solution mappings.
It has two member variables: $I_p$, the current iterator pipeline, and $\Omega_s$, the most recently read set of \texttt{maxMpR} solution mappings.
The \texttt{BasicGraphPatternIterator} creates the pipeline by, at each step, selecting the star pattern with the lowest cardinality (lines~\ref{algo:sels}-\ref{algo:sele}).
For the selected star pattern, a \texttt{StarPatternIterator} is created (line~\ref{algo:createspit}), and for the remaining BGP a new \texttt{BasicGraphPatternIterator} is created (line~\ref{algo:createbgpit}).
The solution mappings returned from this pipeline are then returned (line~\ref{algo:getnextt}).

\begin{algorithm*}[htb]
\caption{Query iterator on an SPF client}
\label{algo:qi}
\begin{algorithmic}[1]
\Statex \textbf{Data:} A BGP $B$; a control $c$ of a $c$-specific SPF collection $F$; a maximum amount of distinct solution mapping per request $\mathtt{maxMpR}$
\Statex \textbf{Output:} The next mapping $\mu'$ such that $\mu'\in[[B]]_F$, or \texttt{nil} if no mappings are left
\Function{QueryIterator.GetNext()}{}
\If{$\mathtt{self}.I_B$ has not been assigned to previously}
  \State $\mathtt{self}.I_B\leftarrow\mathtt{BasicGraphPatternIterator}(\mathtt{RootIterator}(),B,c,\emptyset,\mathtt{maxMpR})$;\label{algo:createbgpi}
\EndIf
\While{$\mathtt{self}.\Omega_s=\emptyset$ or $\mathtt{self}.\Omega_s$ has not been assigned to previously}\label{algo:qis}
  \State $\mathtt{self}.\Omega_s\leftarrow\mathtt{self}.I_B.\mathtt{GetNext}()$;\label{algo:getnexttt}
  \State \textbf{if} $\mathtt{self}.\Omega_s=\mathtt{nil}$ \textbf{then} \Return \texttt{nil};\label{algo:qie}
\EndWhile
\State $\mu\leftarrow$ a mapping such that $\mu\in\mathtt{self}.\Omega_s$;\label{algo:rets}
\State $\mathtt{self}.\Omega_s\leftarrow\mathtt{self}.\Omega_s\setminus\{\mu\}$;
\State \Return $\mu$;\label{algo:rete}
\EndFunction
\end{algorithmic}
\end{algorithm*}

Consider again the BGP query $B$ from Figure~\ref{fig:decomp}. 
Creating a \texttt{BasicGraphPatternIterator} with $B$ as its BGP will require first to look up the cardinalities of each star pattern in $B$.
In this case, $S_\mathtt{?p1}$ has the lowest cardinality of 13, so a \texttt{StarPatternIterator} $I_1$ is created with $S_\mathtt{?p1}$ as its star pattern and an empty solution mapping.
This iterator is then used as the source iterator of the pipeline created on line~\ref{algo:createbgpit}.
However, since the remaining BGP (after removing $S_\mathtt{?p1}$ from $B$) only consists of a single star pattern ($S_\mathtt{?p2}$), a \texttt{StarPatternIterator} is also created for $S_\mathtt{?p2}$ as the pipeline $I_2$ ($I_p=I_2$).
This means, that when calling $I_2.\mathtt{GetNext}()$ on line~\ref{algo:getnextt}, $I_2$ will effectively call $I_1.\mathtt{GetNext}()$, ensuring that the 13 bindings from $S_\mathtt{?p1}$ will be used to obtain the bindings for $S_\mathtt{?p1}\bowtie S_\mathtt{?p2}$.

The \texttt{QueryIterator}, Algorithm~\ref{algo:qi}, creates a \texttt{BasicGraphPatternIterator} (line~\ref{algo:createbgpi}) and iterates over the sets of bindings obtained by calling the \texttt{GetNext()} function on the iterator (lines~\ref{algo:qis}-\ref{algo:qie}).
However, if a non-empty set of solution mappings has already been obtained from the iterator, the \texttt{QueryIterator} instead returns one of those mappings and removes it from the set (lines~\ref{algo:rets}-\ref{algo:rete}).
Consider again the example with the BGP query $B$ from Figure~\ref{fig:decomp}, the \texttt{QueryIterator} will create a \texttt{BasicGraphPatternIterator} with $B$ as its BGP.
This creates a pipeline with $I_2$ from above and $I_1$ as its source iterator.
When calling $I_2.\mathtt{GetNext}()$ on line~\ref{algo:getnexttt}, $I_2$ will call $I_1.\mathtt{GetNext}()$, which will find the 13 bindings for $S_\mathtt{?p1}$.
$I_2$ will then use these bindings to find the first 50 results.
In this example there are just 8 results, so these are returned to $\mathtt{self}.\Omega_s$.

\subsection{Server-Side Query Processing}
\label{sec:server_side}
An SPF server is able to answer any syntactically valid star pattern.
Upon receiving a request for a star pattern, the SPF server matches the star pattern to the knowledge graph using the star pattern-based selector function.
An SPF request includes a star pattern $S$, a finite sequence of distinct solution bindings $\Omega$, and a page number $p$. The server processes such a request over a knowledge graph $\mathcal{G}$ using the following steps:
\begin{enumerate}
\item Given the star pattern $S$, find the set of corresponding stars $s_{(S,\Omega)}(\mathcal{G})$ (Definition~\ref{def:selector}).
\item Return an LDF page $\phi$ that corresponds to the requested page $p$ (LDF pages do not overlap) such that $\phi.\Gamma'$ consists of sets of matching stars. 
\end{enumerate}
These results are then processed by the client, which combines them with results from other star patterns in the query, thereby computing the query answer.
To process star patterns, the SPF server uses similar left-deep join trees as the client.
Therefore, star patterns are as efficiently processed by the SPF server as the client.

An SPF server supports both the TPF and brTPF selectors in addition to the SPF selector.
The server chooses which method to invoke based on the received request. For instance, the SPF method is invoked only if the request contains an SPF selector.
In practice, the TPF and brTPF selectors would only rarely be used with an SPF client. However, having all three methods available in the server has two advantages. First, it makes the server compatible with TPF and brTPF. Second and more importantly, SPF performs as good as brTPF in the worst case where all star patterns have exactly one triple pattern.

\subsection{Implementation Details}
The SPF server was implemented using Java 8 and the SPF client using Node.js.
The source code is available at \texttt{http://relweb.cs.aau.dk/spf}.
%

\textit{Server.}
The SPF server is implemented as an extension of the Java implementation of the TPF server\footnote{\url{https://github.com/LinkedDataFragments/Server.java}}.
The server implementation uses \texttt{HDT}~\cite{DBLP:conf/dcc/Hernandez-Illera15,FMPGPA:13} as backend. \texttt{HDT} is originally proposed to process a single triple pattern over a knowledge graph efficiently. 
However, this implementation was extended to also be able to process the star pattern requests over the \texttt{HDT} backend.
The SPF server uses Characteristic Sets~\cite{DBLP:conf/icde/NeumannM11} to provide cardinality estimations.


\textit{Client.}
The TPF Node.js client\footnote{\url{https://github.com/LinkedDataFragments/Client.js}} was extended to accommodate not only SPF requests but also brTPF requests.
Thus, and in line with TPF~\cite{DBLP:journals/ws/VerborghSHHVMHC16} and brTPF~\cite{DBLP:journals/corr/HartigA16}, the SPF client uses a pipeline of iterators that represent a left-deep join tree.
However, TPF and brTPF define the join operations on triple patterns, whereas SPF defines join operations on star patterns.
The star patterns within a query are ordered based on the cardinality estimations for the star patterns provided by the server.

\section{Experimental Evaluation}
\label{sec:evaluation}
The experimental evaluation compares SPF to TPF~\cite{DBLP:journals/ws/VerborghSHHVMHC16}, brTPF~\cite{DBLP:journals/corr/HartigA16}, SaGe~\cite{sage}, Smart-KG~\cite{smartkg}, and a SPARQL endpoint.
All source code, experimental setup (queries, datasets, etc.), as well as the full experimental results are provided on the SPF website\footnote{\url{http://relweb.cs.aau.dk/spf}}.


\subsection{Experimental Setup}
This section contains a description of the experimental setup, including a characterization of the datasets and queries used, hardware and software setup, and the measured metrics for the evaluation.

\paragraph{Dataset and Queries:}
The experiments were run using both synthetic datasets from the WatDiv benchmark~\cite{watdiv} and a real-world dataset with DBpedia~\cite{dbpedia}.
To test the scalability of SPF, four different sized WatDiv datasets were used.
Furthermore, to test SPF in a real-world setting, the English part of DBpedia 2016-04~\cite{dbpedia} was used.
The characteristics of the used datasets can be seen in Table~\ref{tbl:datasets}.

\begin{table*}[htb]
\centering
\caption{Characteristics of used datasets}\label{tbl:datasets}
\begin{tabular}{lllll}
\hline
\textbf{Dataset}     & \textbf{\#triples}    & \textbf{\#subjects}\hspace{1ex} & \textbf{\#predicates}\hspace{1ex} & \textbf{\#objects}  \\ \hline
\texttt{watdiv10M}   & 10,916,457    & 521,585     & 86            & 1,005,832   \\
\texttt{watdiv100M}  & 108,997,714   & 5,212,385   & 86            & 9,753,266   \\
\texttt{watdiv1B}\hspace{1ex} & 1,092,155,948\hspace{1ex} & 52,120,385  & 86  & 92,220,397  \\
\texttt{watdiv10B}\hspace{1ex} & 10,920,048,634\hspace{1ex} & 521,200,385  & 86  & 837,127,565  \\\hline
\texttt{dbpedia}     & 1,040,358,853 & 58,167,851  & 68,687        & 206,201,072 \\ \hline
\end{tabular}
\end{table*}

\begin{figure*}[htb]
    \centering	
    \subfloat[Triple pattern count (\#TP)]{
        \includegraphics[width=.49\textwidth]{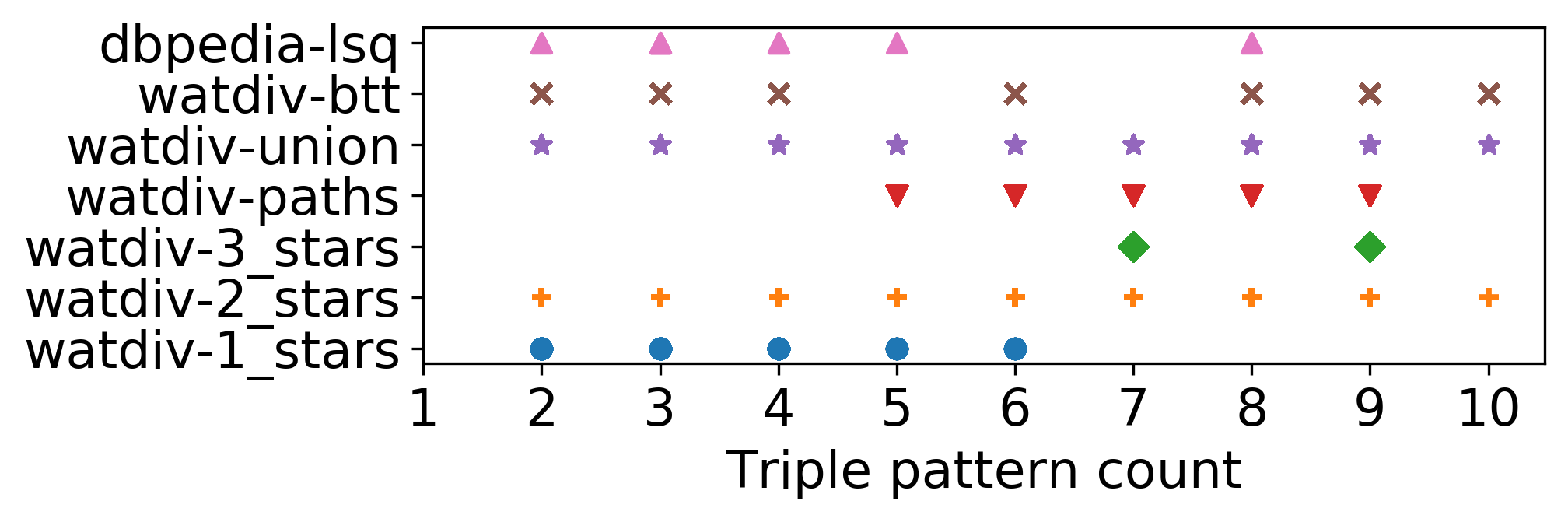}
        \label{fig:ch1}
    }
    \subfloat[Join vertex count (\#JV)]{
        \includegraphics[width=.49\textwidth]{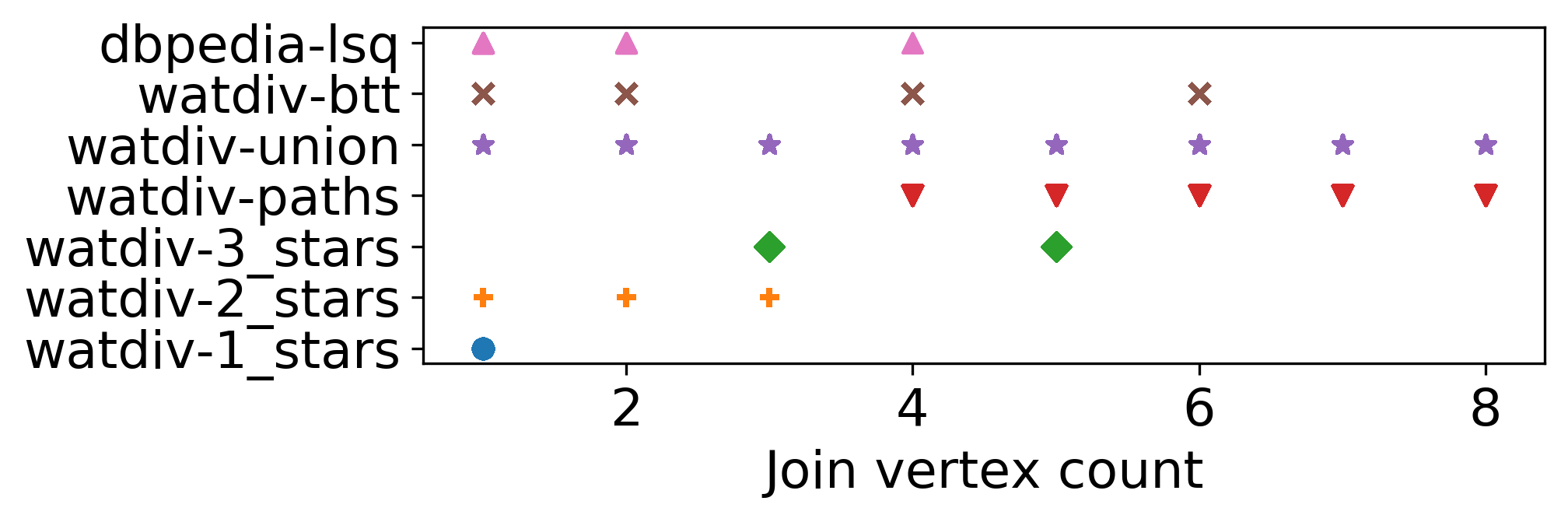}
        \label{fig:ch2}
    }\\
    \subfloat[Join vertex degree ($\textsc{deg}$)]{
        \includegraphics[width=.49\textwidth]{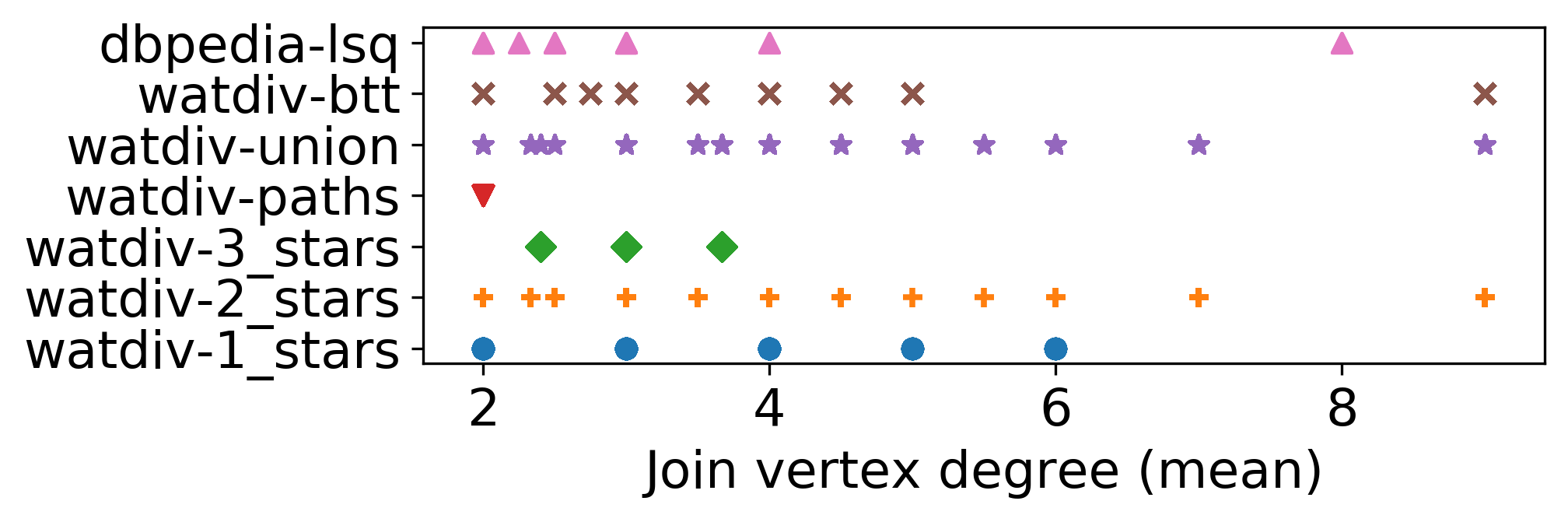}
        \label{fig:ch3}
    }
    \subfloat[Result cardinality (\#Results)]{
        \includegraphics[width=.49\textwidth]{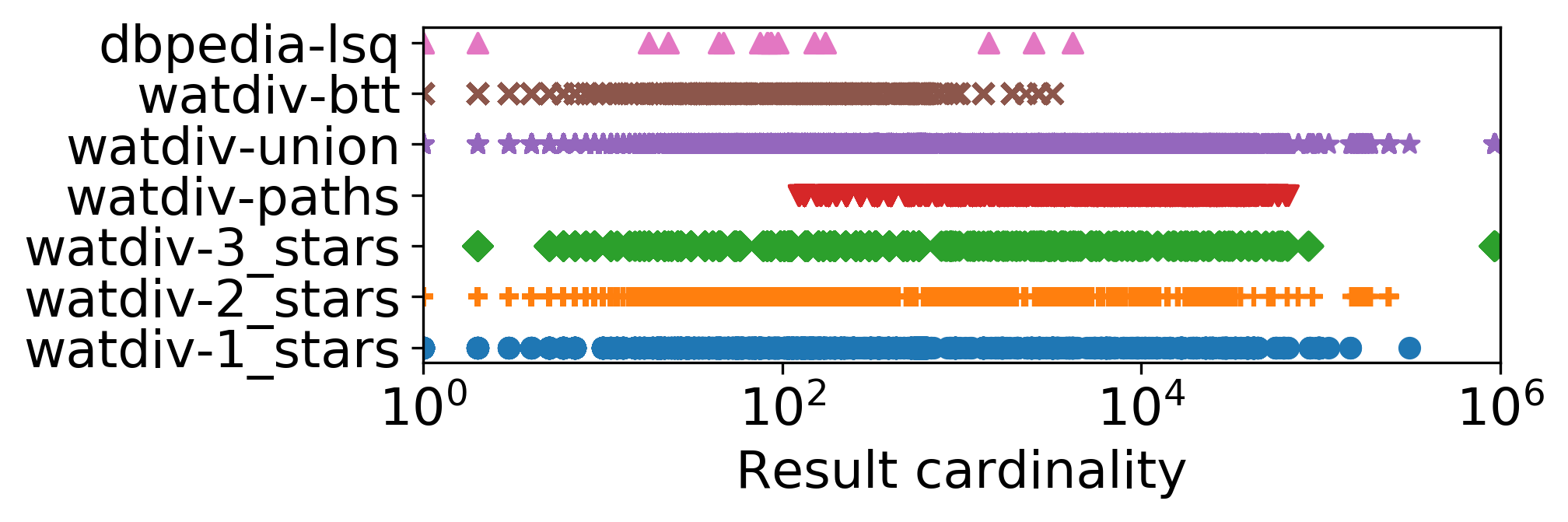}
        \label{fig:ch4}
    }\\
    \subfloat[TP selectivity ($\textsc{sel}_\mathcal{G}(tp)$) mean]{
        \includegraphics[width=.49\textwidth]{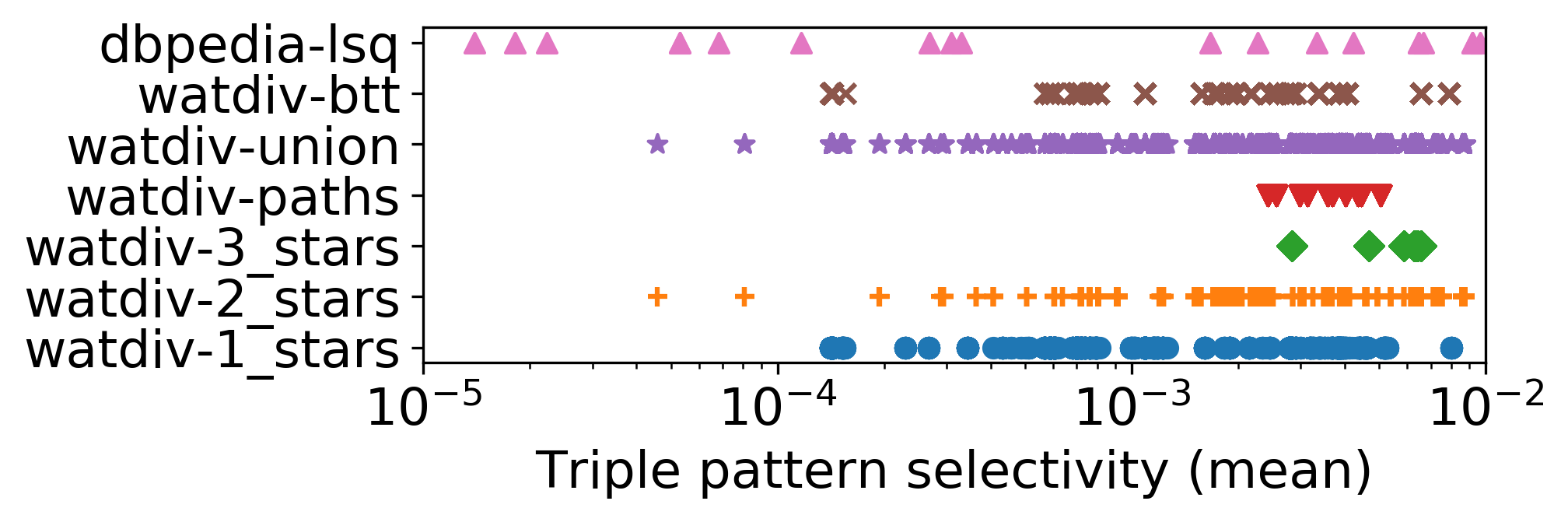}
        \label{fig:ch5}
    }
    \subfloat[TP selectivity ($\textsc{sel}_\mathcal{G}(tp)$) stdev]{
        \includegraphics[width=.49\textwidth]{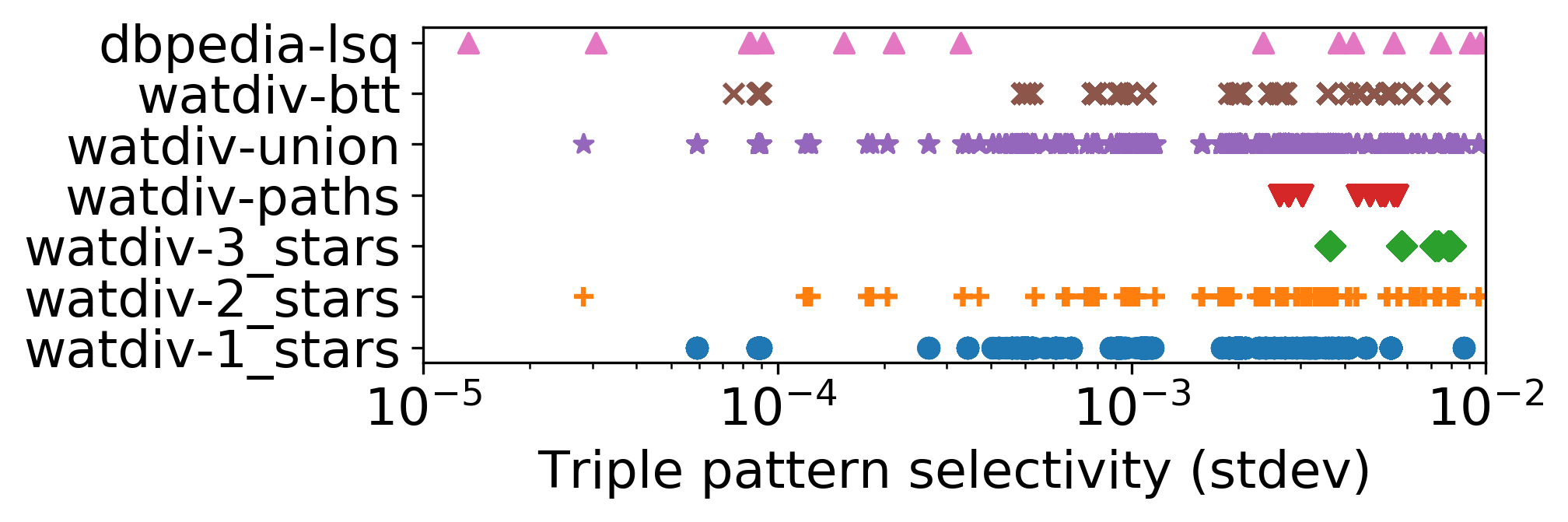}
        \label{fig:ch6}
    }
    \caption{Characteristics of all query loads (WatDiv query loads over \texttt{watdiv100M}).}\label{fig:characteristics}
\end{figure*}

To study the impact of the number of star-shaped subqueries, and to stress-test the approach, the WatDiv template and query generators were used to obtain query loads with no star patterns (i.e. path queries), as well as query loads with up to 3 subject-based star patterns.
Each above mentioned query load contains 6400 queries.
Furthermore, the setup was tested with queries derived from the basic testing templates (BTT) provided by WatDiv\footnote{\url{https://dsg.uwaterloo.ca/watdiv/basic-testing.shtml}}.
The WatDiv basic testing templates provides 20 query templates with relatively diverse characteristics (Figure~\ref{fig:characteristics}).
For the DBpedia dataset, user-issued queries were obtained from the LSQ~\cite{DBLP:conf/semweb/SaleemAHMN15} query log.
As most LSQ queries contain a single triple pattern or return an empty result set, we selected a challenging set of 24 representative queries with diverse characteristics.
The complete set of queries is available on the website. 
This query load, called \texttt{dbpedia-lsq}, was executed in random order by all clients concurrently in each configuration and include queries with the SPARQL operators \texttt{FILTER}, \texttt{UNION} and \texttt{OPTIONAL}.
Such queries are processed by first processing the BGPs then combining them according to operators in the query.

Figure~\ref{fig:characteristics} shows an overview of the characteristics of the query loads~\cite{watdiv}:
Triple pattern count (\#TP),
join vertex count (\#JV),
join vertex degree ($\textsc{deg}$), i.e., the number of triple patterns incident on a join vertex,
result cardinality (\#Results), and
triple pattern selectivity ($\textsc{sel}_\mathcal{G}(tp)$), i.e., the average ratio of cardinality of each triple pattern to the size of the knowledge graph.
High selectivity thus means that the triple patterns in a query have high cardinalities (high ratio of triples).

\begin{table}[h]
\caption{Join vertex types over query loads}
\label{tbl:jvtype}
\begin{tabular}{llll}
\hline
\textbf{Query load} & \textbf{SS} & \textbf{SO} & \textbf{OO} \\ \hline
watdiv-1\_star      & 100\%       & 0\%           & 0\%           \\
watdiv-2\_stars     & 55.38\%          & 34.31\%          & 10.31\%          \\
watdiv-3\_stars     & 57.69\%          & 30.77\%          & 11.53\%          \\
watdiv-paths        & 0\%          & 100\%          & 0\%          \\
watdiv-union        & 53.27\%          & 41.27\%          & 5.46\%          \\
watdiv-btt          & 56\%          & 34\%          & 10\%          \\
dbpedia-lsq         & 64.71\%          & 26.47\%          & 8.82\%          \\ \hline
\end{tabular}
\end{table}

Table~\ref{tbl:jvtype} shows the relative distribution of the types of joins in each query load (SS is subject-subject joins, SO is subject-object or object-subject joins, and OO is object-object joins).
The \texttt{watdiv-union} query load that contains the combined queries from \texttt{watdiv-1\_star}, \texttt{watdiv-2\_stars}, \texttt{watdiv-3\_stars} and \texttt{watdiv-paths} was added as well. 
All query loads (except \texttt{watdiv-1\_star}) include queries with subject-object joins.
All queries in the \texttt{watdiv-paths} query load contain only subject-object joins.

\paragraph{Experimental Configuration:} 
To assess how each approach performs under different loads, experiments were run over eight configurations with $2^i$ clients concurrently issuing queries to the server in each configuration (0 $\leq$ i $\leq$ 7), i.e., up to 128 clients. In the configuration with $2^i$ clients, a total of $244 \times 2^i$ queries are executed and at most $2^i$ queries are executed concurrently, i.e., each client executes one query at a time. 
Each query load was run separately to assess the impact of the query load on the performance of the interfaces.
For the \texttt{watdiv-1\_star}, \texttt{watdiv-2\_stars}, \texttt{watdiv-3\_stars}, and \texttt{watdiv-paths} query loads, 50 distinct queries were executed by each client in the configuration (24 distinct queries for \texttt{watdiv-btt}).
For \texttt{dbpedia-lsq}, the 24 queries in the query load were run on all clients in the configuration in a distinct, random order on each client.

\paragraph{Hardware Setup:}
To run the clients, a virtual machine (VM) running all 128 clients concurrently was used.
The VM had 128 vCPU cores with a clock speed of 2.5GHz, 64KB L1 cache, 512KB L2 cache, 8192KB L3 cache, and 2TB main memory.
Each client was limited to use just one vCPU core and $15$GB RAM.
The LDF server and the SPARQL endpoint were run, at all times, on a server with 32 vCPU cores, with a clock speed of 3GHz, 64KB L1, 4096KB L2, and 16384KB L3 cache, and a main memory of 256GB.
To simulate a realistic setup in terms of the network, each client was limited to a bandwidth of 20 MB/s.

\paragraph{Evaluation Metrics:}
\begin{itemize}
\item \textit{Number of Requests to the Server (NRS)}: The number of requests the client issues to the server while processing a query.
\item \textit{Workload Time}: The average amount of time (in minutes) it takes each client to complete an entire workload including queries that time out.
\item \textit{Throughput}: The number of completed queries divided by the total workload time averaged over all clients (number of queries per minute).
\item \textit{Query Execution Time (QET)}: The amount of time (in milliseconds) elapsed since a query is issued until its processing is finished.
\item \textit{Query Response Time (QRT)}: The amount of time (in milliseconds) elapsed since a query is issued until the first result is computed.
\item \textit{Number of Transferred Bytes (NTB)}: The amount of data transferred (in bytes) between the client and the server while processing a query (both from and to the server).
\item \textit{CPU Load (CPU)}: The average CPU load on the server (in percentage).
\end{itemize}

\paragraph{Software configuration:} 
Virtuoso Open-Source version 7.2.5 
was used to run the SPARQL endpoint, configured to use up to all 32 threads at a time (one per vCPU core on the server) with the following variables set\footnote{\texttt{NumberOfBuffers} and \texttt{MaxDirtyBuffers} uses the recommended configuration from \url{http://vos.openlinksw.com/owiki/wiki/VOS/VirtRDFPerformanceTuning} given the server resources. \texttt{ResultSetMaxRows} was set to the maximum number of rows Virtuoso allows in a 64-bit system, and \texttt{MaxQueryCostEstimationTime} was set to 60,000 seconds to avoid rejection of requests by the server.}:
\begin{itemize}
\item $\mathtt{NumberOfBuffers}=9735000$
\item $\mathtt{MaxDirtyBuffers}=7301250$
\item $\mathtt{ResultSetMaxRows}=2097150$
\item $\mathtt{MaxQueryCostEstimationTime}=60000$
\end{itemize}
We chose Virtuoso since Verborgh et al.~\cite{DBLP:journals/ws/VerborghSHHVMHC16} showed that, this is the endpoint that performed best with respect to high throughput and low CPU usage.

Two LDF server implementations were used; one which was a combined TPF, brTPF, and SPF server\footnote{The combined server implementation is available at \url{http://relweb.cs.aau.dk/spf}.}, as well as the Smart-KG\footnote\url{https://git.ai.wu.ac.at/beno/smartkg} server implementation.
The LDF servers were configured to use up to all 32 cores concurrently.
The LDF page size was, throughout the experiments, set to 100 results, and the maximum number of elements in $\Omega$ was set to 30 for both brTPF and SPF, i.e., they can send up to 30 bindings with each request.
The original TPF\footnote{\url{https://github.com/LinkedDataFragments/Client.js}} and brTPF\footnote{\url{http://olafhartig.de/brTPF-ODBASE2016/}} Node.js clients were used.
The SaGe server was configured with 65 workers\footnote{As recommended for 32 cores by \url{https://docs.gunicorn.org/en/stable/design.html\#how-many-workers}}. and a time quantum of 75 milliseconds as recommended by~\cite{sage}.
Any query that takes longer than the time quantum to execute is suspended at least once.
The timeout for executing a query was set to 600 seconds, i.e., 10 minutes.

\paragraph{Experimental Results:}
The objective is to assess whether SPF can execute SPARQL queries containing star patterns more efficiently in terms of response time and network traffic without incurring too much additional load on the server.
Furthermore, the experiments investigate if SPF is, in the case of path queries, still as good in terms of performance as brTPF.

The SPARQL endpoint became unresponsive (i.e., all queries timed out) for certain configurations due to high server load.
Moreover, some Smart-KG clients ran out of memory for some configurations over the large datasets due to very large partitions being transferred and loaded into memory.
These cases are clearly marked in the figures within this section with \textit{"Unresponsive"} and \textit{"Out of memory"}, respectively.
A full list of configurations that did not finish is available on the website.

\begin{figure*}[htb]
    \centering	
    \subfloat[Scalability of \texttt{watdiv-1\_star} (\textit{log})]{
        \includegraphics[width=.49\textwidth]{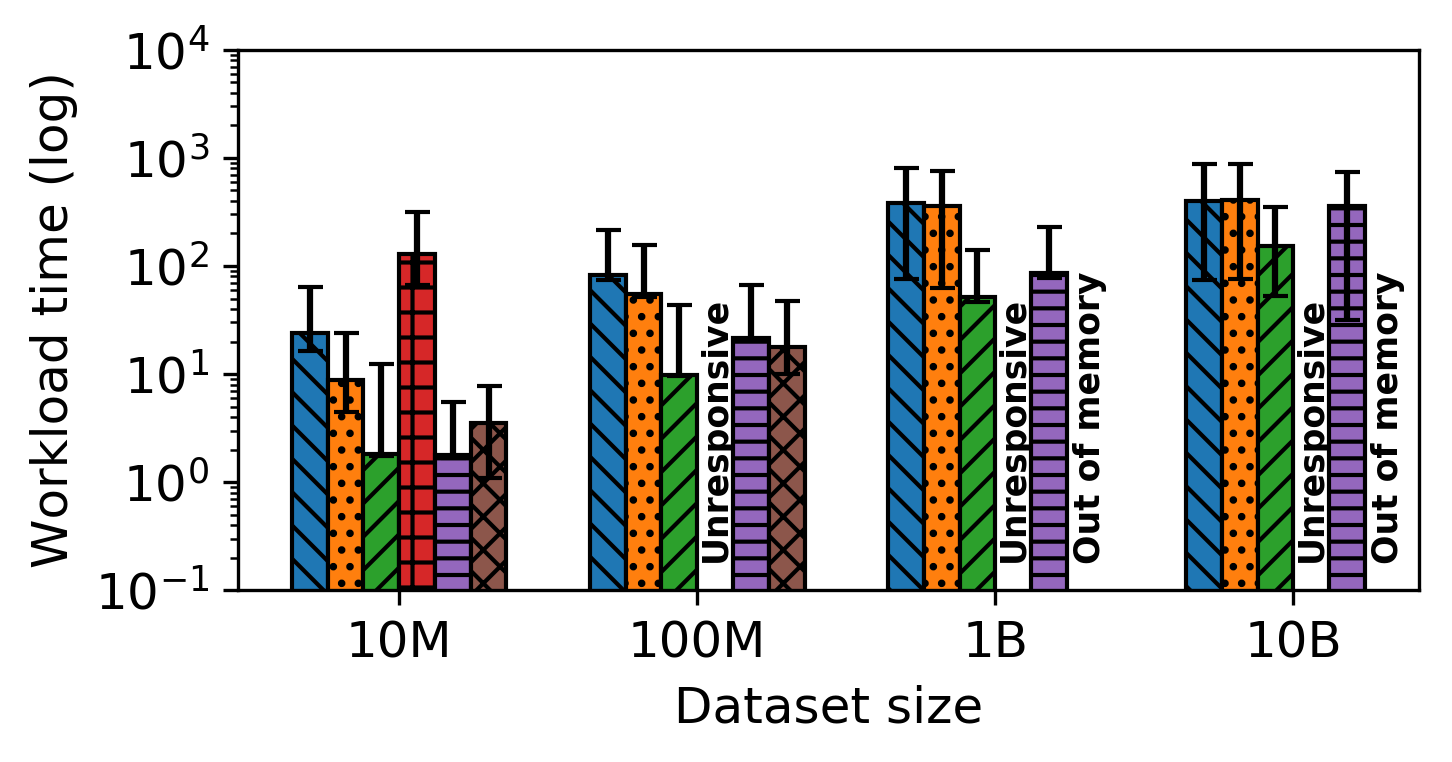}
        \label{fig:sc1}
    }
    \subfloat[Scalability of \texttt{watdiv-2\_stars} (\textit{log})]{
        \includegraphics[width=.49\textwidth]{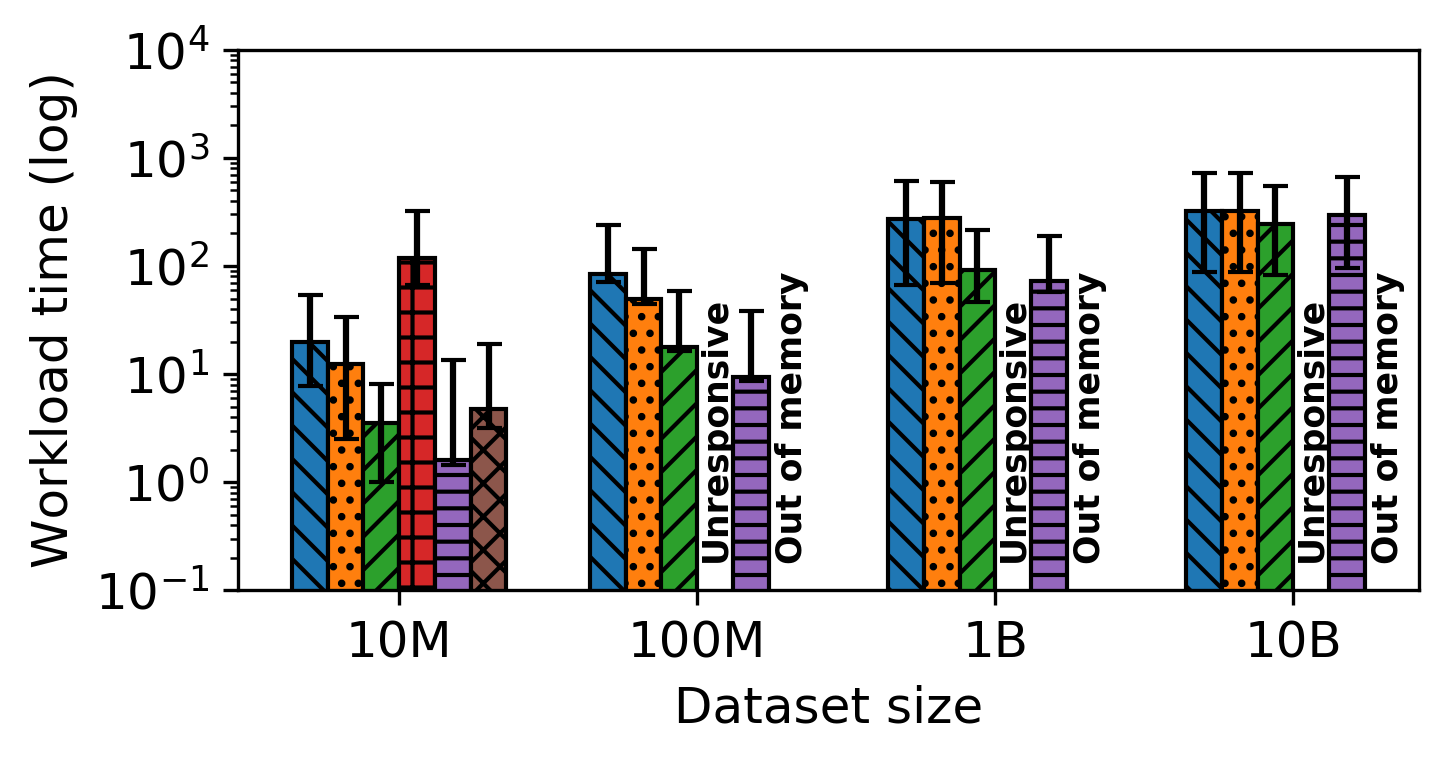}
        \label{fig:sc2}
    }\\
    \subfloat[Scalability of \texttt{watdiv-3\_stars} (\textit{log})]{
        \includegraphics[width=.49\textwidth]{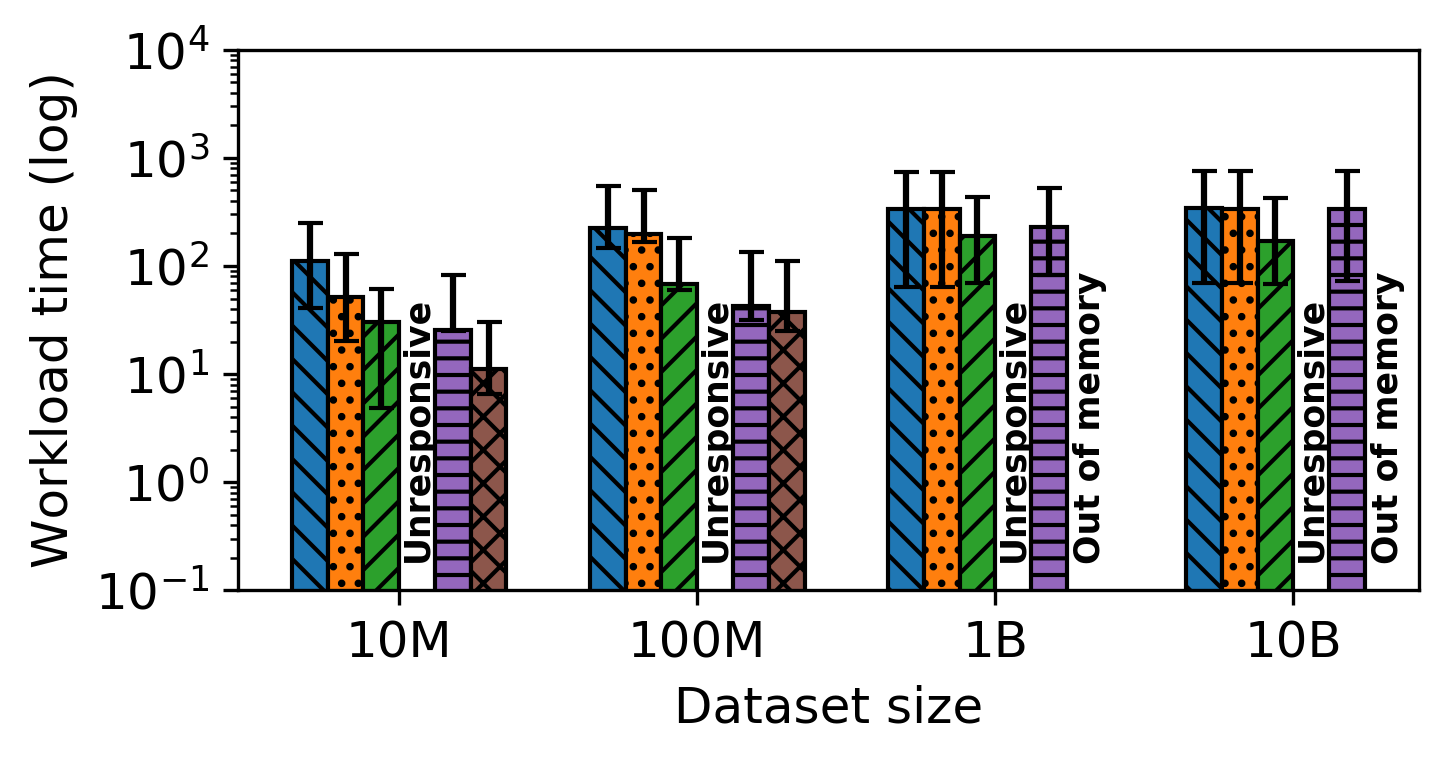}
        \label{fig:sc3}
    }
    \subfloat[Scalability of \texttt{watdiv-paths} (\textit{log})]{
        \includegraphics[width=.49\textwidth]{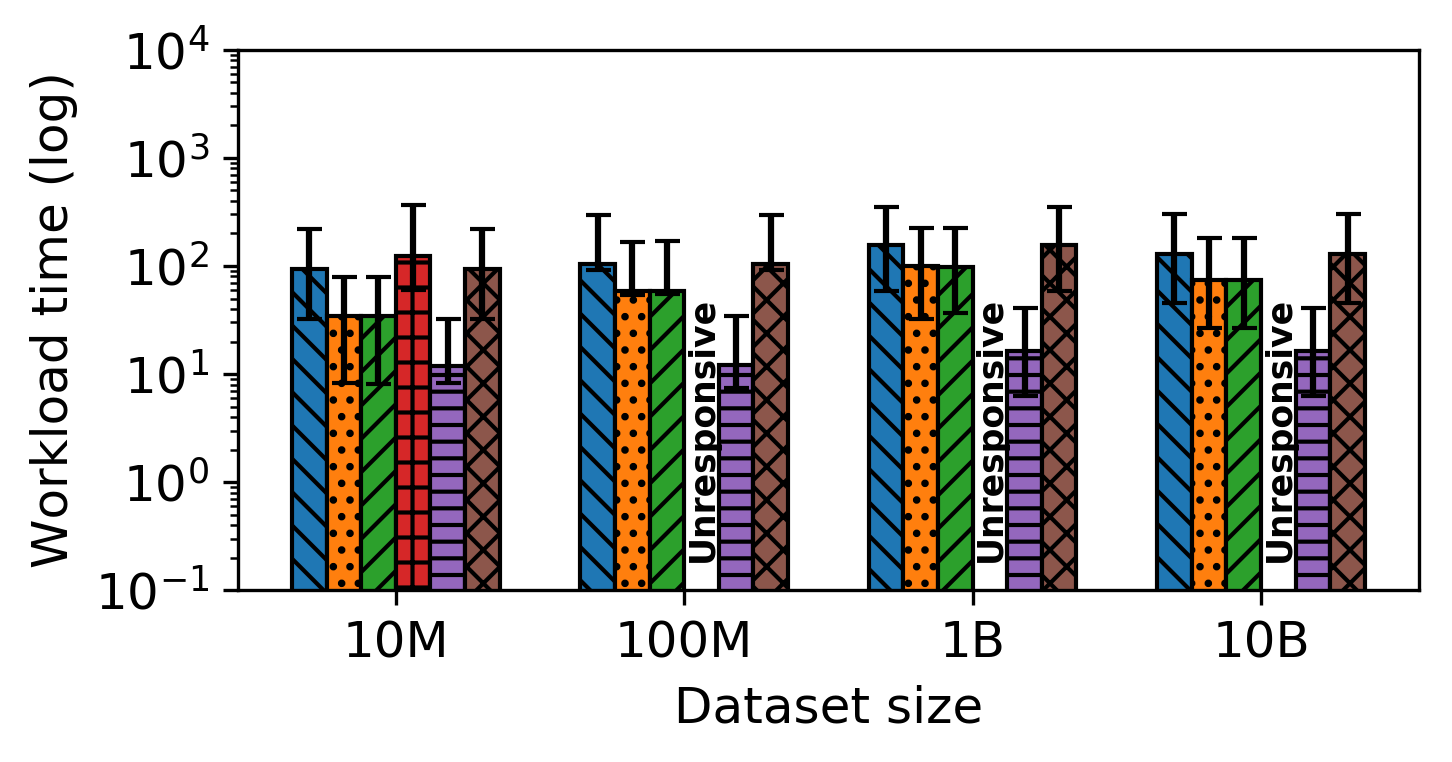}
        \label{fig:sc4}
    }\\
    \subfloat[Scalability of \texttt{watdiv-union} (\textit{log})]{
        \includegraphics[width=.49\textwidth]{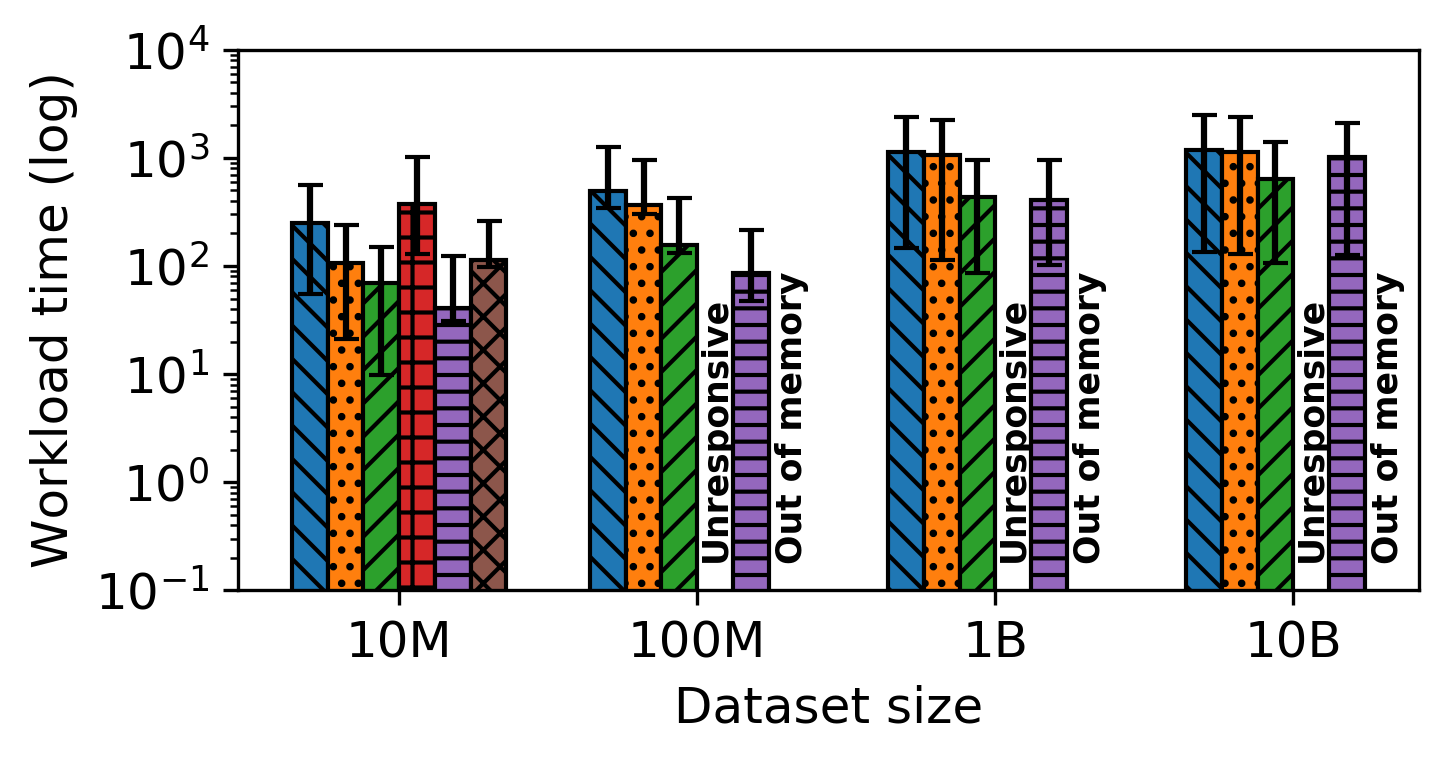}
        \label{fig:sc5}
    }
    \subfloat[Scalability of \texttt{watdiv-btt} (\textit{log})]{
        \includegraphics[width=.49\textwidth]{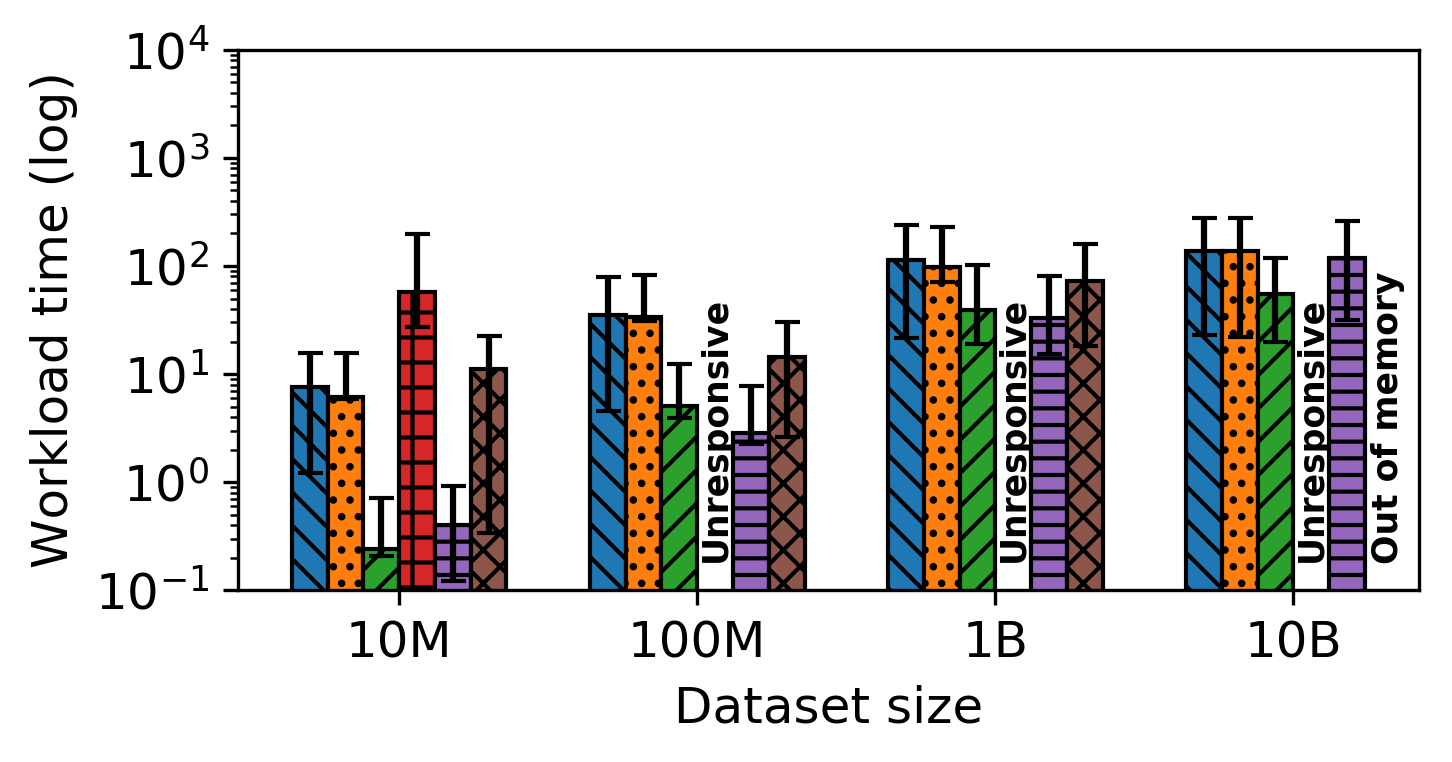}
        \label{fig:sc6}
    }\\
    \begin{minipage}{.6\textwidth}
    \vspace{-235ex}
	\includegraphics[width=\textwidth]{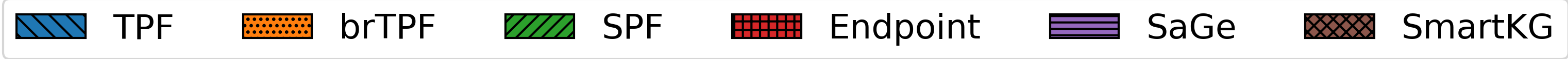}
	\end{minipage}
    \caption{The workload time (in minutes) averaged over each client for all WatDiv query loads over \texttt{watdiv10M}, \texttt{watdiv100M}, \texttt{watdiv1B}, and \texttt{watdiv10B} with 128 clients concurrently issuing queries.}\label{fig:scalability}
\end{figure*}

\subsection{Scalability}\label{subsec:scalability}
Figure~\ref{fig:scalability} shows the average workload time for each approach over all WatDiv query loads and each dataset size with 128 concurrent clients.
This includes the queries that timed out.
Evidently, SPF, SaGe, and Smart-KG have significantly better performance in all cases compared to TPF, brTPF, and the endpoint.
The only exception to this is for \texttt{watdiv-paths} (Figure~\ref{fig:sc4}), where SPF has similar performance as brTPF, and Smart-KG has similar performance as TPF.
This is expected and is due to SPF using the brTPF selector for star patterns with only one triple pattern and Smart-KG using the TPF selector for such star patterns.
Furthermore, most of the Smart-KG clients over the \texttt{watdiv1B} and \texttt{watdiv10B} datasets and some clients for the \texttt{watdiv-2\_stars} (Figure~\ref{fig:sc2}) query load over the \texttt{watdiv100M} dataset exhausted the main memory during the experiments.
These results suggest that if client resources are limited, SPF and SaGe seem to be better choices than Smart-KG, given that they are able to process queries with less memory usage on the client side.

SPF has comparable or better performance compared to SaGe for all query loads except \texttt{watdiv-paths} (Figure~\ref{fig:sc4}).
It was expected that SPF would perform worse than SaGe for this query load, since using the brTPF selector for path queries results in a high number of requests to the server that SaGe does not have to incur.
For \texttt{watdiv-1\_star} (Figure~\ref{fig:sc1}), SPF overall has the best performance out of all tested approaches.
This was expected, since SPF only has to send a single request to the server per 100 results (given the page size of 100).
While SaGe has slightly better performance than SPF for the remaining query loads over the smallest datasets, SPF scales better with the size of the dataset than SaGe (Figure~\ref{fig:scalability}).
As a result, SPF has better performance over \texttt{watdiv10B} for all query loads except \texttt{watdiv-paths}.
This is also evident by the fact that SPF has the best performance out of all systems for \texttt{watdiv-union} (Figure~\ref{fig:sc5}) over the largest dataset.

Overall, the experimental results suggest that for smaller datasets, SaGe has a better performance than SPF; 
however, SPF evidently scales better with the size of the dataset than any other approach. 
In particular, for the largest dataset with over 10 billion triples, SPF has the best performance for all query loads with the exception of \texttt{watdiv-paths}.
This shows that SPF is generally able to increase query processing performance compared to state-of-the-art interfaces for very large datasets and when a large number of clients send queries to the server concurrently.

\begin{figure*}[htb]
    \centering	
    \subfloat[Throughput for \texttt{watdiv10M} (\textit{log})]{
        \includegraphics[width=.49\textwidth]{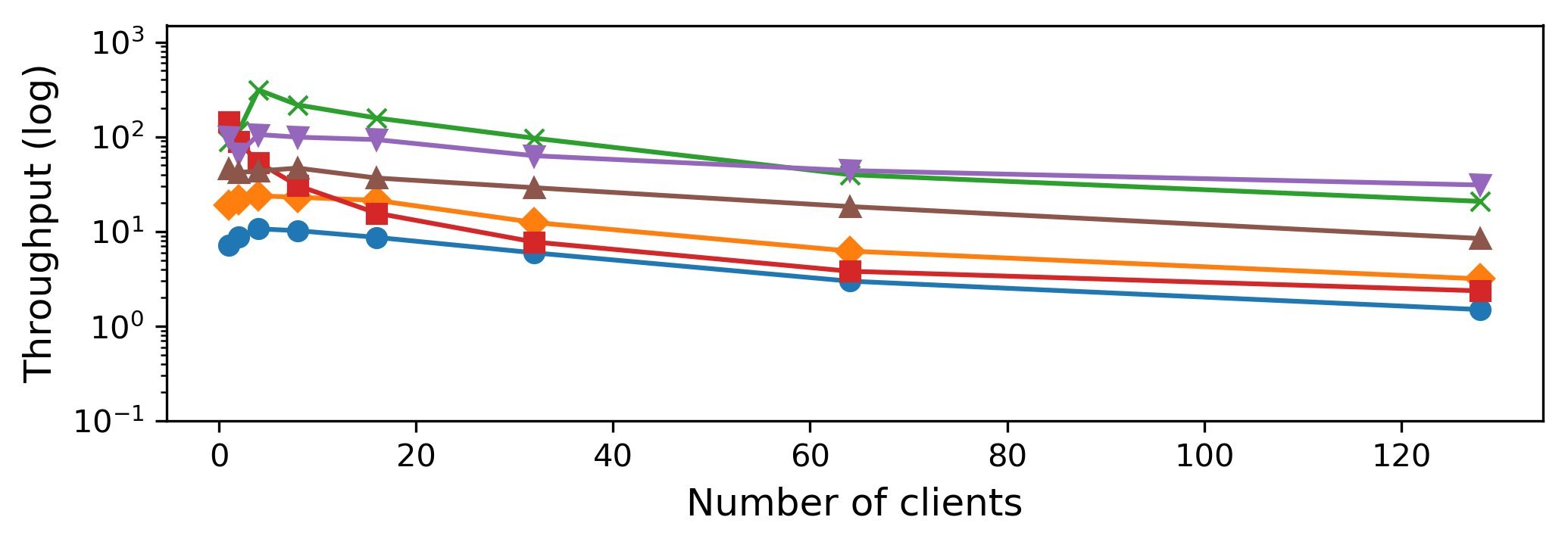}
        \label{fig:tp1}
    }
    \subfloat[Throughput for \texttt{watdiv100M} (\textit{log})]{
        \includegraphics[width=.49\textwidth]{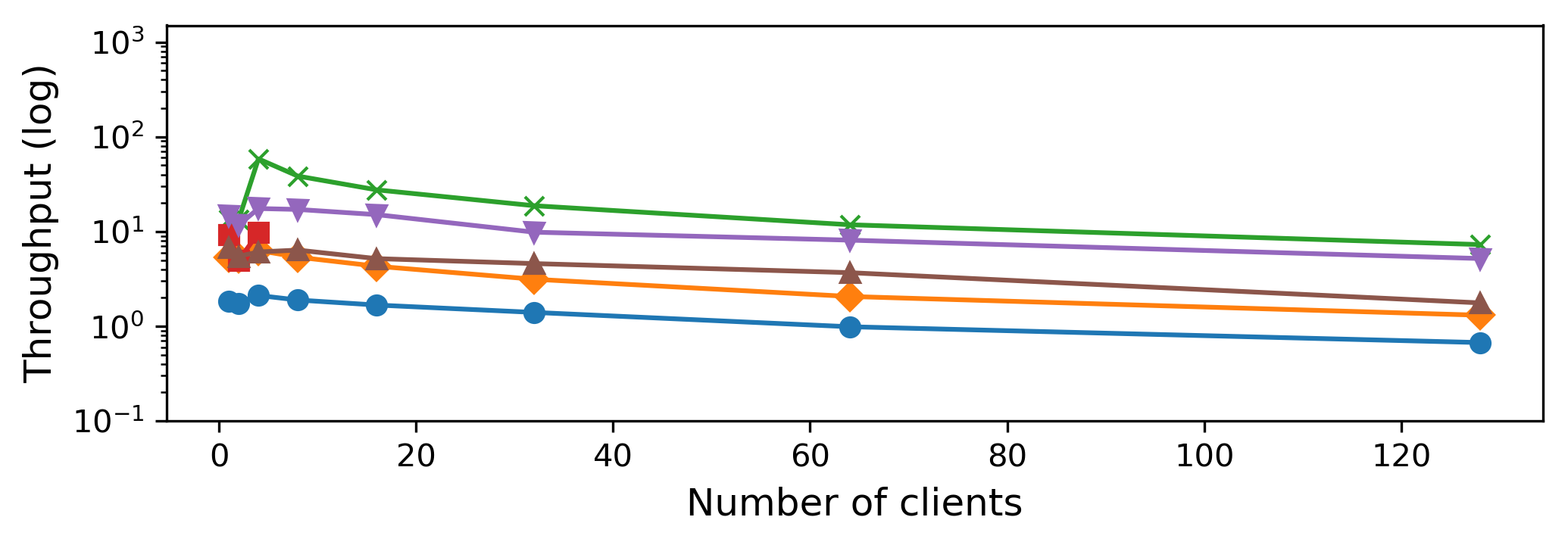}
        \label{fig:tp2}
    }\\
    \subfloat[Throughput for \texttt{watdiv1B} (\textit{log})]{
        \includegraphics[width=.49\textwidth]{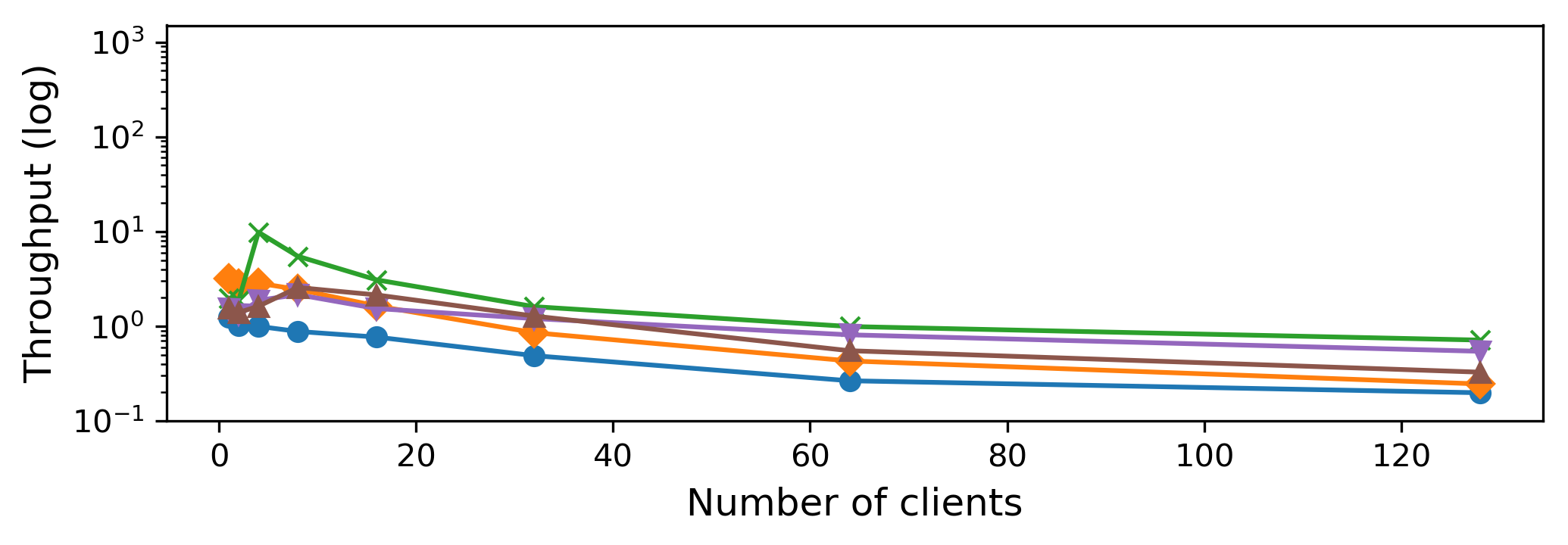}
        \label{fig:tp3}
    }
    \subfloat[Throughput for \texttt{watdiv10B} (\textit{log})]{
        \includegraphics[width=.49\textwidth]{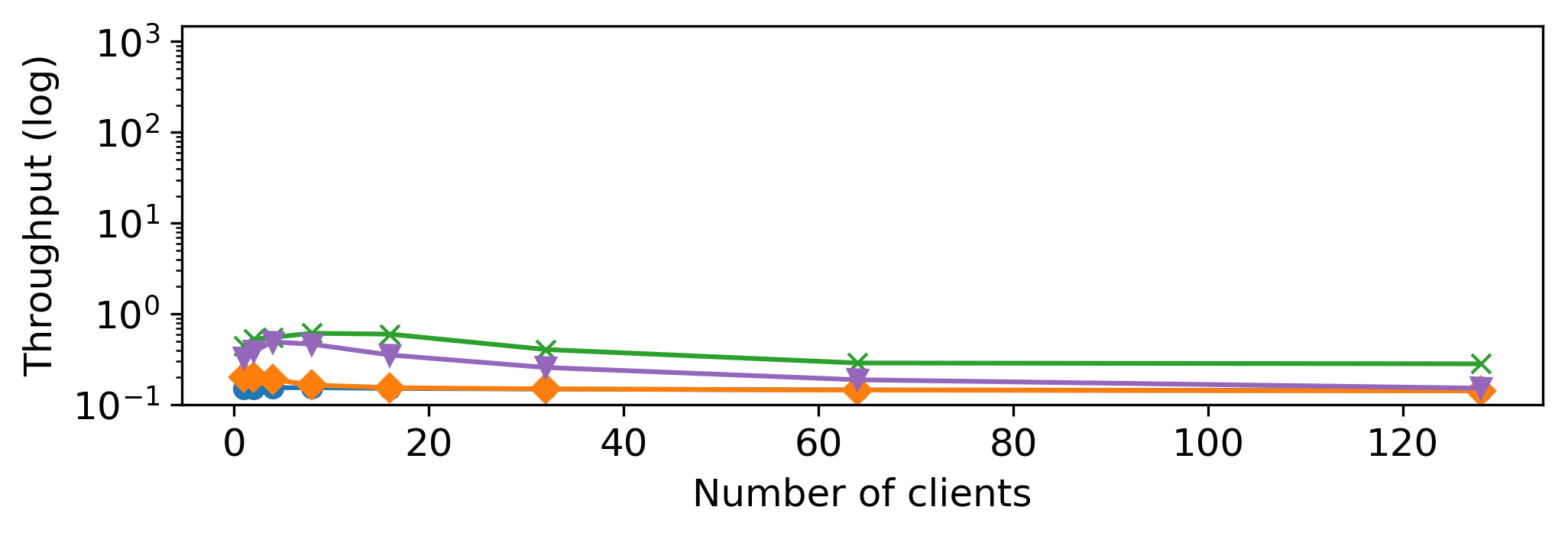}
        \label{fig:tp5}
    }\\
    \subfloat[Throughput for \texttt{dbpedia} (\textit{log})]{
        \includegraphics[width=.49\textwidth]{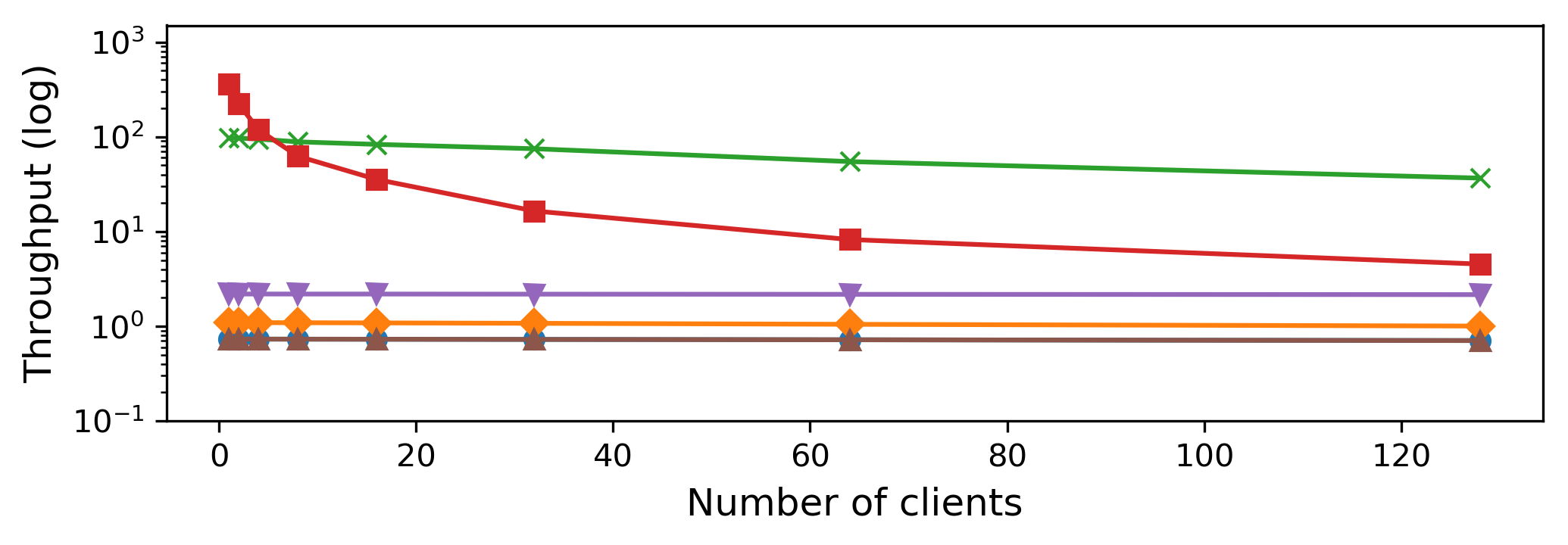}
        \label{fig:tp4}
    }\\
    \begin{minipage}{.6\textwidth}
    \vspace{-165ex}
	\includegraphics[width=\textwidth]{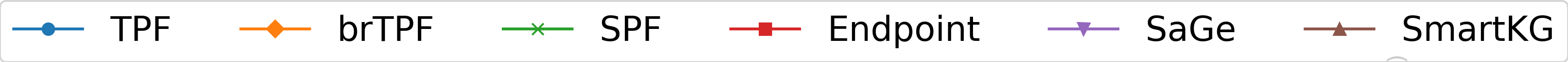}
	\end{minipage}
    \caption{Throughput (\# queries/m) for \texttt{watdiv-union} over the different WatDiv datasets and throughput for \texttt{dbpedia-lsq}. Missing values are due to unresponsiveness or the clients exhausting memory.}\label{fig:throughput}
\end{figure*}

\subsection{Performance Under Load}\label{subsec:load}
Figure~\ref{fig:throughput} shows the throughput for different numbers of concurrent clients for \texttt{watdiv-union} over each WatDiv dataset and for \texttt{dbpedia-lsq} over \texttt{dbpedia}, including queries that timed out.
This means that the figures include the entire execution time for the approaches that did not time out (e.g., SPF) while they include only partial execution times for the approaches that did time out (e.g., TPF).
Note that for \texttt{watdiv-union} over \texttt{watdiv10M}, \texttt{watdiv100M}, \texttt{watdiv1B}, and \texttt{watdiv10B}, the throughput increases for SPF, brTPF, TPF, SaGe and Smart-KG until four concurrent clients, but it decreases afterwards.
This is due to the fact that when running more concurrent clients, more queries are processed in total. 
However, the increased server load does not significantly affect the throughput until after the four concurrent clients.
Although the throughput (Figures~\ref{fig:tp1}-\ref{fig:tp4}) of all the interfaces decreases as the number of concurrent clients increases, 
SPF maintains between 3-7 times higher throughput compared to brTPF for WatDiv, and 96 times higher throughput for \texttt{dbpedia}.

\begin{figure*}[htb]
    \centering	
    \subfloat[No. timeouts for \texttt{watdiv10M}]{
        \includegraphics[width=.49\textwidth]{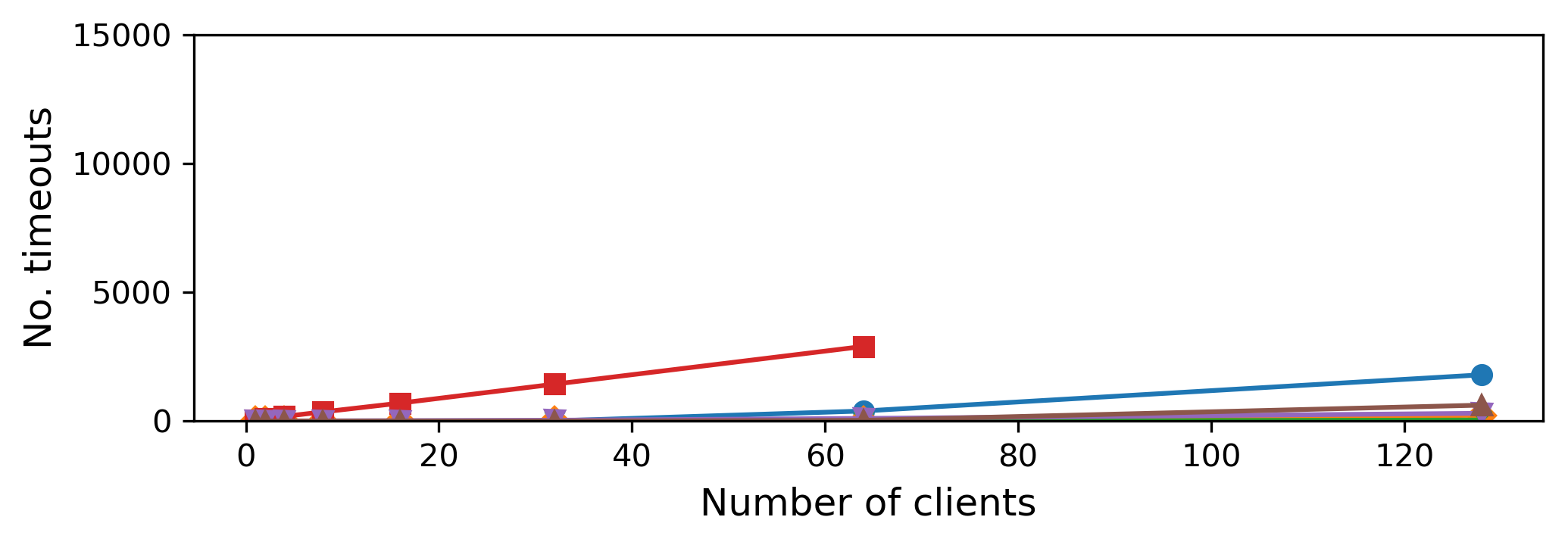}
        \label{fig:to1}
    }
    \subfloat[No. timeouts for \texttt{watdiv100M}]{
        \includegraphics[width=.49\textwidth]{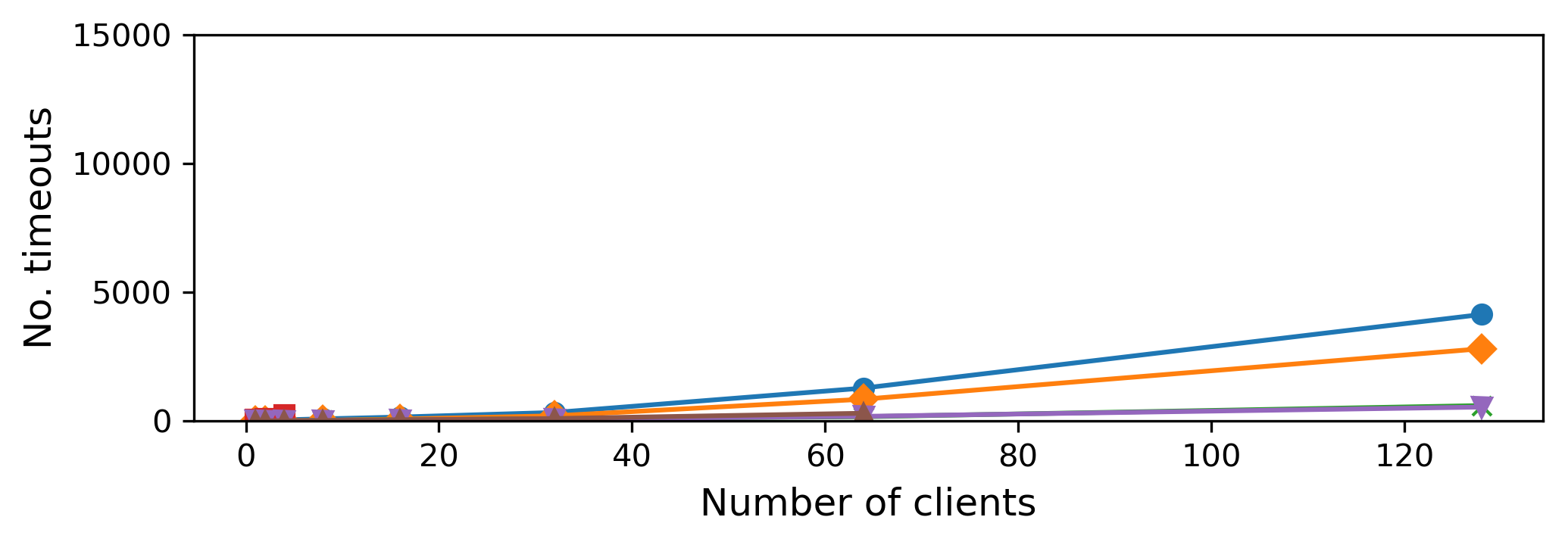}
        \label{fig:to2}
    }\\
    \subfloat[No. timeouts for \texttt{watdiv1B}]{
        \includegraphics[width=.49\textwidth]{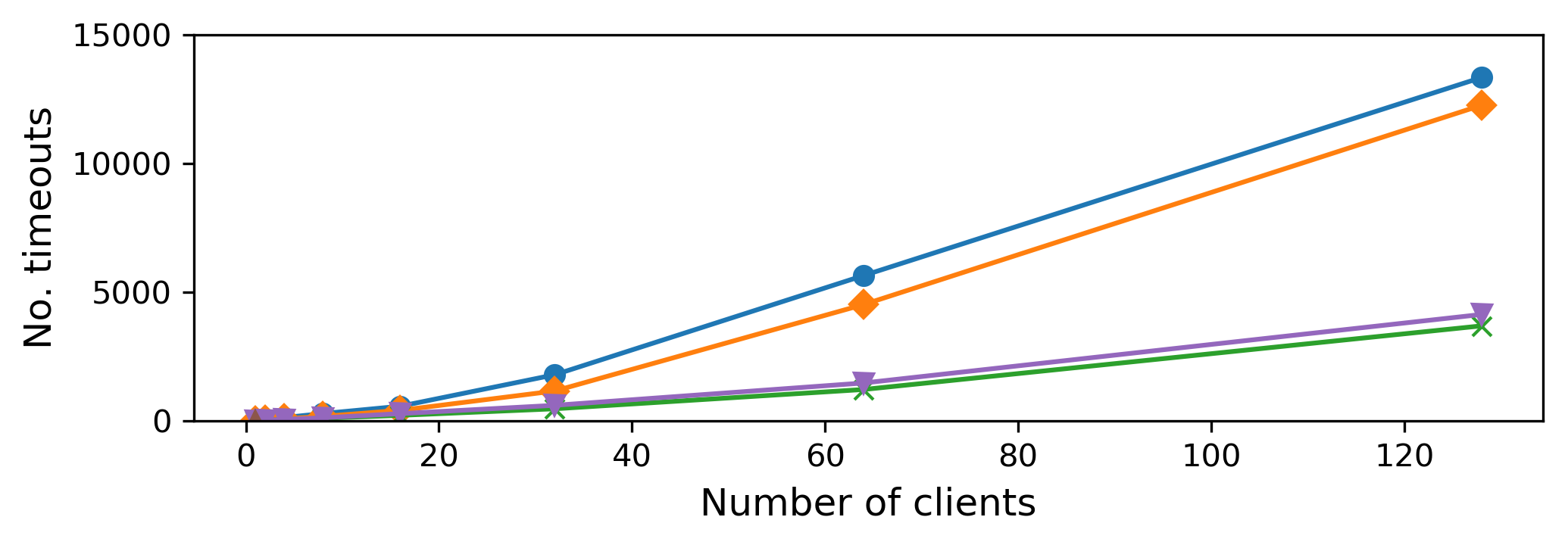}
        \label{fig:to3}
    }
    \subfloat[No. timeouts for \texttt{watdiv10B}]{
        \includegraphics[width=.49\textwidth]{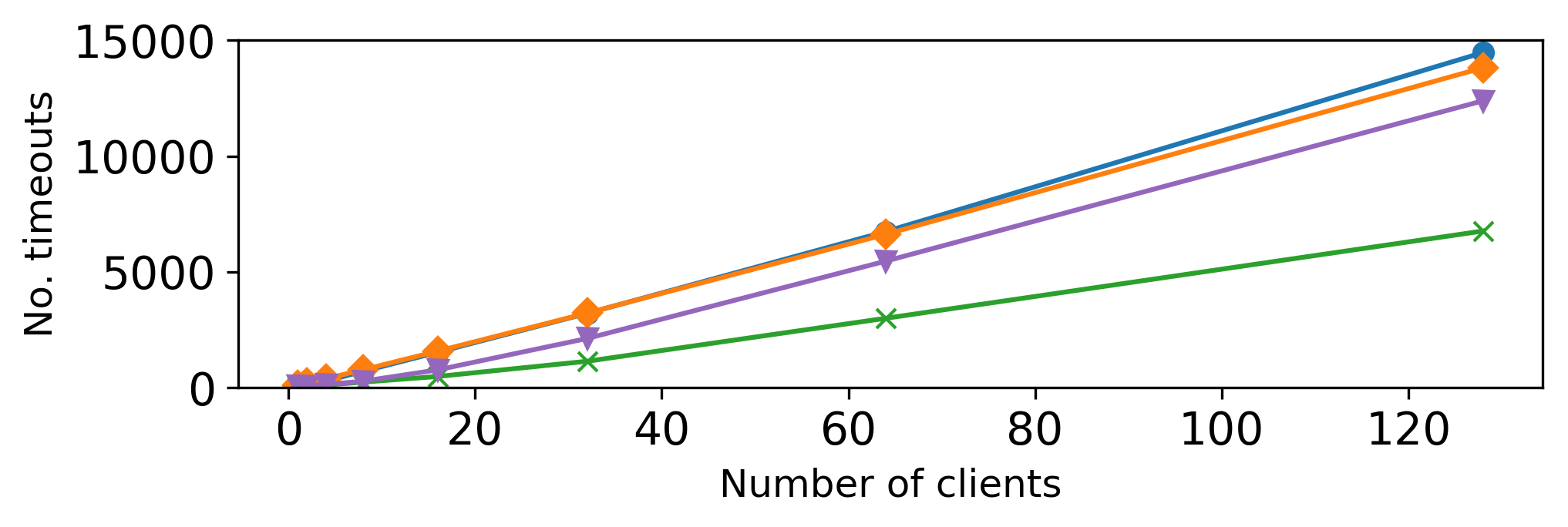}
        \label{fig:to4}
    }\\
    \begin{minipage}{.6\textwidth}
    \vspace{-108ex}
	\includegraphics[width=\textwidth]{results/graphs/legend1.png}
	\end{minipage}
    \caption{Number of timeouts for \texttt{watdiv-union} over each WatDiv query load and all WatDiv dataset sizes. Missing values are due to unresponsiveness or the clients exhausting memory.}\label{fig:timeout}
\end{figure*}

While SPF has slightly higher workload execution times compared to SaGe for the smallest WatDiv datasets (Figure~\ref{fig:sc5}), the throughput is slightly higher for the 100 million triples dataset and above (Figure~\ref{fig:tp2}).
This is due to the larger numbers of queries that time out for SaGe (541 timeouts for SaGe compared to 366 timeouts for SPF for 128 clients over \texttt{watdiv100M}, Figure~\ref{fig:to2}) that contribute to the overall workload time but do not add to the number of completed queries that the throughput relies upon.
In any case, SPF has up to 5 times higher throughput than SaGe for the WatDiv datasets (Figures~\ref{fig:tp1}-\ref{fig:tp5}).
%
Furthermore, as evident by Figure~\ref{fig:scalability}, SPF scales better with the size of the dataset for all query loads except \texttt{watdiv-paths} compared to SaGe and Smart-KG.
This is due to the increased load on the server caused by processing the higher amounts of intermediate results on the server that SaGe incurs, and the larger partitions Smart-KG has to transfer over the network as the dataset size increases.
Last, SPF has up to 45 times higher throughput compared to SaGe and up to 137 higher throughput than Smart-KG (Figure~\ref{fig:tp4}) for \texttt{dbpedia-lsq}.

Compared to TPF and brTPF, the relative gain in performance provided by SPF is slightly lower for \texttt{watdiv10B} compared to \texttt{watdiv10M} (3 times higher throughput compared to 7 times higher throughput for 128 clients with respect to brTPF). 
The same tendencies are also supported by the performance for the remaining query loads (Appendix~\ref{app:a}, Figures~\ref{fig:throughput1_app1}-\ref{fig:cpu_dbp_app}).
This is explained by the fact that larger datasets that share characteristics generally incur a higher number of intermediate results for each star pattern that have to be processed by the server.
In addition, the higher number of timeouts that especially TPF and brTPF incurs means that more partial times are included than for SPF and SaGe.
Moreover, DBpedia is roughly the same size in terms of number of triples as \texttt{watdiv1B} and presents a larger relative gain in performance for SPF (96 times for 128 clients in comparison to brTPF).
Section~\ref{subsec:ioqp} shows that the query loads with larger join vertex degrees (Figure~\ref{fig:ch3}) and higher selectivity (Figure~\ref{fig:ch5}), e.g., \texttt{watdiv-3\_stars}, result in lower relative gain in performance for SPF.
Nevertheless, SPF has higher throughput than TPF and brTPF even for the largest dataset.

\begin{figure*}[htb]
    \centering	
    \subfloat[CPU load \texttt{watdiv10M}]{
        \includegraphics[width=.49\textwidth]{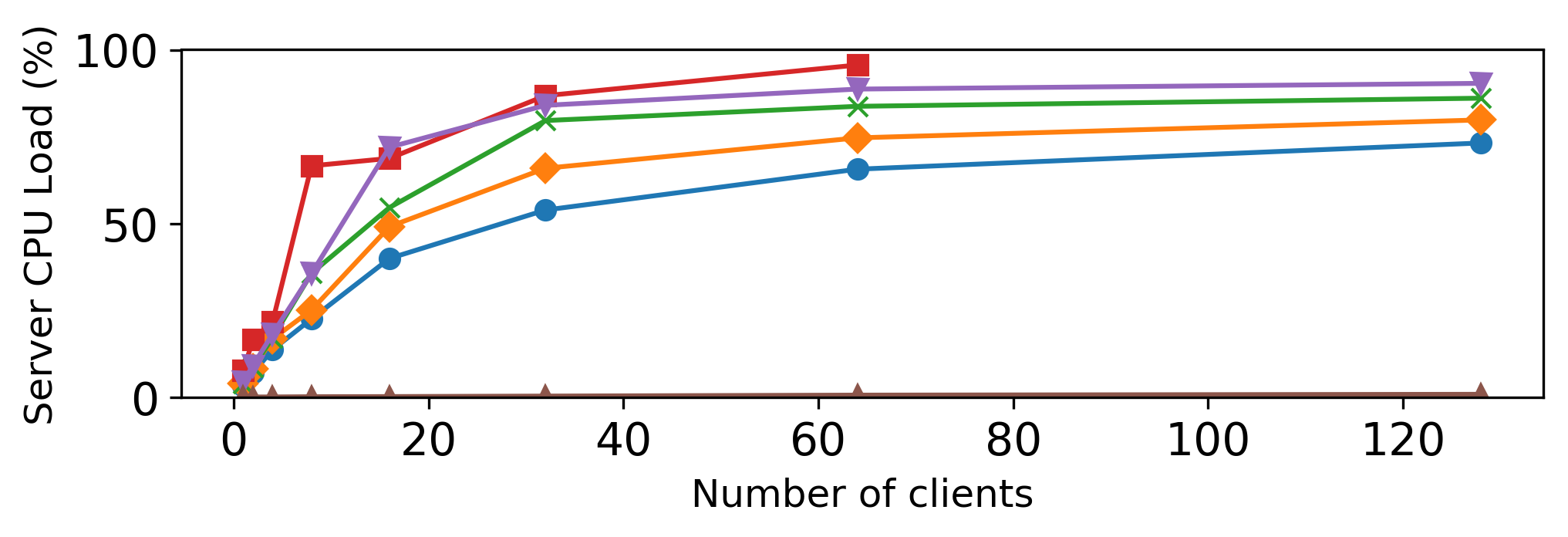}
        \label{fig:cpu1}
    }
    \subfloat[CPU load \texttt{watdiv100M}]{
        \includegraphics[width=.49\textwidth]{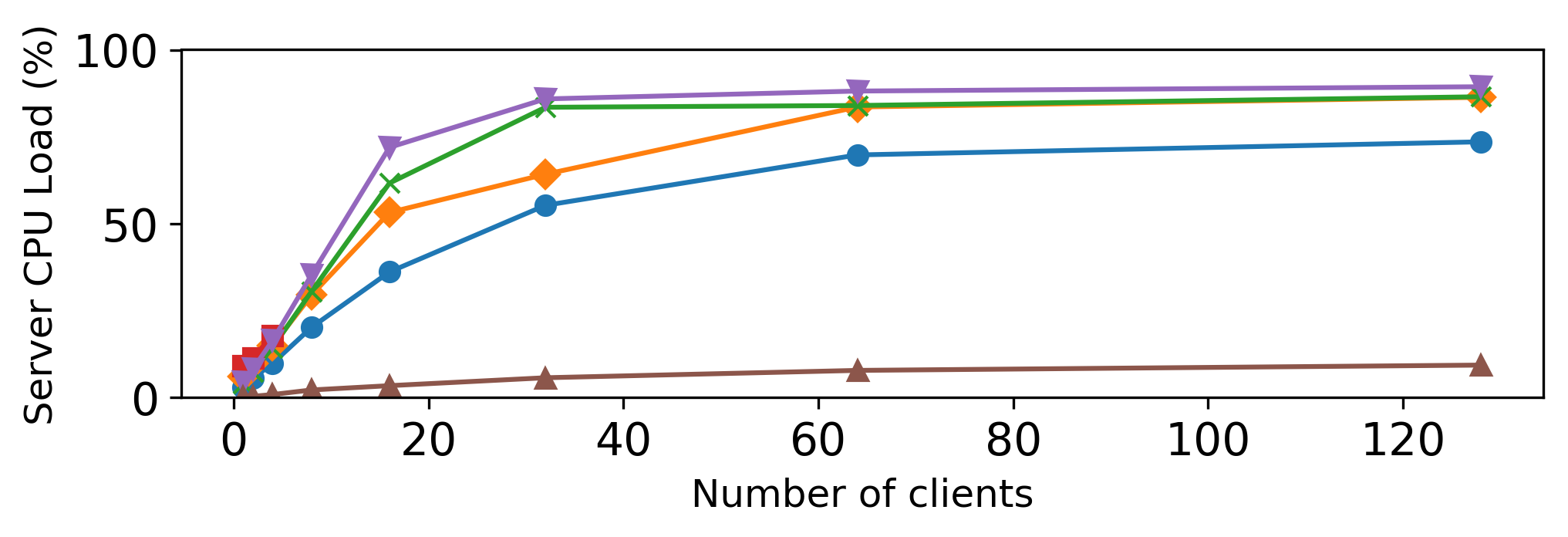}
        \label{fig:cpu2}
    }\\
    \subfloat[CPU load \texttt{watdiv1B}]{
        \includegraphics[width=.49\textwidth]{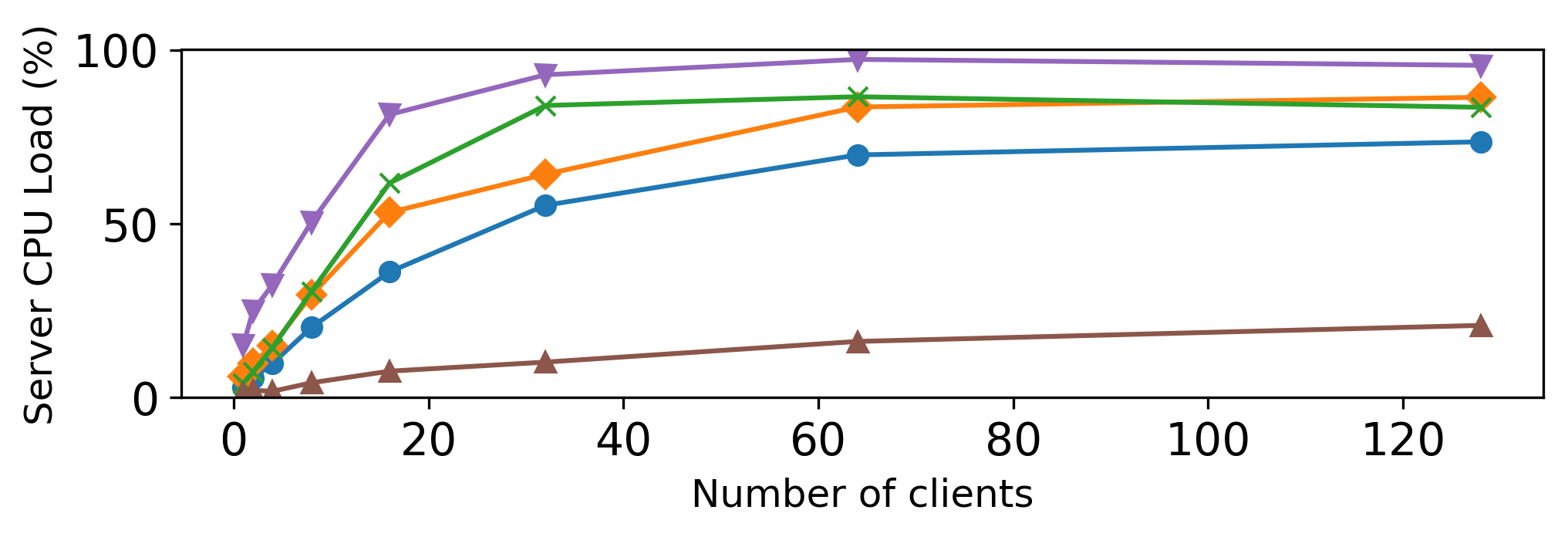}
        \label{fig:cpu3}
    }
    \subfloat[CPU load \texttt{watdiv10B}]{
        \includegraphics[width=.49\textwidth]{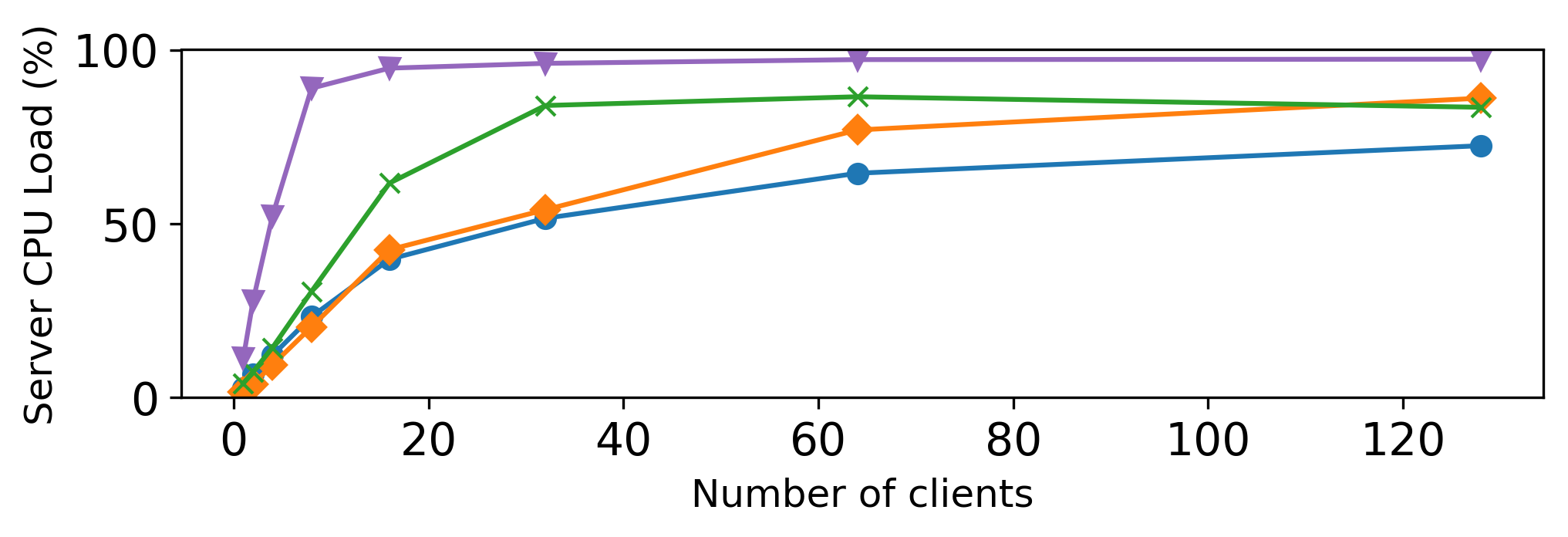}
        \label{fig:cpu4}
    }\\
    \begin{minipage}{.6\textwidth}
    \vspace{-108ex}
	\includegraphics[width=\textwidth]{results/graphs/legend1.png}
	\end{minipage}
    \caption{CPU load for \texttt{watdiv-union} over each WatDiv query load and all WatDiv dataset sizes. Includes queries that timed out. Missing values are due to unresponsiveness or the clients exhausting memory.}\label{fig:cpu}
\end{figure*}

Even though SPF servers compute star patterns, Figure~\ref{fig:cpu} shows that SPF only incurs up to 1.08 times as much CPU load in comparison to brTPF for 128 clients.
Moreover, SaGe has significantly higher CPU load than SPF in all configurations.
The CPU load is relatively similar across all WatDiv datasets.
This is due to the fact that each client makes at most one server request at a time, meaning at most 128 requests at a time have to be processed concurrently.
The CPU load, for all configurations, is below 100\% for all systems.
While SPF has a slightly higher CPU load than brTPF even under high load (1.08 times higher for 128 clients), the relative increase in throughput for SPF remains the same for increased numbers of clients for all WatDiv datasets.
Smart-KG has the lowest CPU load overall; this is expected, since the Smart-KG server only processes singular and infrequent triple patterns on the server and instead transfers the remaining predicate-family partitions to the client.
Overall, the CPU load suggests that SPF is able to utilize the stronger server resources better than approaches like Smart-KG and TPF, while not increasing the load on the server enough to significantly affect performance even for the largest datasets.

Due to the more efficient query processing, SPF has fewer timeouts than all other approaches (Figure~\ref{fig:timeout}).
Comparing SPF to brTPF, it is clear that the number of timeouts rises faster for TPF, brTPF, and SaGe than for SPF for the larger WatDiv datasets (Figure~\ref{fig:to4}).
As the dataset grows larger, brTPF becomes more similar to TPF in the number of timeouts, while it stays relatively low for SPF.
This is due to the increased sizes of intermediate results that TPF and brTPF have to deal with and the further limited amount of intermediate results of the server-side star join that reduces network traffic that only SPF can benefit from.
SPF further has a lower number of timeouts than both SaGe and Smart-KG throughout the experiments, showing that SPF is able to process queries in all configurations that SaGe and Smart-KG are not able to process.

Comparing the throughput to the number of timeouts over the different datasets, it is not clear why the relative gain in throughput that SPF provides decreases, while the ratio of timed out queries gets relatively better for SPF compared to brTPF and TPF for the larger datasets.
However, when looking at the individual results, this becomes more clear.
The increased size in the dataset means that it takes longer time to process each star pattern on the server (since each triple pattern has more intermediate bindings).
This is mostly mitigated by the limited amount of intermediate bindings for each star pattern.
However, for queries with high selectivity this can cause processing subqueries on the server to have slightly lower performance.
Due to the limited server requests, SPF is able to process more queries within the timeout.
This also helps explaining why SPF has such an improved performance for DBpedia compared to \texttt{watdiv1B} (Figure~\ref{fig:tp3} and \ref{fig:tp4}); brTPF and TPF have quite similar throughputs for boths datasets given that they are roughly the same size.
However, Figure~\ref{fig:tp4} shows that SPF increases throughput by up to two orders of magnitude for DBpedia (SPF has 96 times higher throughput for 128 clients compared to brTPF).
The lower number of results (Figure~\ref{fig:ch4}) and lower selectivity (Figure~\ref{fig:ch5}) for \texttt{dbpedia-lsq} means that SPF is able to decrease the number of intermediate results more significantly compared to brTPF and thus improve the throughput.
Furthermore, the larger numbers of timeouts for TPF and brTPF (Figure~\ref{fig:timeout}) means that the throughput for these systems include more incomplete results, lowering the reported throughput overall.

The endpoint is the best performing interface for few concurrent clients and small dataset.
However, its performance deteriorates faster when the number of clients increases.
All other approaches are able to handle the increased load more efficiently than the endpoint (Figures~\ref{fig:tp1} and~\ref{fig:tp2}).
This is in line with the experiments shown in~\cite{DBLP:journals/ws/VerborghSHHVMHC16} and shows that SPF seems to be a suitable alternative to handle large query loads.

The experimental results thus confirm that, while the relative performance of SPF compared to alternative systems depends on the characteristics of the queries, SPF is able to, for most query loads, maintain better performance overall compared to the state of the art for large datasets and under high load.

\begin{figure*}[htb]
    \centering
    \subfloat[NRS \texttt{watdiv10M} (\textit{y-axis in log scale}).]{
        \includegraphics[width=.49\textwidth]{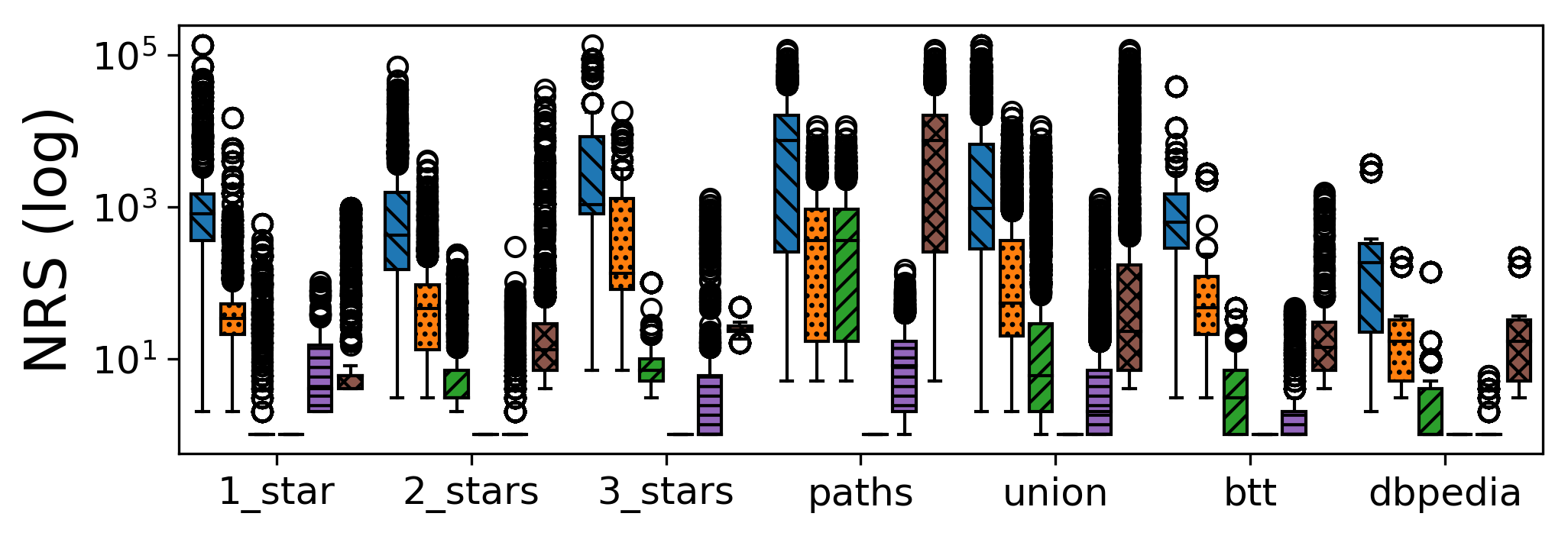}
        \label{fig:req1}
    }
    \subfloat[NRS \texttt{watdiv100M} (\textit{y-axis in log scale}).]{
        \includegraphics[width=.49\textwidth]{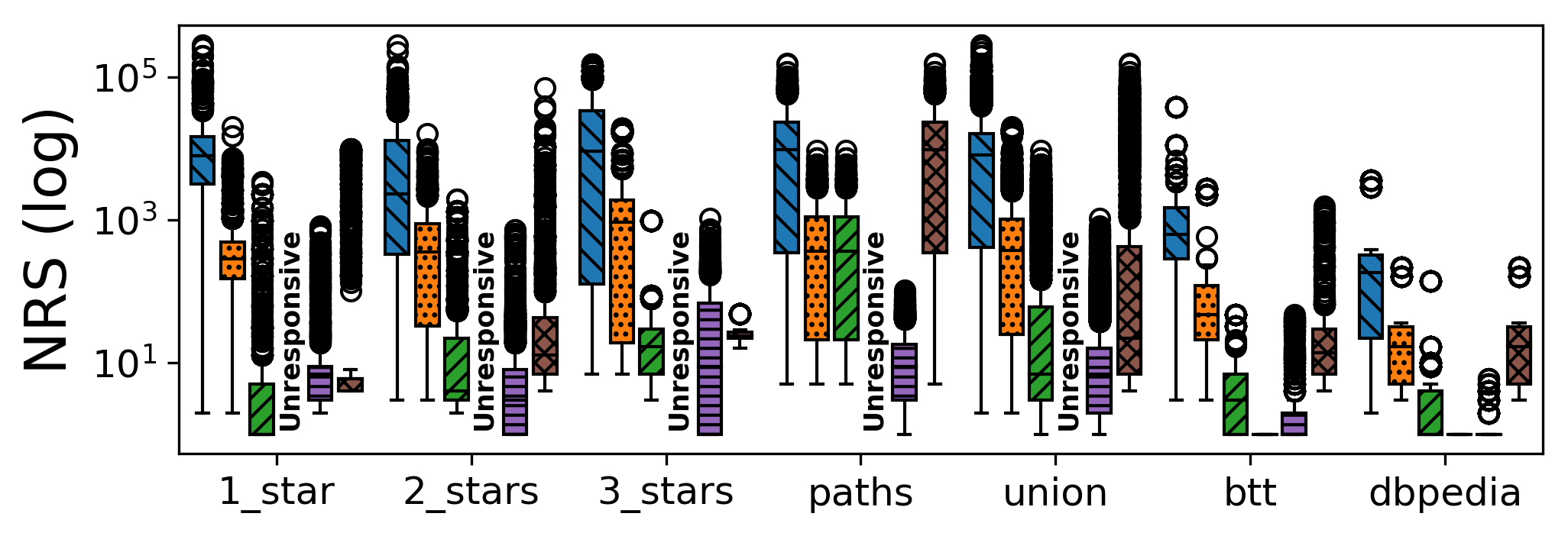}
        \label{fig:req2}
    }\\
    \subfloat[NRS \texttt{watdiv1B} (\textit{y-axis in log scale}).]{
        \includegraphics[width=.49\textwidth]{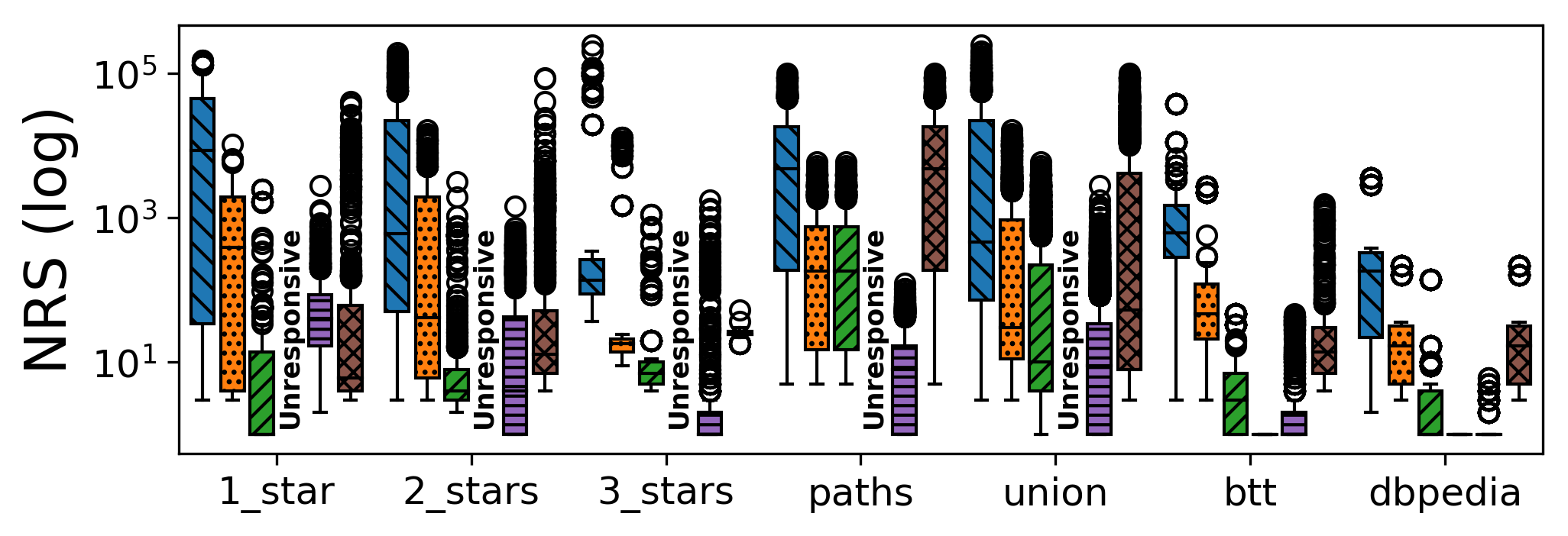}
        \label{fig:req3}
    }
    \subfloat[NRS \texttt{watdiv10B} (\textit{y-axis in log scale}).]{
        \includegraphics[width=.49\textwidth]{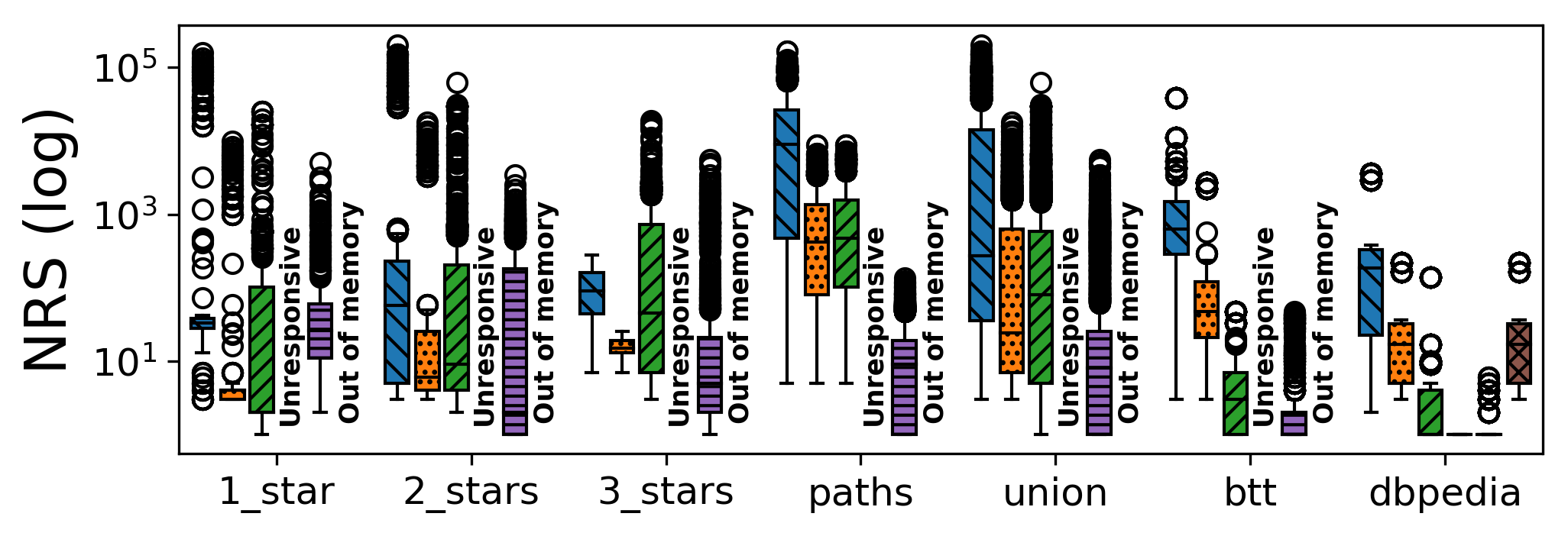}
        \label{fig:req4}
    }\\
    \subfloat[NTB \texttt{watdiv10M} (\textit{y-axis in log scale}).]{
        \includegraphics[width=.49\textwidth]{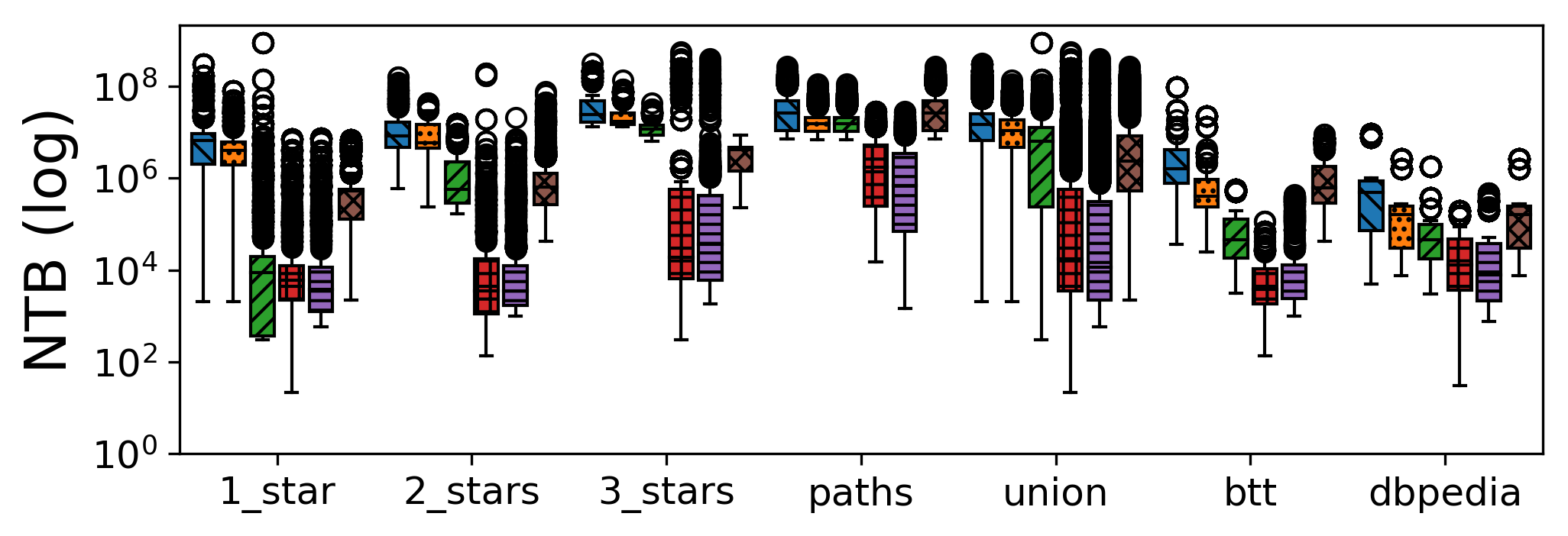}
        \label{fig:ntb1}
    }
    \subfloat[NTB \texttt{watdiv100M} (\textit{y-axis in log scale}).]{
        \includegraphics[width=.49\textwidth]{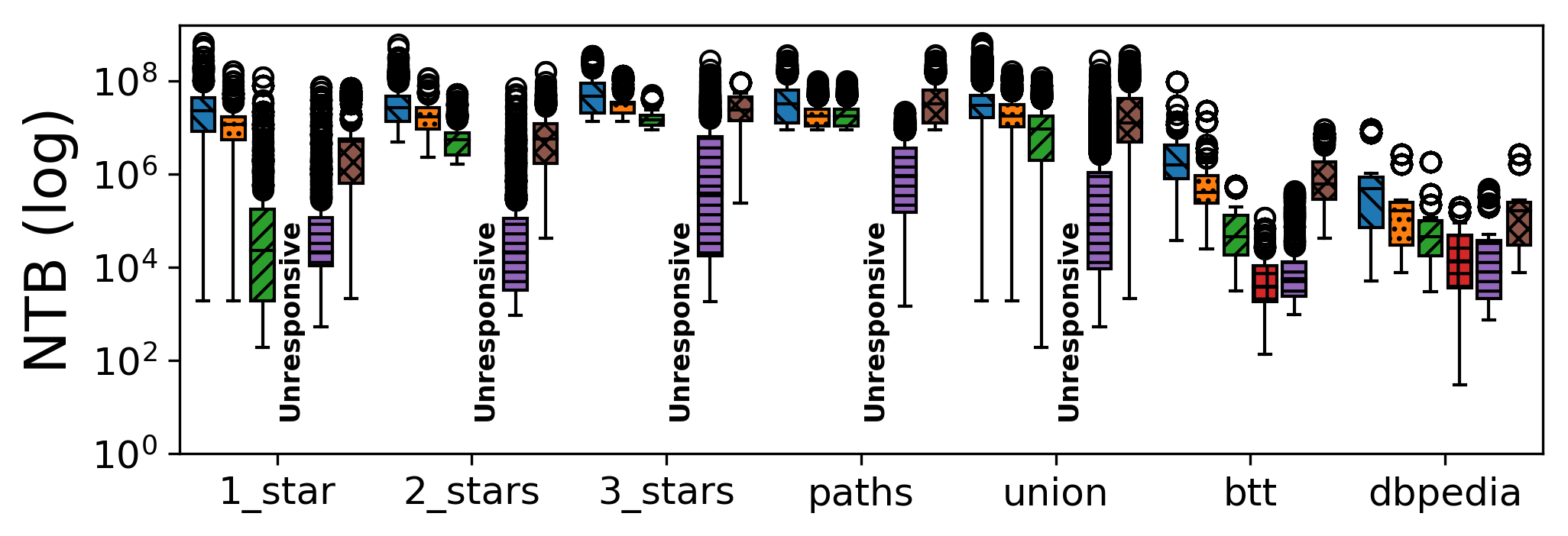}
        \label{fig:ntb2}
    }\\
    \subfloat[NTB \texttt{watdiv1B} (\textit{y-axis in log scale}).]{
        \includegraphics[width=.49\textwidth]{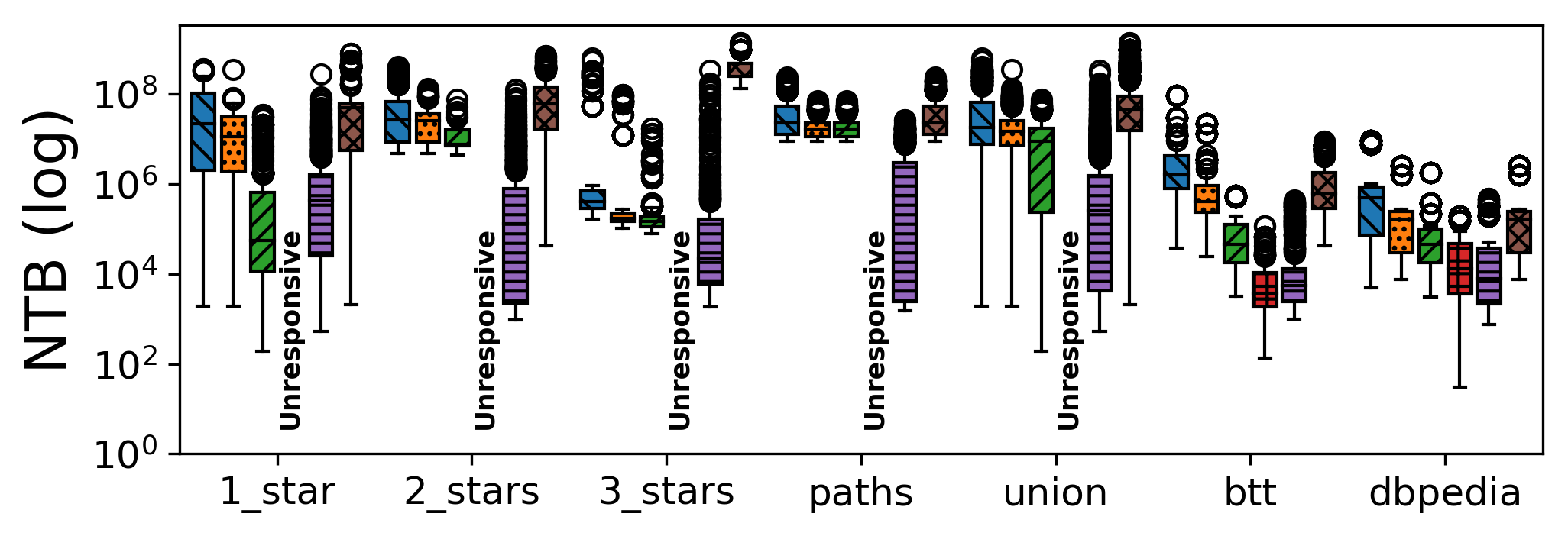}
        \label{fig:ntb3}
    }
    \subfloat[NTB \texttt{watdiv10B} (\textit{y-axis in log scale}).]{
        \includegraphics[width=.49\textwidth]{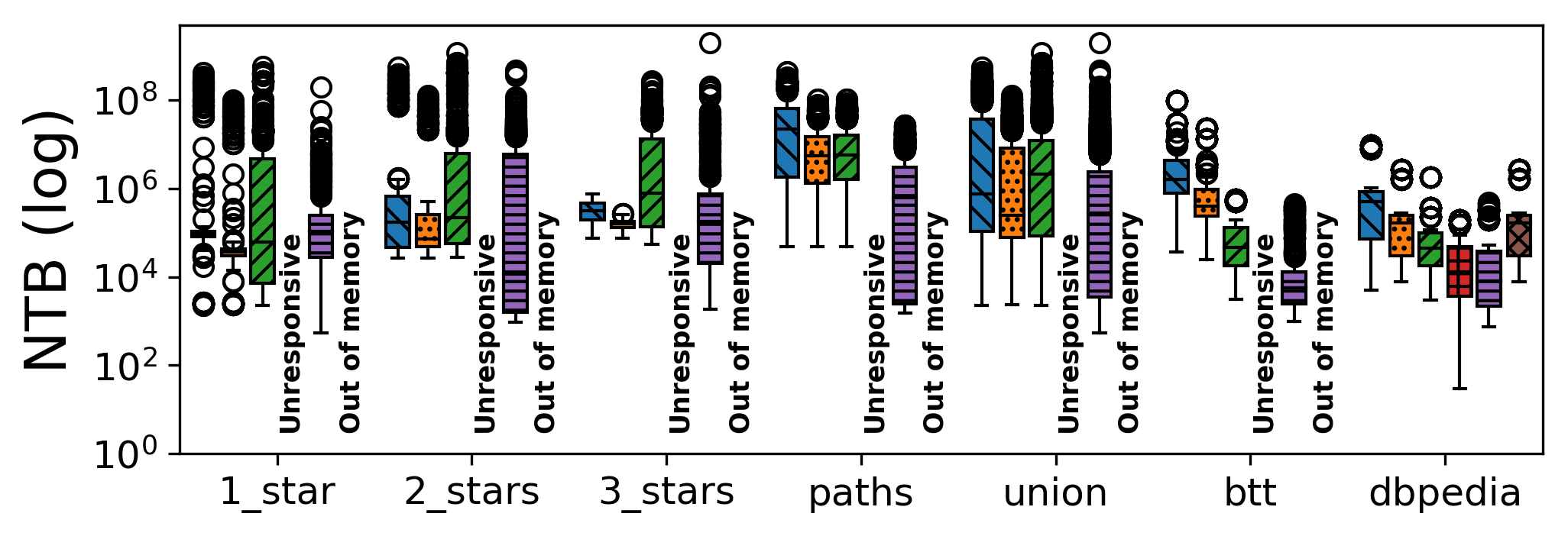}
        \label{fig:ntb4}
    }
    \begin{minipage}{.6\textwidth}
    \vspace{-214ex}
    \hspace{-100ex}
	\includegraphics[width=\textwidth]{results/graphs/legend2.png}
	\end{minipage}
    \caption{NRS and NTB with 64 clients excluding queries that timed out for any approach.}\label{fig:networkusage}
\end{figure*}

\begin{figure*}[htb]
\label{fig:rtandetql}
    \centering
    \subfloat[QET \texttt{watdiv10M} (\textit{y-axis in log scale}).]{
        \includegraphics[width=.49\textwidth]{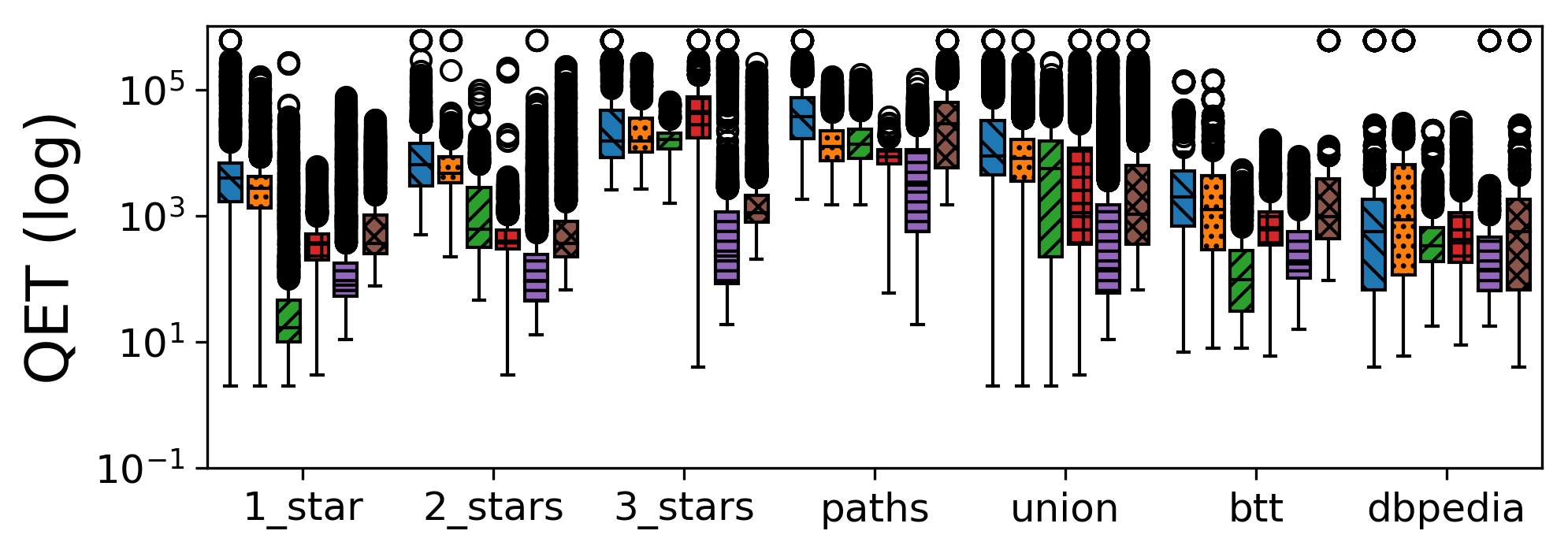}
        \label{fig:etql1}
    }
	\subfloat[QET \texttt{watdiv100M} (\textit{y-axis in log scale}).]{
        \includegraphics[width=.49\textwidth]{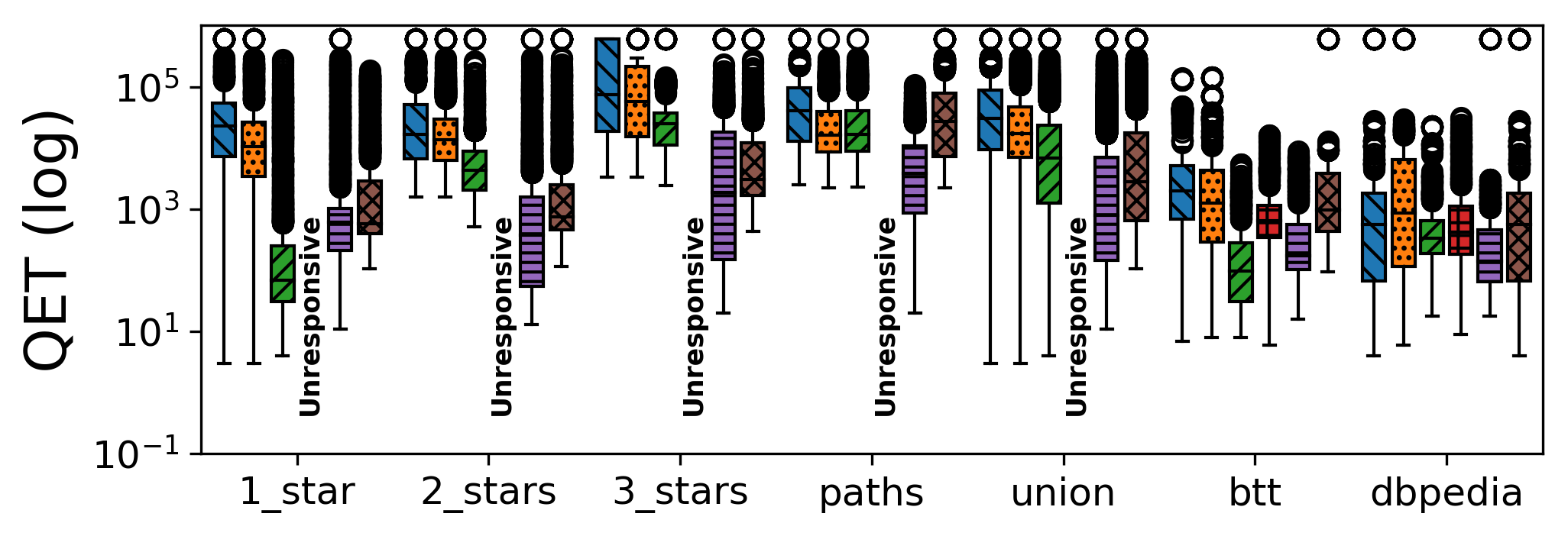}
        \label{fig:etql2}
    }\\
    \subfloat[QET \texttt{watdiv1B} (\textit{y-axis in log scale}).]{
        \includegraphics[width=.49\textwidth]{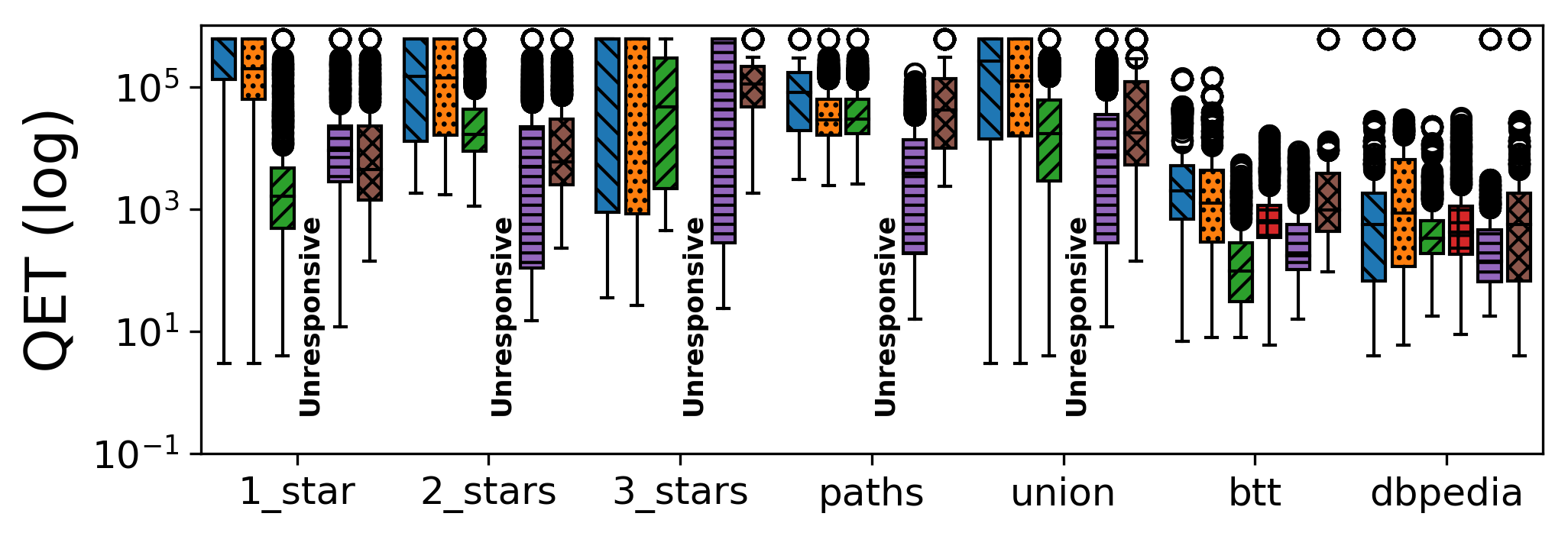}
        \label{fig:etql3}
    } 
    \subfloat[QET \texttt{watdiv10B} (\textit{y-axis in log scale}).]{
        \includegraphics[width=.49\textwidth]{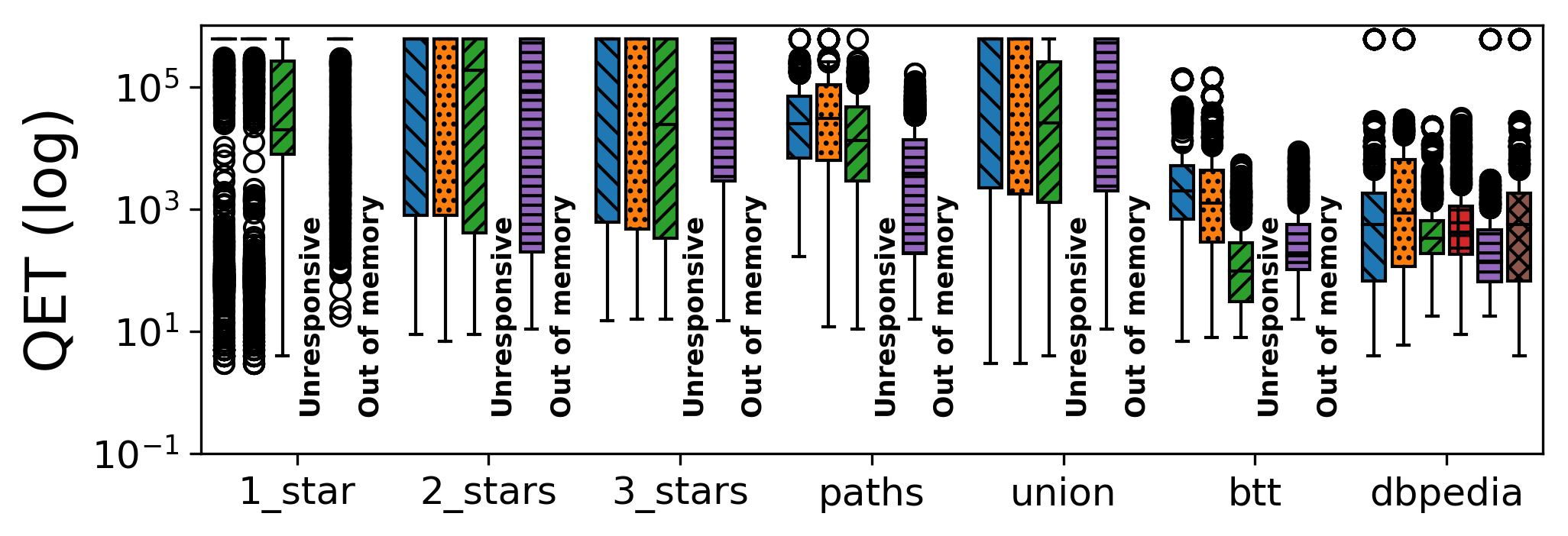}
        \label{fig:etql4}
    }\\
    \subfloat[QRT \texttt{watdiv10M} (\textit{y-axis in log scale}).]{
        \includegraphics[width=.49\textwidth]{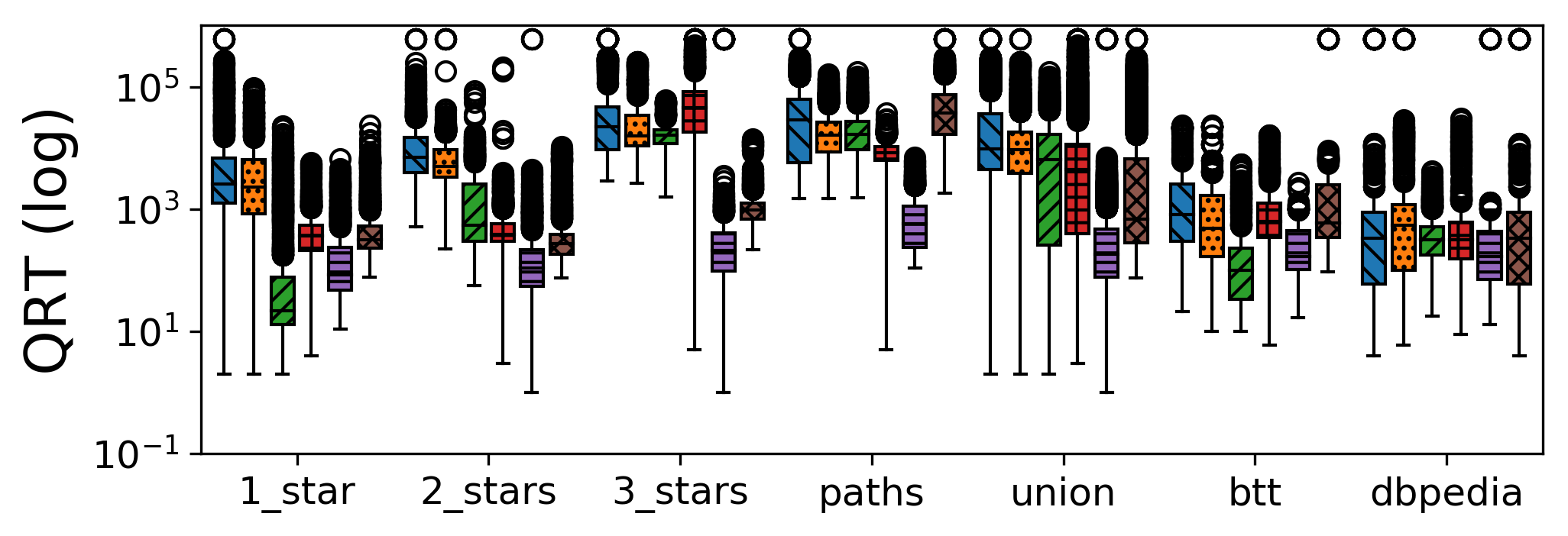}
        \label{fig:rt1}
    }
    \subfloat[QRT \texttt{watdiv100M} (\textit{y-axis in log scale}).]{
        \includegraphics[width=.49\textwidth]{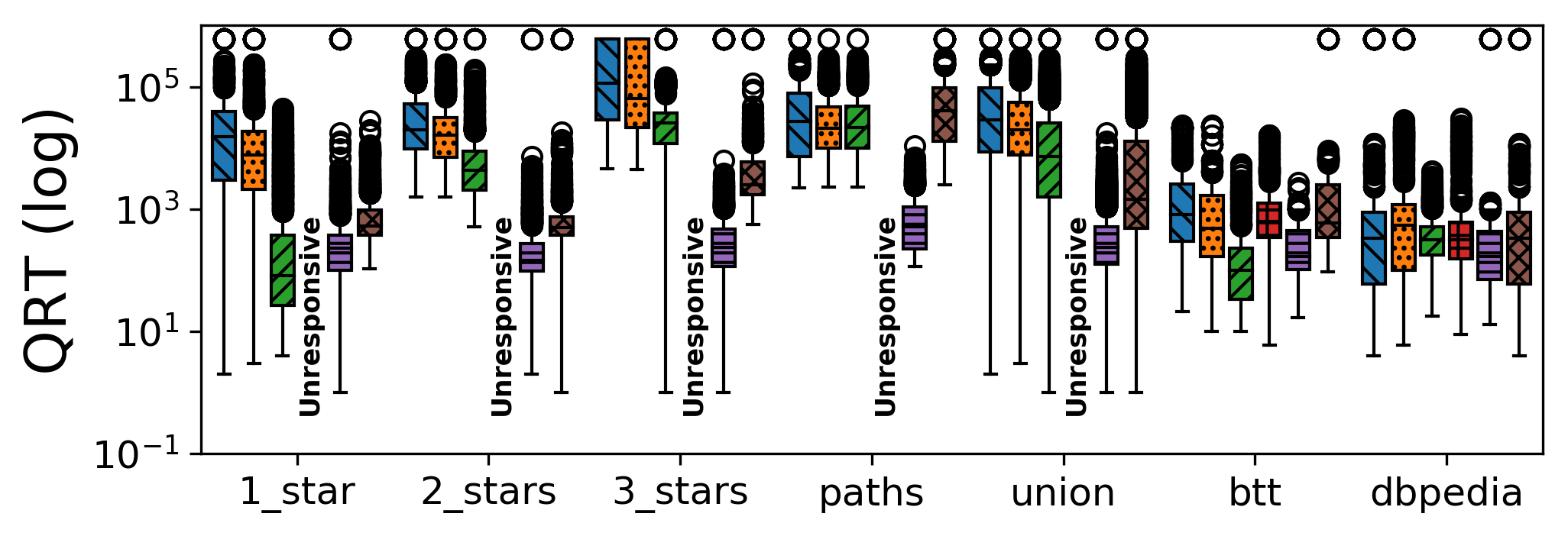}
        \label{fig:rt2}
    }\\
    \subfloat[QRT \texttt{watdiv1B} (\textit{y-axis in log scale}).]{
        \includegraphics[width=.49\textwidth]{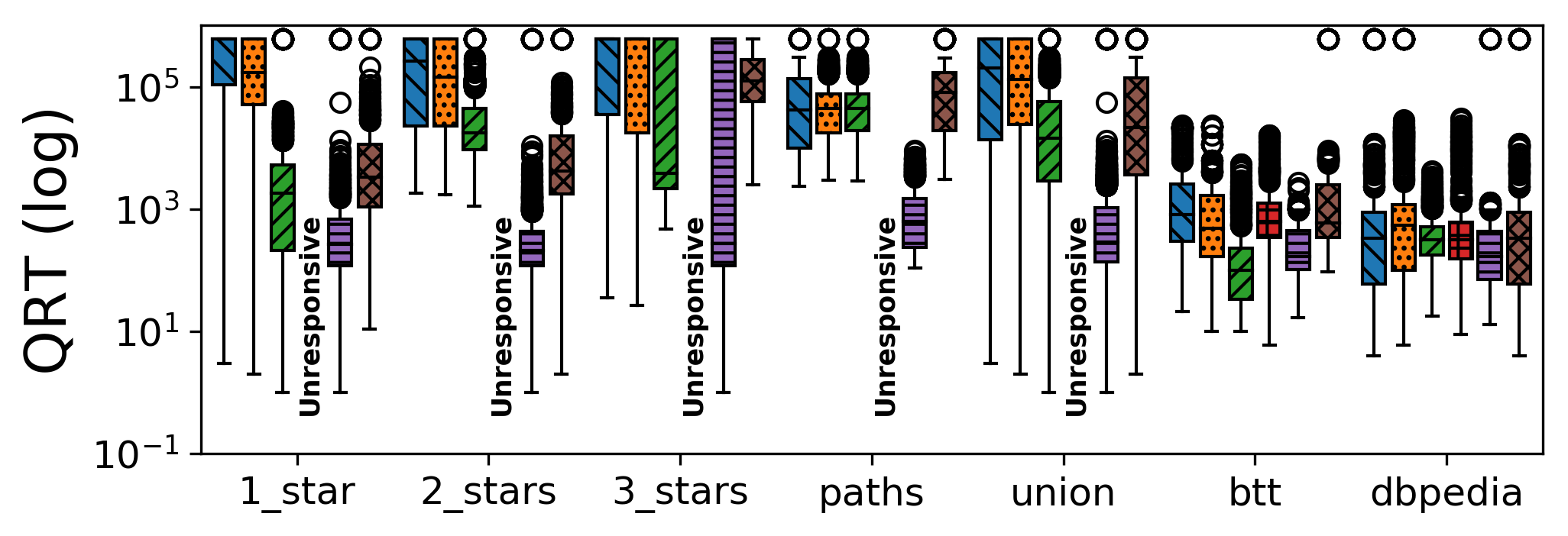}
        \label{fig:rt3}
    }
    \subfloat[QRT \texttt{watdiv10B} (\textit{y-axis in log scale}).]{
        \includegraphics[width=.49\textwidth]{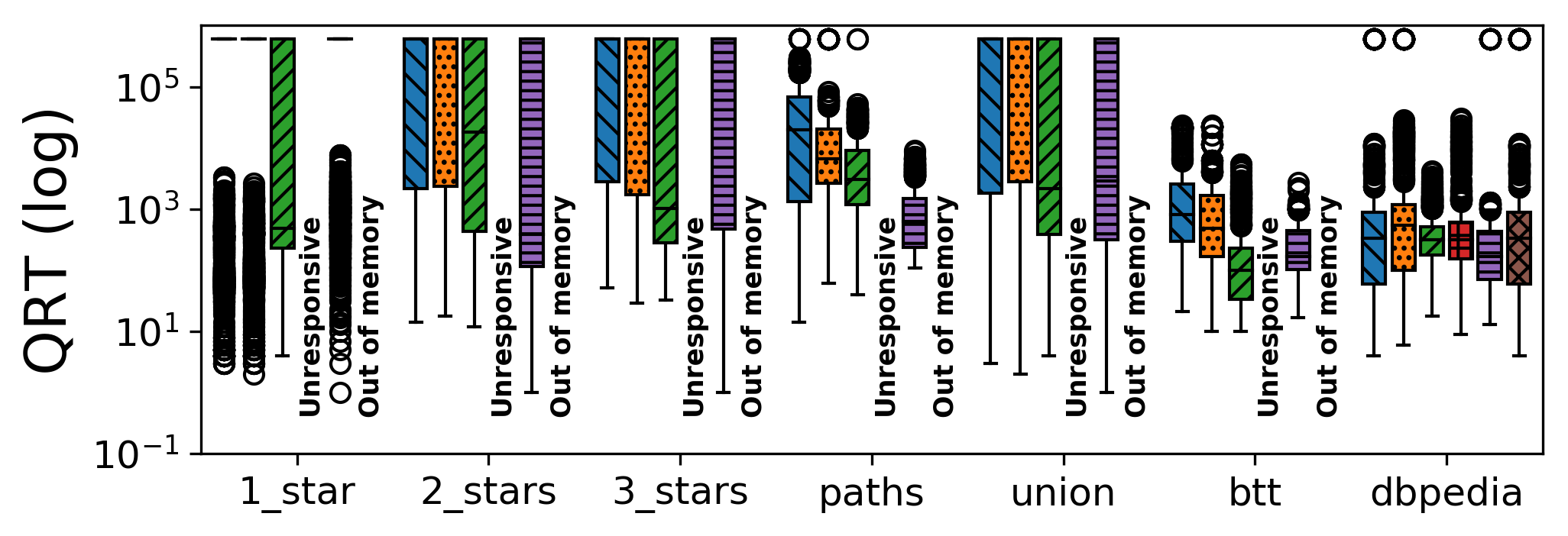}
        \label{fig:rt4}
    }
    \begin{minipage}{.6\textwidth}
    \vspace{-214ex}
    \hspace{-100ex}
	\includegraphics[width=\textwidth]{results/graphs/legend2.png}
	\end{minipage}
    \caption{QET (in ms) and QRT (in ms) with 64 clients including queries that timed out for any approach.}\label{fig:qert}
\end{figure*}

\subsection{Network Traffic} 
As previously highlighted, this section assesses whether or not sending more selective requests, i.e., subqueries that may be composed of more than one triple pattern, has an impact on the network traffic.
Especially for queries with large star patterns, it was expected that utilizing such subqueries results in fewer requests to the server and less data transfer (i.e., intermediate results) between server and client. 

Figures~\ref{fig:req1}-\ref{fig:req4} show NRS for the experiments with 64 clients (since 64 was the highest number of clients all approaches were able to finish for \texttt{watdiv10M}) over all WatDiv datasets for all WatDiv query loads as well as \texttt{dbpedia-lsq} over \texttt{dbpedia}.
Note, that the figures in Figure~\ref{fig:networkusage} exclude queries that timed out for any approach, since such queries have incomplete values for the approaches it timed out for.
Figure~\ref{fig:networkusage_app} in Appendix~\ref{app:a} contains figures that include queries that timed out.
In any case, the figures show that SPF sends significantly fewer requests to the server than both brTPF and TPF over all datasets and query loads.
This is due to the fact that in order to process a triple pattern, TPF sends one request for each intermediate binding while brTPF sends one request per 30 intermediate bindings (since $|\Omega|=30$).
SPF, however, sends considerably fewer requests since the intermediate results for the triple patterns within a star pattern are processed by the server.
This is especially the case for \texttt{watdiv-1\_star}, since SPF only has to make one request per 100 results (since the page size was set to 100).
As the queries include more star patterns, SPF sends more requests, although at all times fewer than brTPF and TPF.
SPF sends the same number of requests to the server as brTPF for the \texttt{watdiv-paths} query load as SPF's query processing is the same as brTPF's query processing when no stars are included in the query.
While Smart-KG normally only sends one request per star pattern, the presence of infrequent triple patterns often means that it ends up sending a larger number of requests than SPF.
In fact, this is the case for all query loads containing star patterns, and SPF clearly has a lower number of requests than Smart-KG.
SaGe was expected to issue fewer requests than SPF since it only sends more than one request when the query has been suspended after the time quantum; however, in some cases, SPF actually has a comparable number of requests to the server.
This is mostly due to the more complex queries taking longer, thus being suspended a higher number of times.

Similarly, since the SPF server processes larger parts of the queries, fewer intermediate results are returned to the clients, resulting in a lower NTB (Figures~\ref{fig:ntb1}-\ref{fig:ntb4}).
The NTB is significantly lower for SPF in comparison to TPF, brTPF, and Smart-KG throughout all query loads except \texttt{watdiv-paths}, where the results are similar for SPF and brTPF. This shows that compared to TPF, brTPF, and Smart-KG, SPF significantly reduces the network traffic.
SaGe, however, transfers less data overall than all other LDF interfaces.
This is due to the fact that SaGe, in contrast to SPF and brTPF, does not transfer any intermediate results between the server and client.
The endpoint has the lowest NTB and NRS since only one request per query is sent to the server and only the final results are transferred back to the client.
However, as shown in Figure~\ref{fig:throughput}, this results in higher CPU usage on the server and lower throughput under load overall.
The experimental results are consistent across all dataset sizes.
Given that SPF clearly has a low network usage compared to other LDF interfaces while improving performance under load, SPF seems to be a suitable alternative to handle large query loads.

\subsection{Impact of Query Pattern}\label{subsec:ioqp}
Figure~\ref{fig:qert} shows QET and QRT for all WatDiv query loads over all WatDiv datasets in the configuration with 64 concurrent clients, and includes queries that timed out.
For queries with star patterns, it is clear that SPF has better performance than both TPF and brTPF over all configurations.
%
The difference between SPF and other interfaces is more significant for the 1-star query load. This is expected since fewer requests are made for these queries. 
In fact, some queries in the \texttt{watdiv-1\_star} query load can be answered with just a single call to the server. 
As shown in Figure~\ref{fig:qert}, SPF outperforms other interfaces more significantly for the \texttt{watdiv-1\_star} and \texttt{watdiv-2\_stars} query loads. 
These two query loads have larger star patterns than the other query loads (Figure~\ref{fig:ch3})  
and therefore TPF and brTPF have to make more requests to the server for these queries and Smart-KG has to transfer larger partitions over the network, whereas SPF still only makes one request to the server per 100 bindings to each star pattern (cf. the page size was set to 100).
For \texttt{watdiv-3\_stars} over \texttt{watdiv1B} and \texttt{watdiv10B}, while the mean QET and QRT is lower for SPF than brTPF and TPF, few queries have a slightly higher QET and QRT.
This supports the earlier point that for the larger datasets and with queries with high selectivity, each star pattern request takes a little longer to process.
This means that queries with more star patterns and higher selectivity are more heavily affected.
Moreover, while SaGe actually has lower QET and QRT for \texttt{watdiv-2\_stars} and \texttt{watdiv-3\_stars} for the smaller datasets, SPF generally outperforms SaGe for \texttt{watdiv1B} and \texttt{watdiv10B}.
This is in line with the workload time in Figure~\ref{fig:scalability} and shows that SPF is able to more efficiently scale with the size of the dataset than SaGe and Smart-KG.
Note also that SPF was able to finish many more queries (since fewer queries timed out, Figure~\ref{fig:timeout}) than the other approaches, and thus SPF outperforms brTPF, TPF, Smart-KG, and SaGe for large datasets under load.
For queries with no star patterns, it was expected that SPF does not have a worse performance than brTPF.
This is in line with the experimental results, as SPF has similar performance as brTPF for \texttt{watdiv-paths}, and better performance for all other query loads.
%

With the exception of SaGe, all approaches have response times quite similar to execution times.
They all receive their first result only slightly earlier than obtaining the full result.
For TPF, brTPF, Smart-KG, and SPF this is most likely due to the fact that most of the joins in the query are already processed upon receiving the first result.
For the endpoint, QRT and QET are the same since it processes the entire query on the server before returning the result.
SaGe, on the other hand, has slightly lower QRT than QET since the first result in some cases is obtained upon query suspension before it has finished execution.
Like QET, the improvement in QRT is more significant for queries with fewer star patterns since fewer calls to the server are needed.
Moreover, SPF and brTPF have quite similar QRT for the \texttt{paths} query load, as expected.

\subsection{Summary}
Overall, the experimental evaluation shows that SPF achieves a novel, and in most cases better, tradeoff between performance and server load than TPF, brTPF, SaGe, Smart-KG, and a SPARQL endpoint.
SPF does this by significantly reducing the network traffic without incurring too much extra load on the server.
For queries without star patterns, SPF still performs as good as brTPF, both in terms of execution time and network traffic.

Moreover, SPF is able to provide better scalability with the size of the dataset by outperforming all of its competitors for the largest datasets.
While SPF does have slightly higher CPU load on the server (SPF increases server usage by up to 1.08 times compared to brTPF and 1.18 times compared to TPF), it is still significantly more efficient than the alternative approaches for the largest datasets in the presence of high load (SPF increases throughput by up to 45 times compared to SaGe and 96 times compared to brTPF for the \texttt{dbpedia} dataset over 128 clients).
This is true both for large-scale synthetic datasets and real-world datasets, and suggests that SPF is able to combine a lower network load with a higher query throughput at a comparatively low server load.

\section{Conclusions}
\label{sec:conclusion}
In this paper, Star Pattern Fragments (SPF), a new RDF interface that exploits a different tradeoff for distribution of the workload between the server and client, was presented.
The SPF client processes queries by processing SPARQL operators and decomposing each BGP into star-shaped subqueries and sending these subqueries, along with intermediate bindings, to the server. 
An SPF server that is able to answer HTTP requests containing star patterns was implemented as well as an SPF client that is able to answer SPARQL queries.
The experimental results show that SPF reduces the network traffic, both in terms of the number of requests to the server and the amount of transferred data between the client and server, while it increases the query throughput by a factor of up to 45 times compared to SaGe, 96 times compared to brTPF, and 137 times compared to TPF and Smart-KG.
The evaluation also demonstrates that SPF increases the overall performance while only increasing the CPU load on the server by a factor of 1.08 compared to brTPF and 1.18 compared to TPF when 128 clients issue queries. 
%

While a novel distribution of the workload between the client and server was presented, SPF presents an opportunity to explore different ways to utilize this distribution of workload.
While relatively few queries include many object-based star patterns (Table~\ref{tbl:jvtype}), investigating the tradeoff between including such star patterns on the server and the expense of a more complex query decomposition strategy on the client (and overhead of such a strategy) is part of the future work for SPF.
Furthermore, in some cases performance could be increased by  using a query decomposition that does not necessarily include the largest possible star patterns, in order to ensure the optimal join order on the triple pattern level.
In that sense, it could also be interesting to assess whether other query decomposition techniques, not focused on star patterns, could provide any benefits.
Furthermore, it could be interesting to include an SPF-specific cache on the server and to assess its impact on the performance of SPF, as well as accommodating adaptive query processing that considers the complexity of the query and the available resources on the server.
Another interesting aspect of future work would be to consider more complex query types, such as the support of aggregation and analytical queries in the context of semantic data warehousing~\cite{DBLP:conf/esws/IbragimovHPZ15,DBLP:conf/semweb/IbragimovHPZ16,DBLP:conf/semweb/JakobsenAHP15,DBLP:conf/semweb/GalarragaJHP18,DBLP:conf/esws/KampgenH13,DBLP:conf/jist/JakobsenHP16,DBLP:conf/birte/IbragimovHPZ14,DBLP:conf/dolap/NathHP15,DBLP:conf/www/NathHP0B20}.
Lastly, it could be interesting to integrate SPF into systems, such as~\cite{DBLP:conf/semweb/MontoyaAH18,ppbfs} that rely on the different strengths of different RDF interfaces to process SPARQL queries more efficiently.

\section{Epilog}
\label{sec:epilog}
The first version of this paper was uploaded in February 2020 -- the updated version now contains additional experiments and insights. 
In the meantime, SPF has become an integral part of WiseKG~\cite{DBLP:conf/www/AzzamAMKPH21}, which combines the strengths of SPF and SmartKG~\cite{smartkg}.
Like SPF, WiseKG divides the queries into star patterns.
However, for each star pattern, WiseKG uses a cost model to determine whether it is most efficient to process the star on the client (SmartKG-like query processing) or on the server (SPF-like query processing).
As shown in~\cite{DBLP:conf/www/AzzamAMKPH21}, WiseKG increases query processing performance on top of the already increased performance achieved by SPF, and is therefore currently, to the best of our knowledge, the most performant LDF system.

Furthermore, we also introduced ColChain~\cite{DBLP:conf/www/AebeloeMH21}, which improves the availability of RDF datasets by replicating the data across multiple nodes in a P2P network while also allowing nodes to collaborate on keeping the data up-to-date and process queries dynamically over earlier versions.
While ColChain was demonstrated in~\cite{DBLP:conf/semweb/AebeloeMH21}, we plan to extended it with SPF to improve query processing performance as well as provenance capabilities.

\balance

\vspace{2ex}
\noindent
\textbf{Acknowledgments}
This research was partially funded by the Danish Council for Independent Research (DFF) under grant agreement no. DFF-8048-00051B, Aalborg University's Talent Programme, and the Poul Due Jensen Foundation.

\bibliographystyle{ACM-Reference-Format}
\bibliography{references}

\begin{appendix}
\section{Additional Experimental Results}\label{app:a}

\begin{figure*}[h]
    \centering	
    \subfloat[Throughput for \texttt{watdiv-1\_star} over \texttt{watdiv10M} (\textit{log})]{
        \includegraphics[width=.49\textwidth]{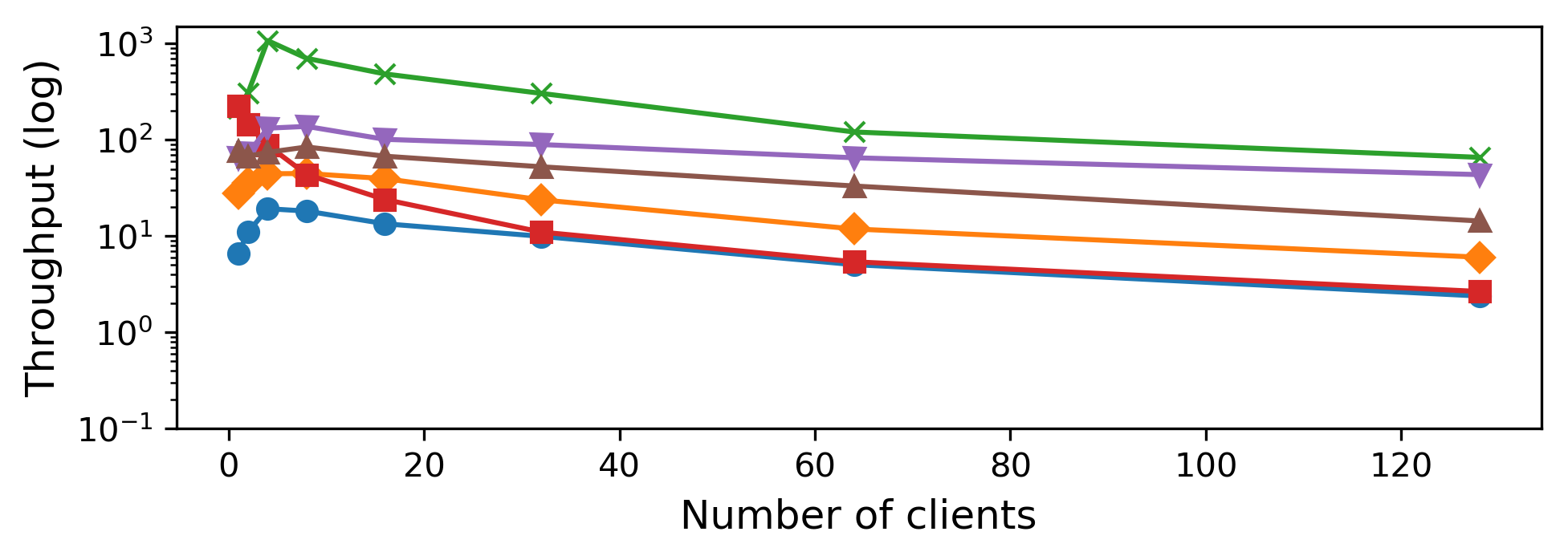}
        \label{fig:tp1_1s_app1}
    }
    \subfloat[Throughput for \texttt{watdiv-1\_star} over \texttt{watdiv100M} (\textit{log})]{
        \includegraphics[width=.49\textwidth]{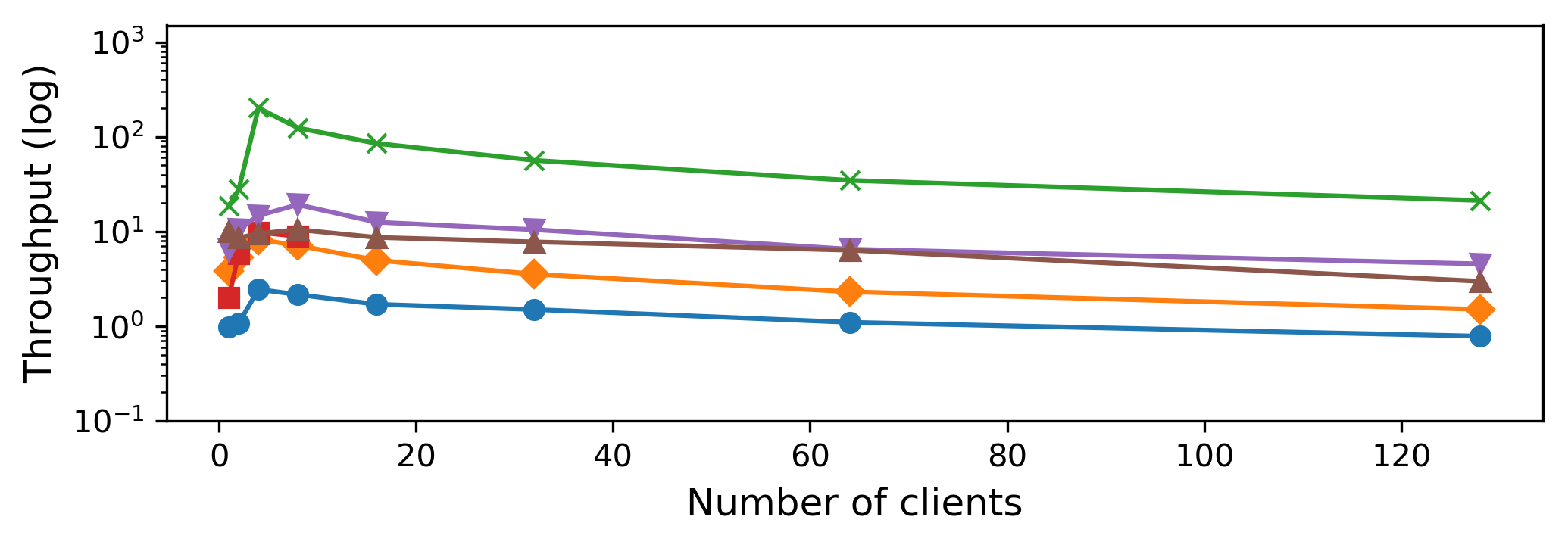}
        \label{fig:tp2_1s_app1}
    }\\\vspace{-1ex}
    \subfloat[Throughput for \texttt{watdiv-1\_star} over \texttt{watdiv1B} (\textit{log})]{
        \includegraphics[width=.49\textwidth]{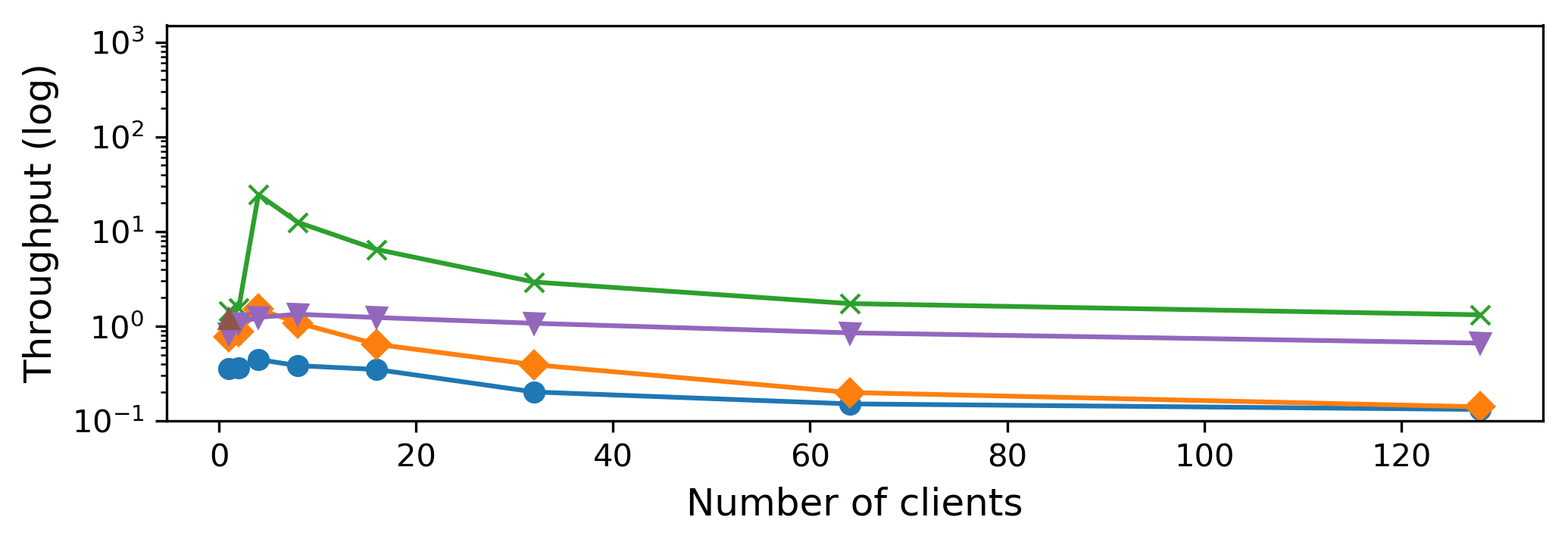}
        \label{fig:tp3_1s_app1}
    }
    \subfloat[Throughput for \texttt{watdiv-1\_star} over \texttt{watdiv10B} (\textit{log})]{
        \includegraphics[width=.49\textwidth]{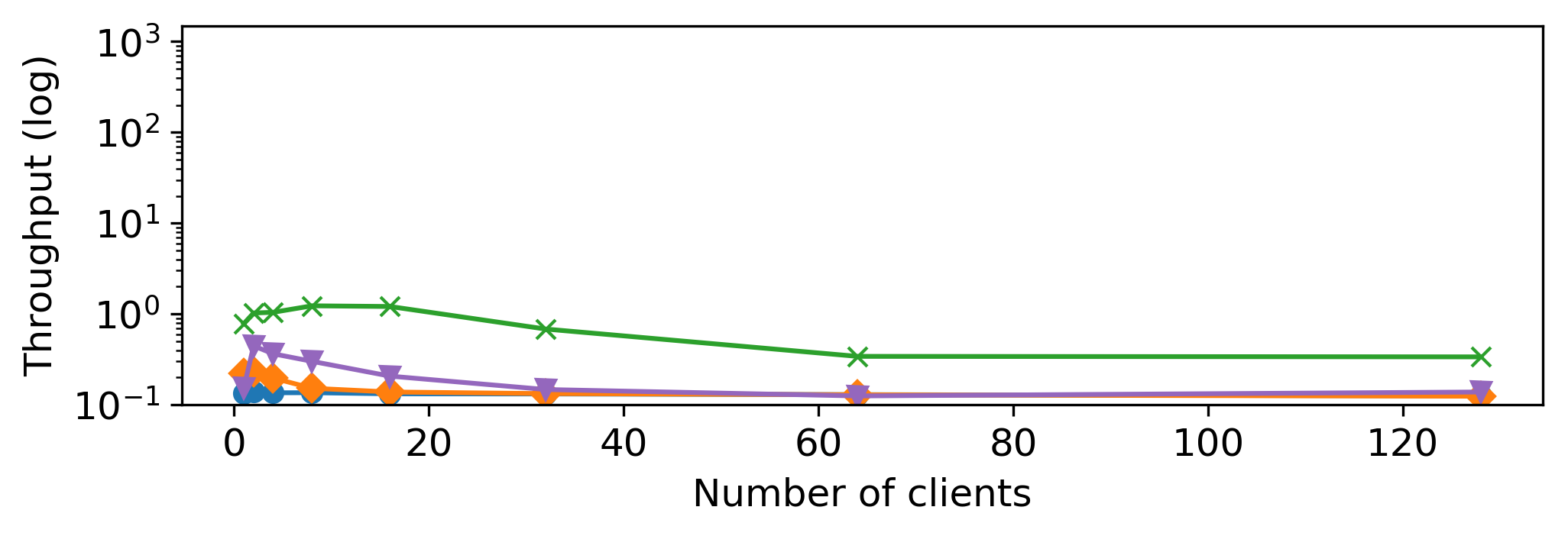}
        \label{fig:tp4_1s_app1}
    }\\\vspace{-1ex}
    \subfloat[Timeouts for \texttt{watdiv-1\_star} over \texttt{watdiv10M}]{
        \includegraphics[width=.49\textwidth]{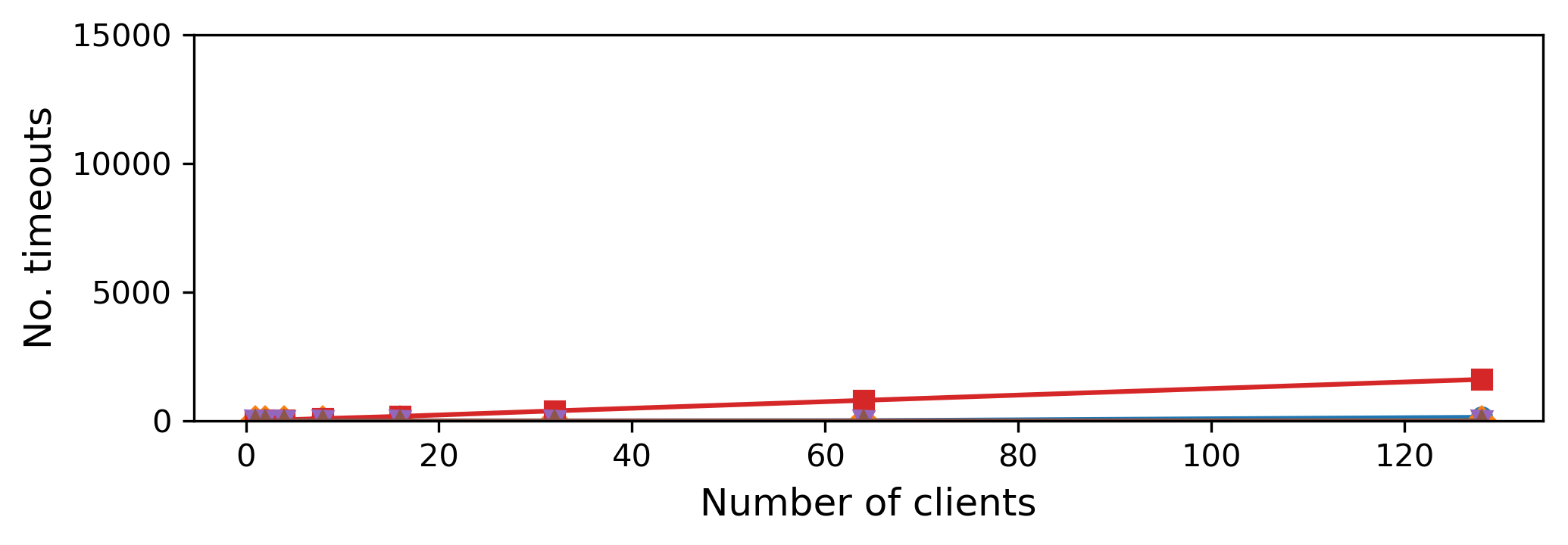}
        \label{fig:to1_1s_app1}
    }
    \subfloat[Timeouts for \texttt{watdiv-1\_star} over \texttt{watdiv100M}]{
        \includegraphics[width=.49\textwidth]{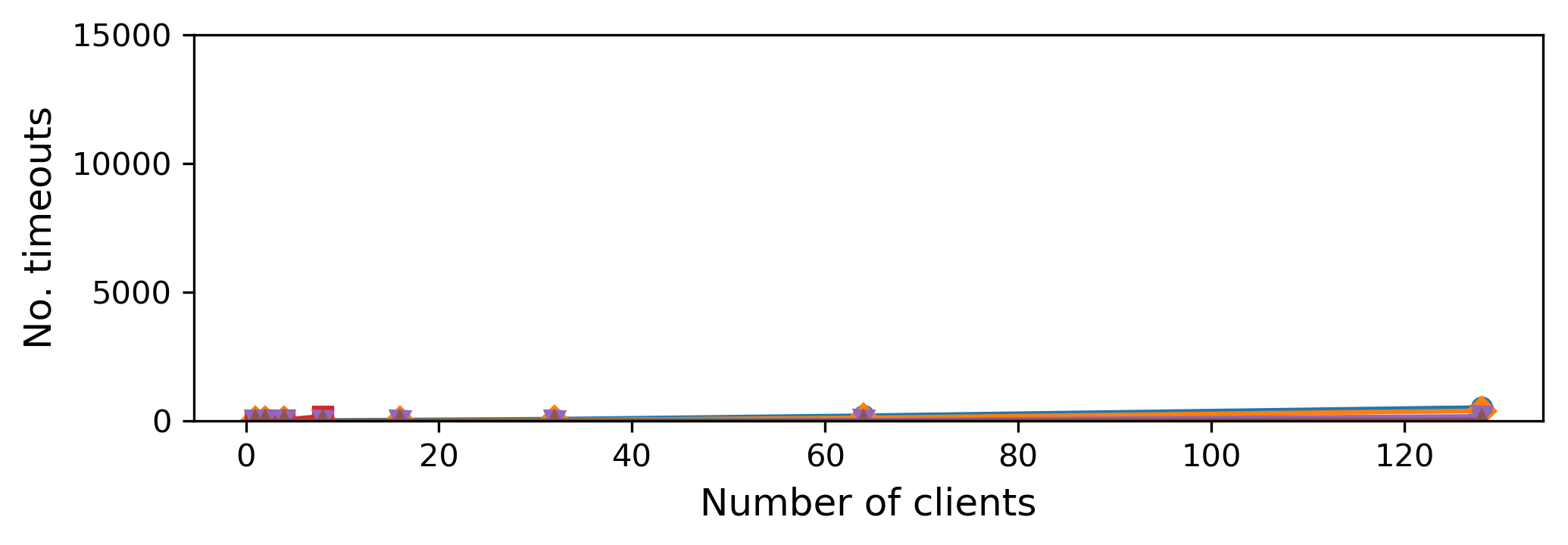}
        \label{fig:to2_1s_app1}
    }\\\vspace{-1ex}
    \subfloat[Timeouts for \texttt{watdiv-1\_star} over \texttt{watdiv1B}]{
        \includegraphics[width=.49\textwidth]{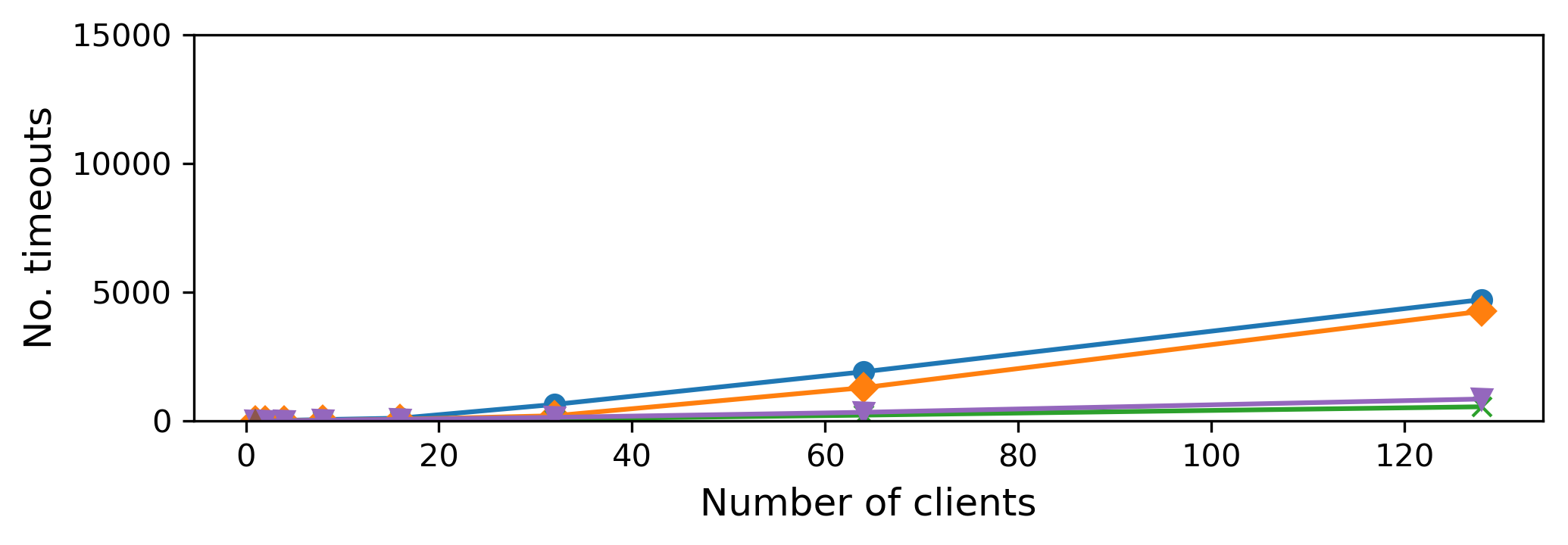}
        \label{fig:to3_1s_app1}
    }
    \subfloat[Timeouts for \texttt{watdiv-1\_star} over \texttt{watdiv10B}]{
        \includegraphics[width=.49\textwidth]{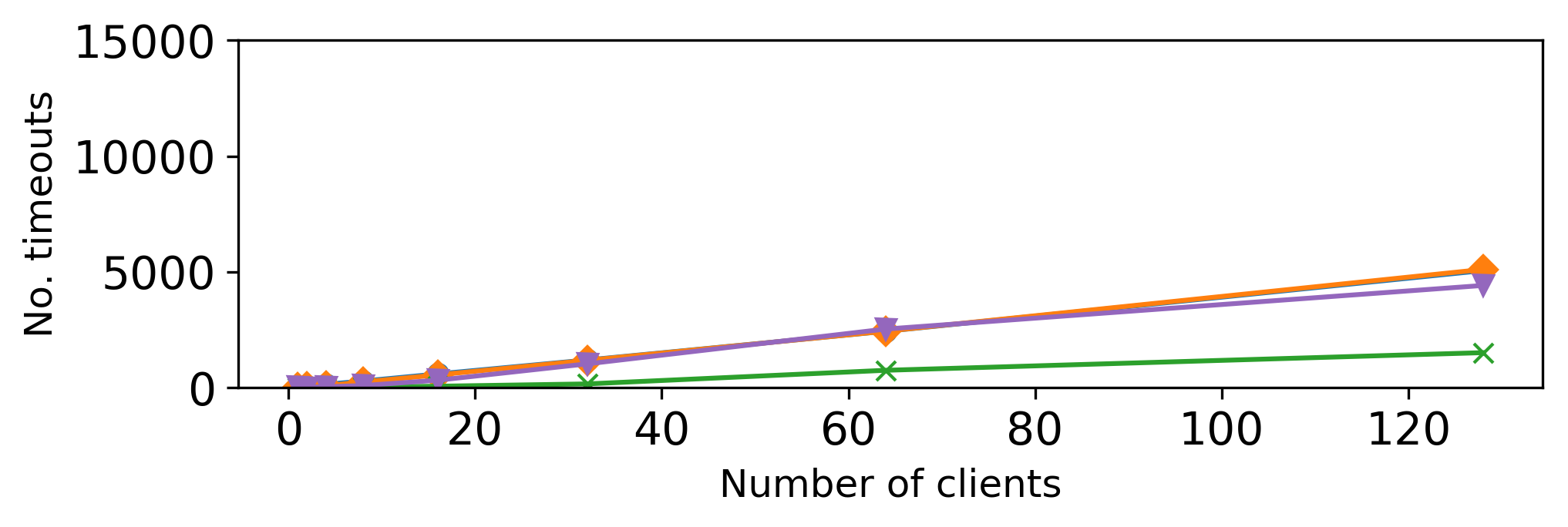}
        \label{fig:to4_1s_app1}
    }
    \begin{minipage}{.6\textwidth}
    \vspace{-214ex}
    \hspace{-100ex}
	\includegraphics[width=\textwidth]{results/graphs/legend1.png}
	\end{minipage}
	\vspace{-1ex}
    \caption{Throughput (\# queries/m) and timeouts for \texttt{watdiv-1\_star} over the different WatDiv datasets. Includes queries that timed out.}\label{fig:throughput1_app1}
\end{figure*}

\begin{figure*}[p]
    \centering	
    \subfloat[Throughput for \texttt{watdiv-2\_stars} over \texttt{watdiv10M} (\textit{log})]{
        \includegraphics[width=.49\textwidth]{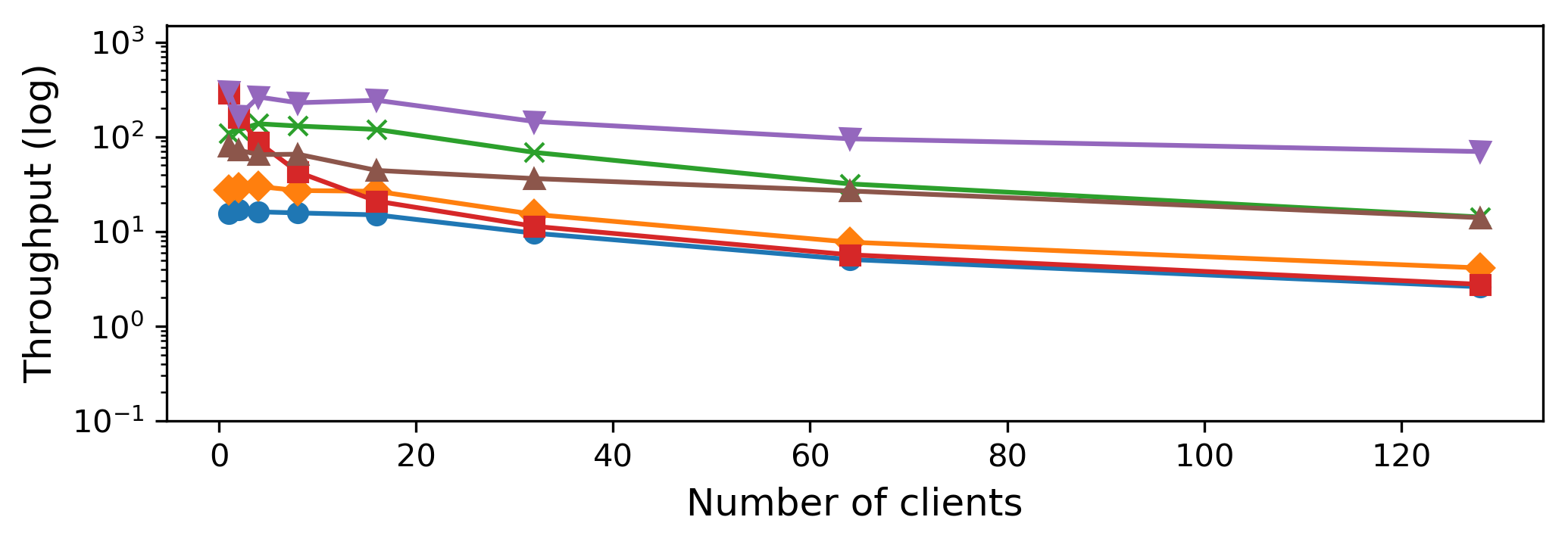}
        \label{fig:tp1_2s_app1}
    }
    \subfloat[Throughput for \texttt{watdiv-2\_stars} over \texttt{watdiv100M} (\textit{log})]{
        \includegraphics[width=.49\textwidth]{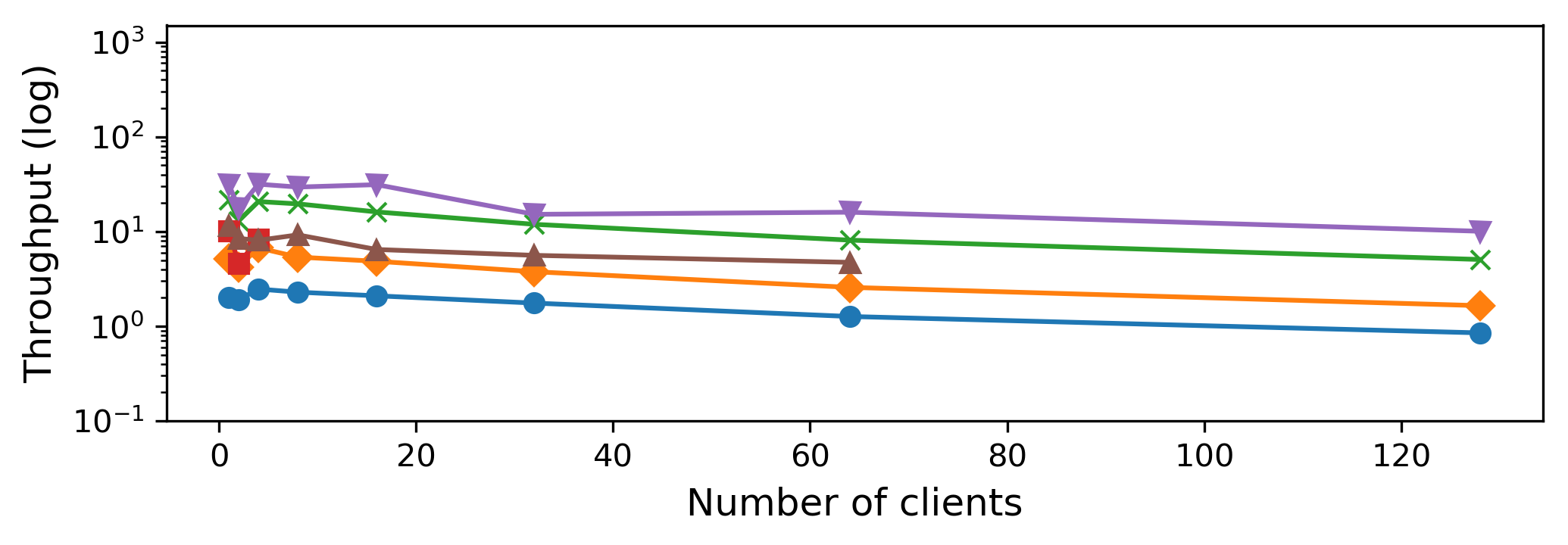}
        \label{fig:tp2_2s_app1}
    }\\\vspace{-1ex}
    \subfloat[Throughput for \texttt{watdiv-2\_stars} over \texttt{watdiv1B} (\textit{log})]{
        \includegraphics[width=.49\textwidth]{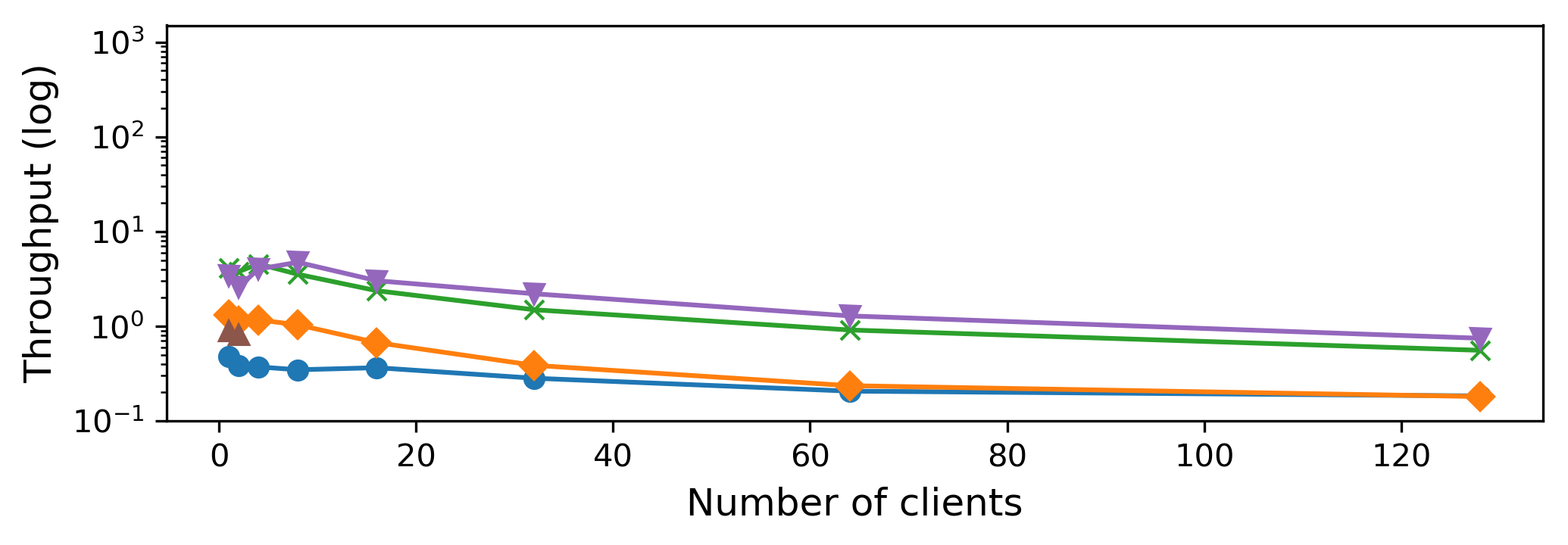}
        \label{fig:tp3_2s_app1}
    }
    \subfloat[Throughput for \texttt{watdiv-2\_stars} over \texttt{watdiv10B} (\textit{log})]{
        \includegraphics[width=.49\textwidth]{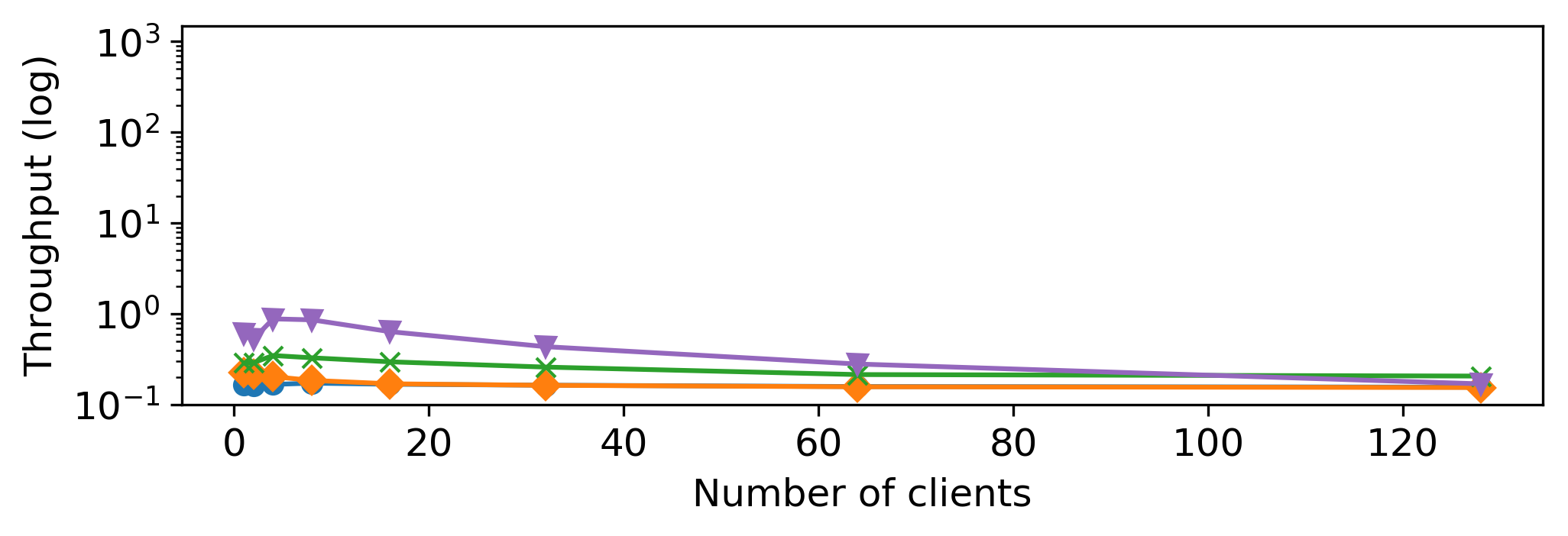}
        \label{fig:tp4_2s_app1}
    }\\\vspace{-1ex}
    \subfloat[Timeouts for \texttt{watdiv-2\_stars} over \texttt{watdiv10M}]{
        \includegraphics[width=.49\textwidth]{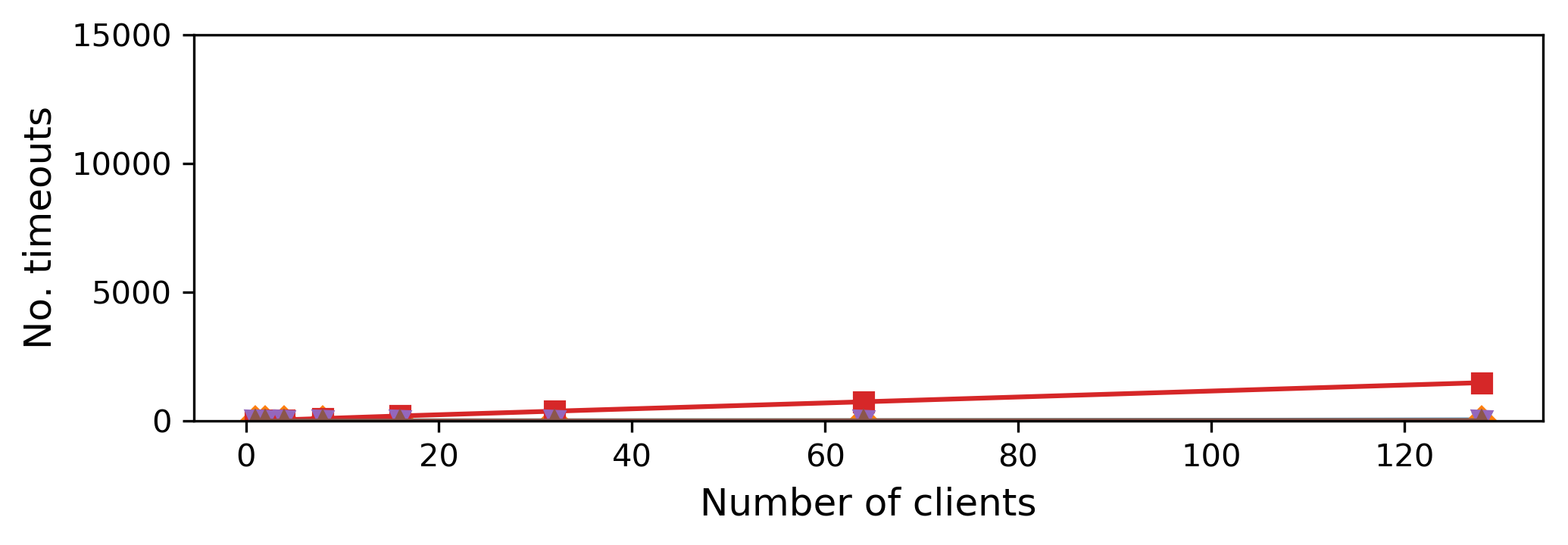}
        \label{fig:to1_2s_app1}
    }
    \subfloat[Timeouts for \texttt{watdiv-2\_stars} over \texttt{watdiv100M}]{
        \includegraphics[width=.49\textwidth]{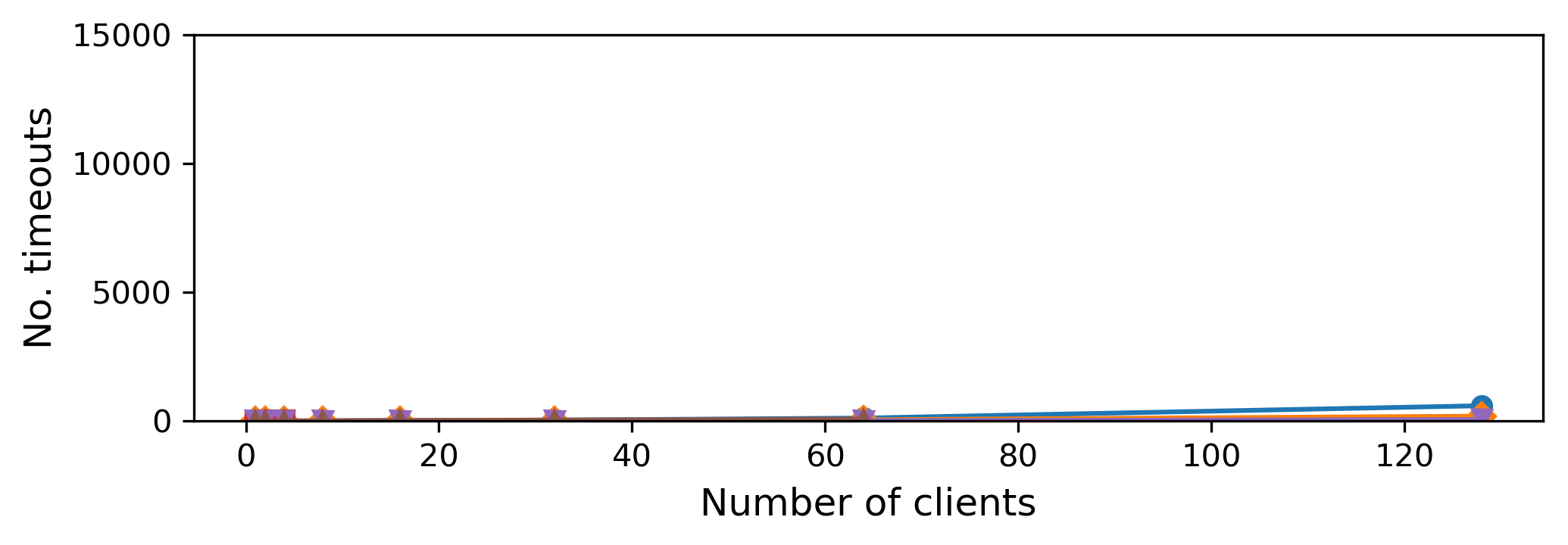}
        \label{fig:to2_2s_app1}
    }\\\vspace{-1ex}
    \subfloat[Timeouts for \texttt{watdiv-2\_stars} over \texttt{watdiv1B}]{
        \includegraphics[width=.49\textwidth]{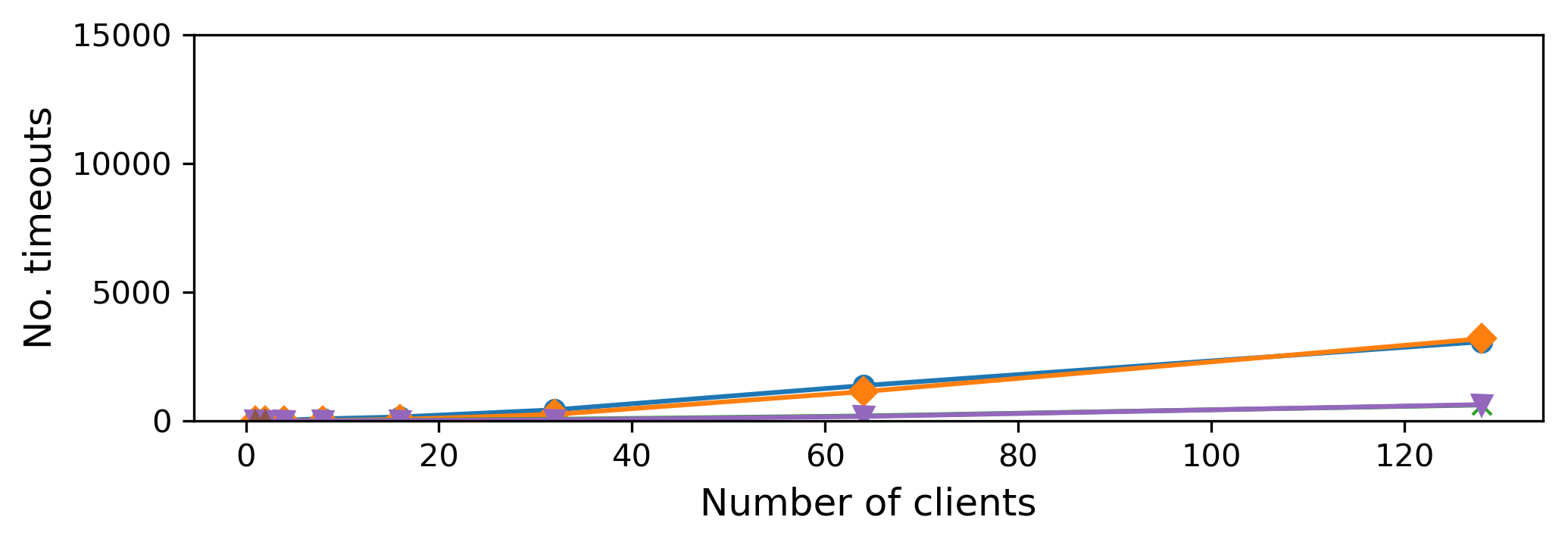}
        \label{fig:to3_2s_app1}
    }
    \subfloat[Timeouts for \texttt{watdiv-2\_stars} over \texttt{watdiv10B}]{
        \includegraphics[width=.49\textwidth]{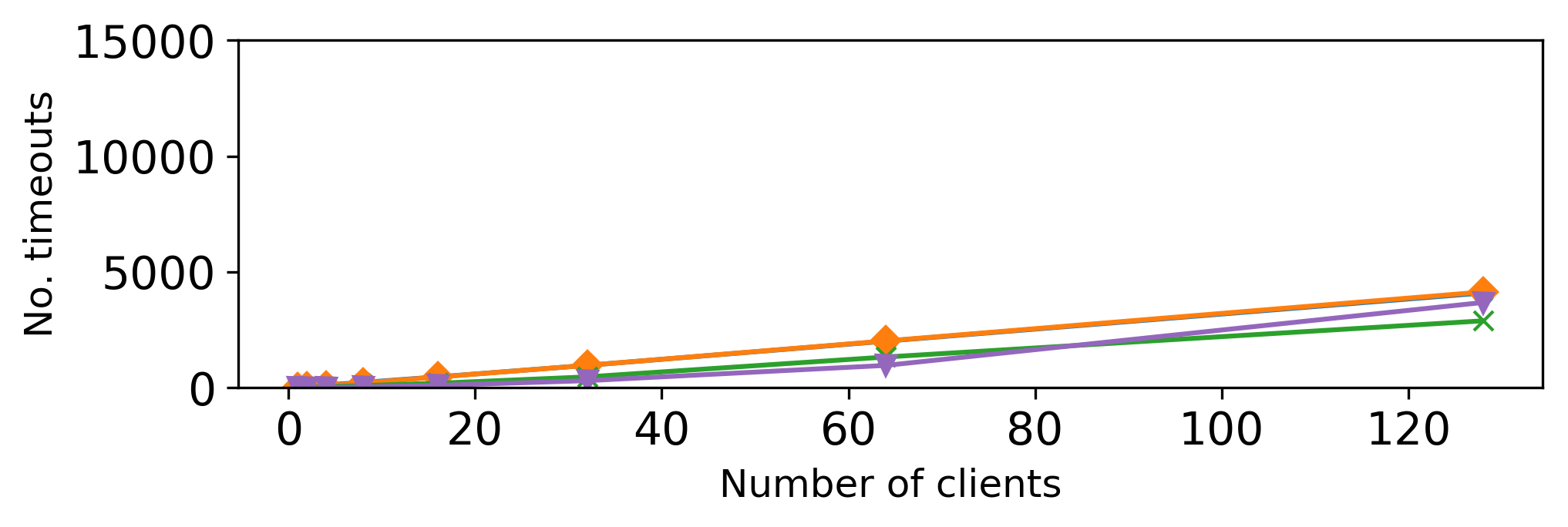}
        \label{fig:to4_2s_app1}
    }
    \begin{minipage}{.6\textwidth}
    \vspace{-214ex}
    \hspace{-100ex}
	\includegraphics[width=\textwidth]{results/graphs/legend1.png}
	\end{minipage}
	\vspace{-1ex}
    \caption{Throughput (\# queries/m) and timeouts for \texttt{watdiv-2\_stars} over the different WatDiv datasets. Includes queries that timed out.}\label{fig:throughput2_app1}
\end{figure*}

\begin{figure*}[p]
    \centering	
    \subfloat[Throughput for \texttt{watdiv-3\_stars} over \texttt{watdiv10M} (\textit{log})]{
        \includegraphics[width=.49\textwidth]{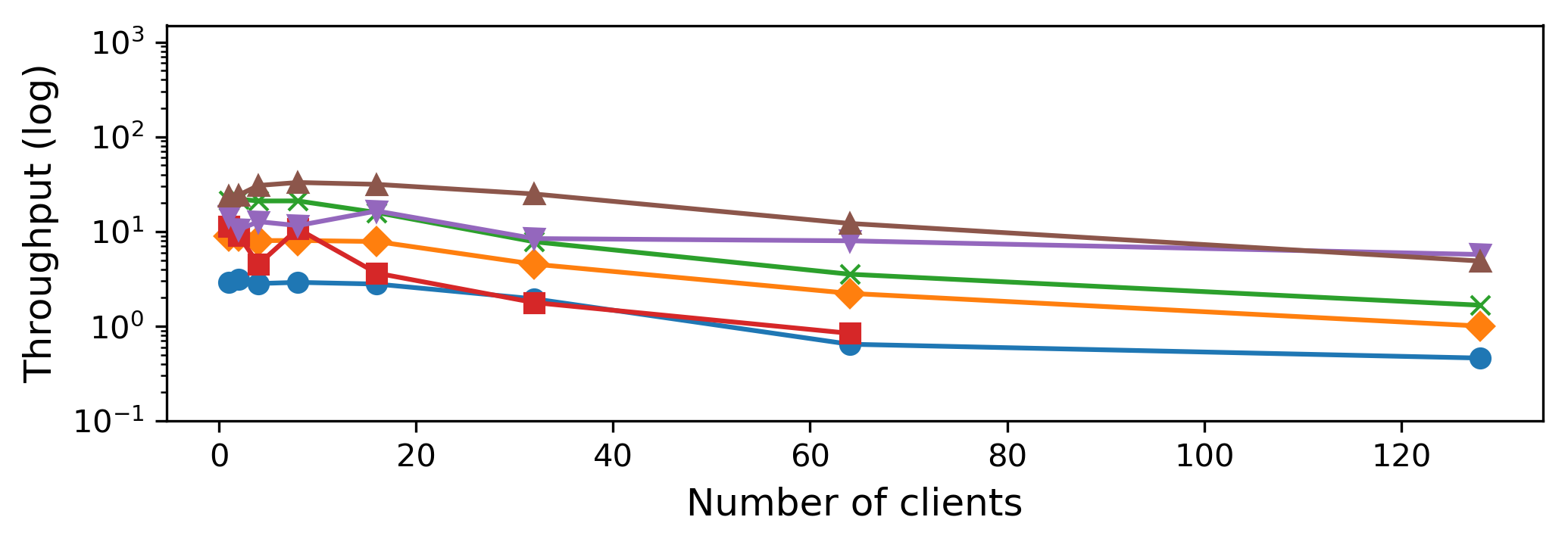}
        \label{fig:tp1_3s_app1}
    }
    \subfloat[Throughput for \texttt{watdiv-3\_stars} over \texttt{watdiv100M} (\textit{log})]{
        \includegraphics[width=.49\textwidth]{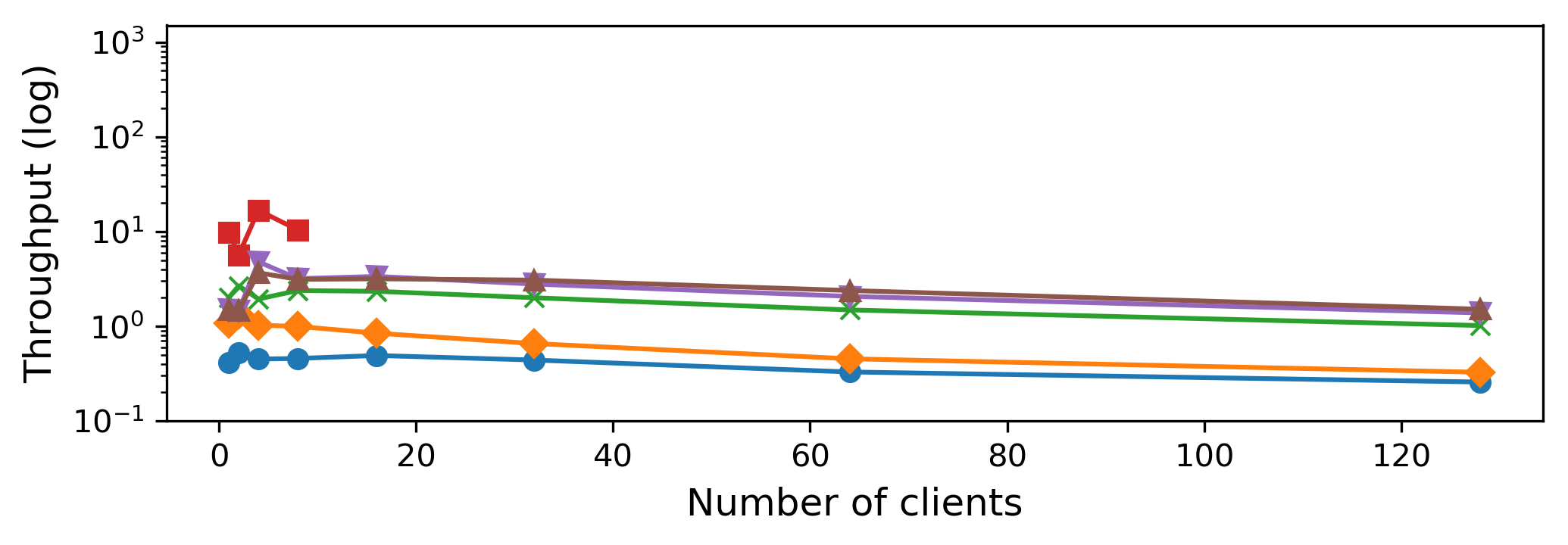}
        \label{fig:tp2_3s_app1}
    }\\\vspace{-1ex}
    \subfloat[Throughput for \texttt{watdiv-3\_stars} over \texttt{watdiv1B} (\textit{log})]{
        \includegraphics[width=.49\textwidth]{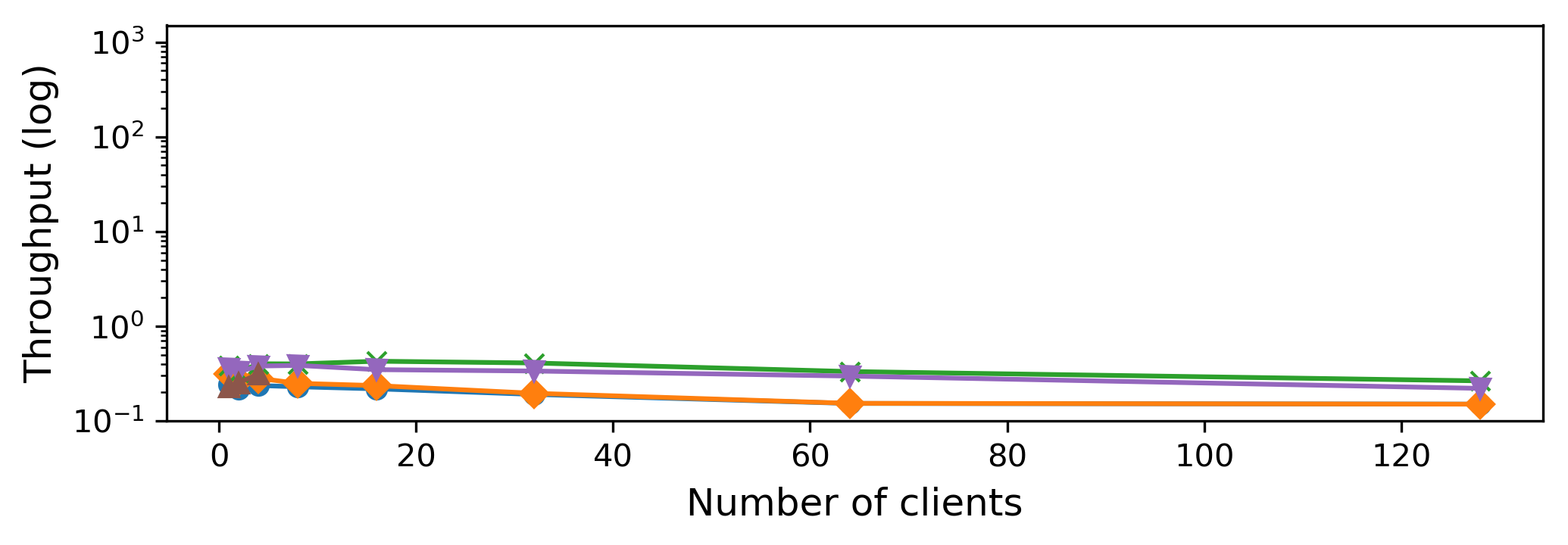}
        \label{fig:tp3_3s_app1}
    }
    \subfloat[Throughput for \texttt{watdiv-3\_stars} over \texttt{watdiv10B} (\textit{log})]{
        \includegraphics[width=.49\textwidth]{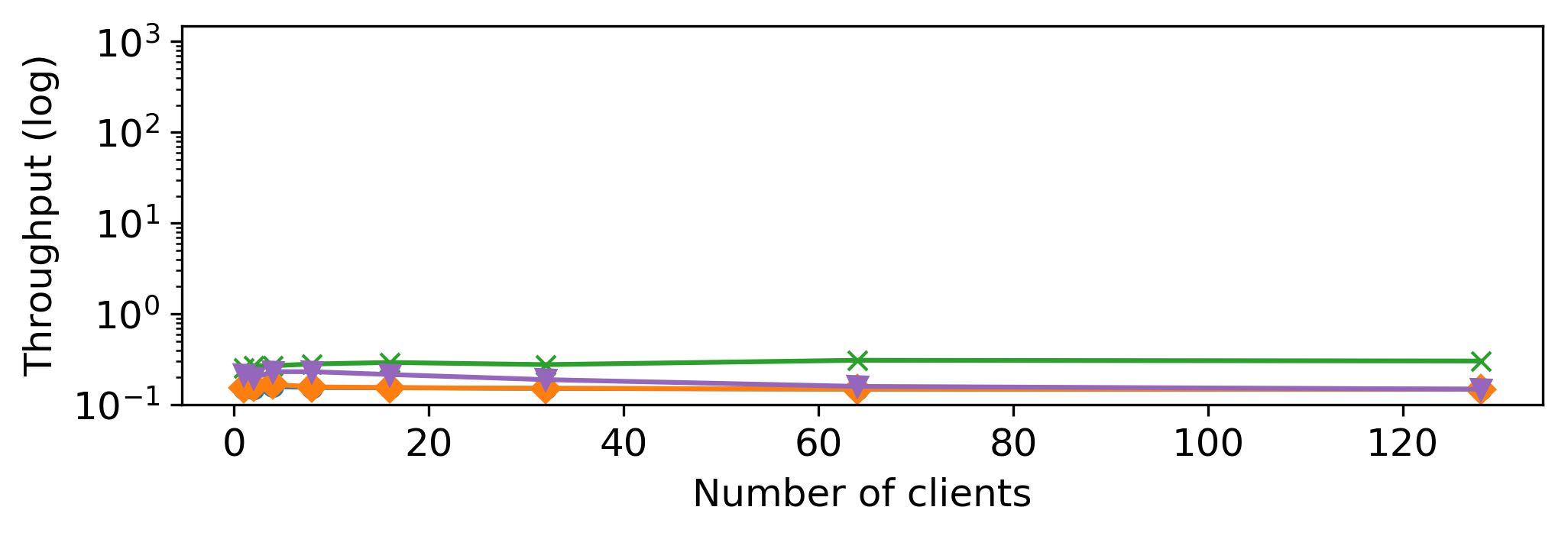}
        \label{fig:tp4_3s_app1}
    }\\\vspace{-1ex}
    \subfloat[Timeouts for \texttt{watdiv-3\_stars} over \texttt{watdiv10M}]{
        \includegraphics[width=.49\textwidth]{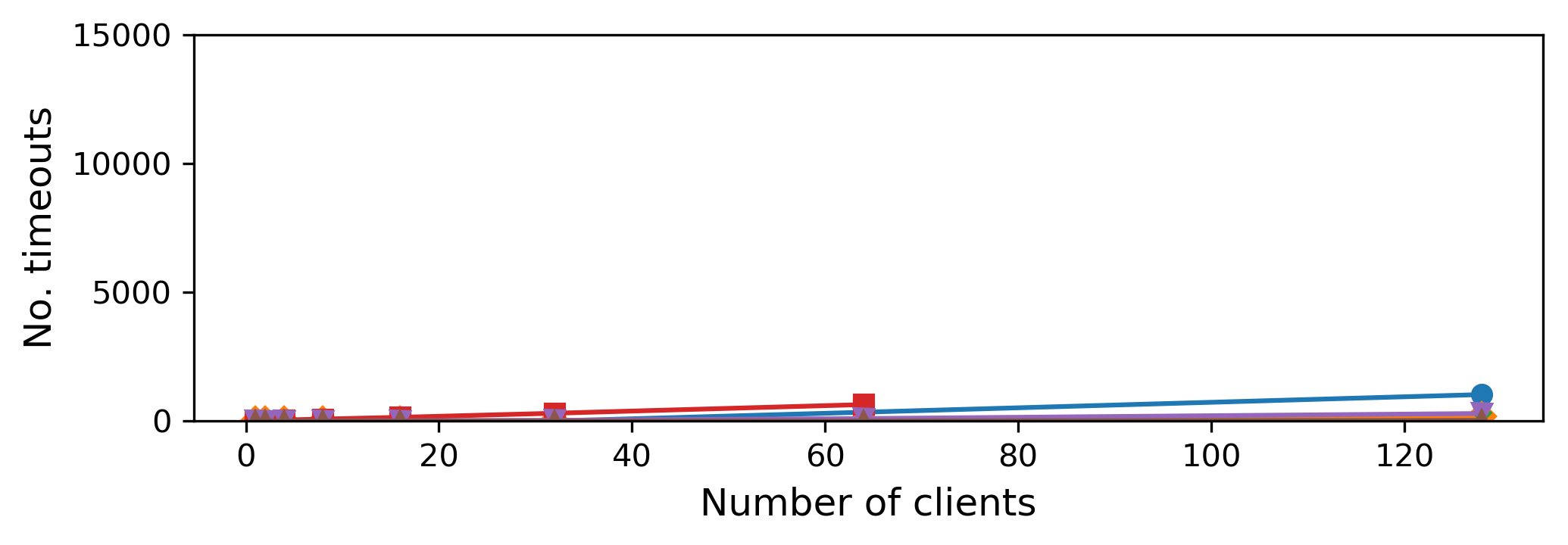}
        \label{fig:to1_3s_app1}
    }
    \subfloat[Timeouts for \texttt{watdiv-3\_stars} over \texttt{watdiv100M}]{
        \includegraphics[width=.49\textwidth]{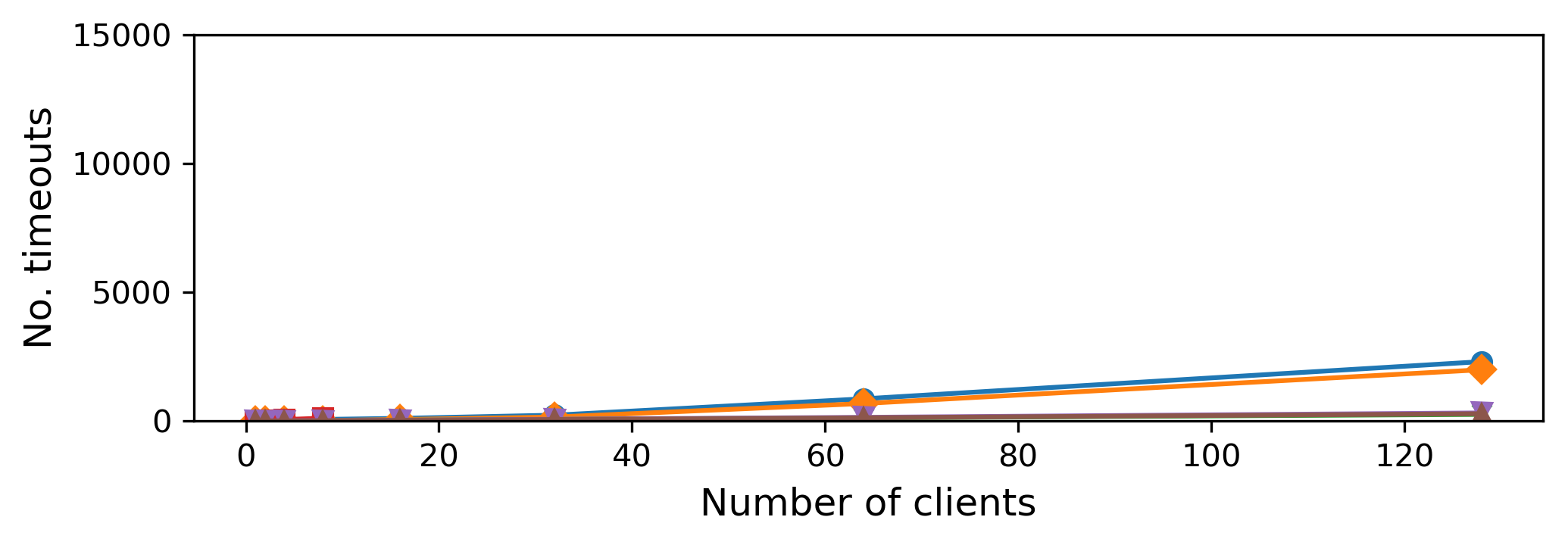}
        \label{fig:to2_3s_app1}
    }\\\vspace{-1ex}
    \subfloat[Timeouts for \texttt{watdiv-3\_stars} over \texttt{watdiv1B}]{
        \includegraphics[width=.49\textwidth]{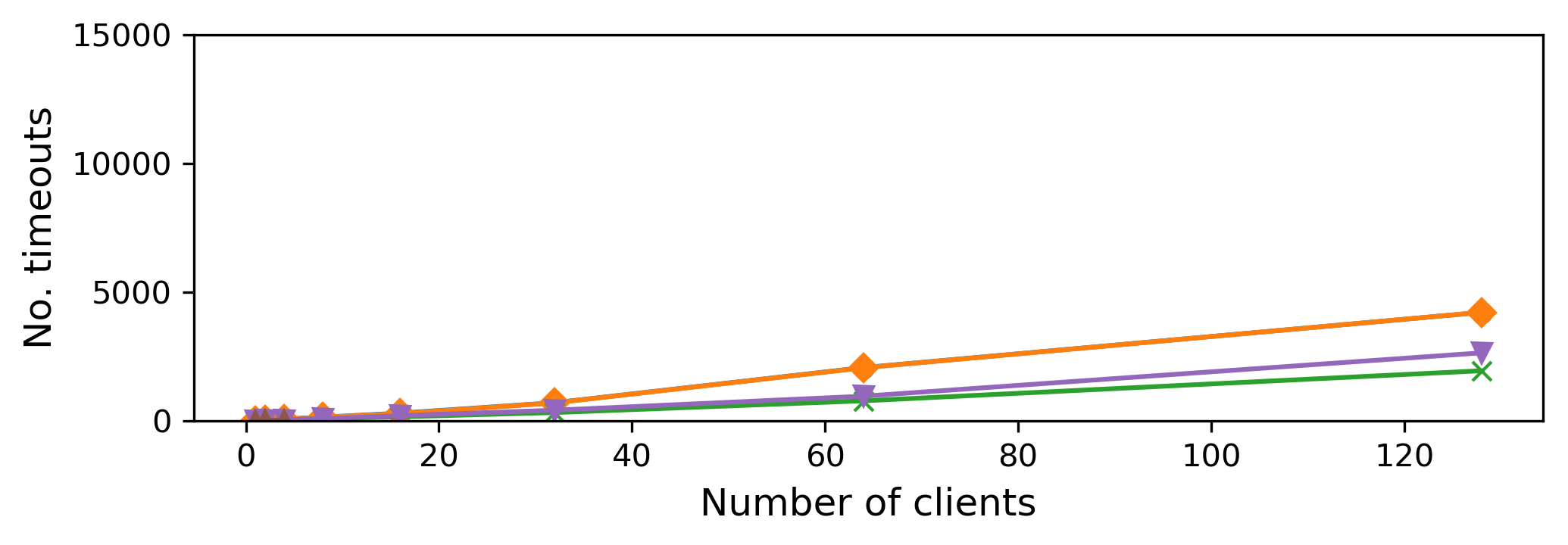}
        \label{fig:to3_3s_app1}
    }
    \subfloat[Timeouts for \texttt{watdiv-3\_stars} over \texttt{watdiv10B}]{
        \includegraphics[width=.49\textwidth]{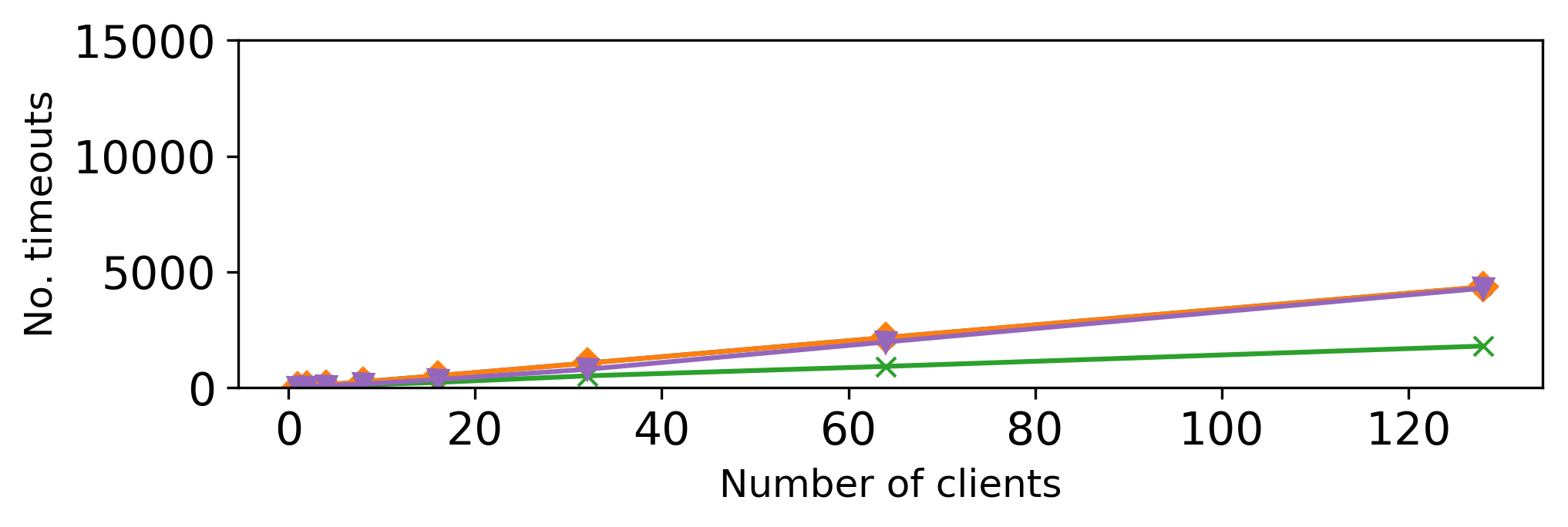}
        \label{fig:to4_3s_app1}
    }
    \begin{minipage}{.6\textwidth}
    \vspace{-214ex}
    \hspace{-100ex}
	\includegraphics[width=\textwidth]{results/graphs/legend1.png}
	\end{minipage}
	\vspace{-1ex}
    \caption{Throughput (\# queries/m) and timeouts for \texttt{watdiv-3\_stars} over the different WatDiv datasets. Includes queries that timed out.}\label{fig:throughput3_app1}
\end{figure*}

\begin{figure*}[p]
    \centering	
    \subfloat[Throughput for \texttt{watdiv-paths} over \texttt{watdiv10M} (\textit{log})]{
        \includegraphics[width=.49\textwidth]{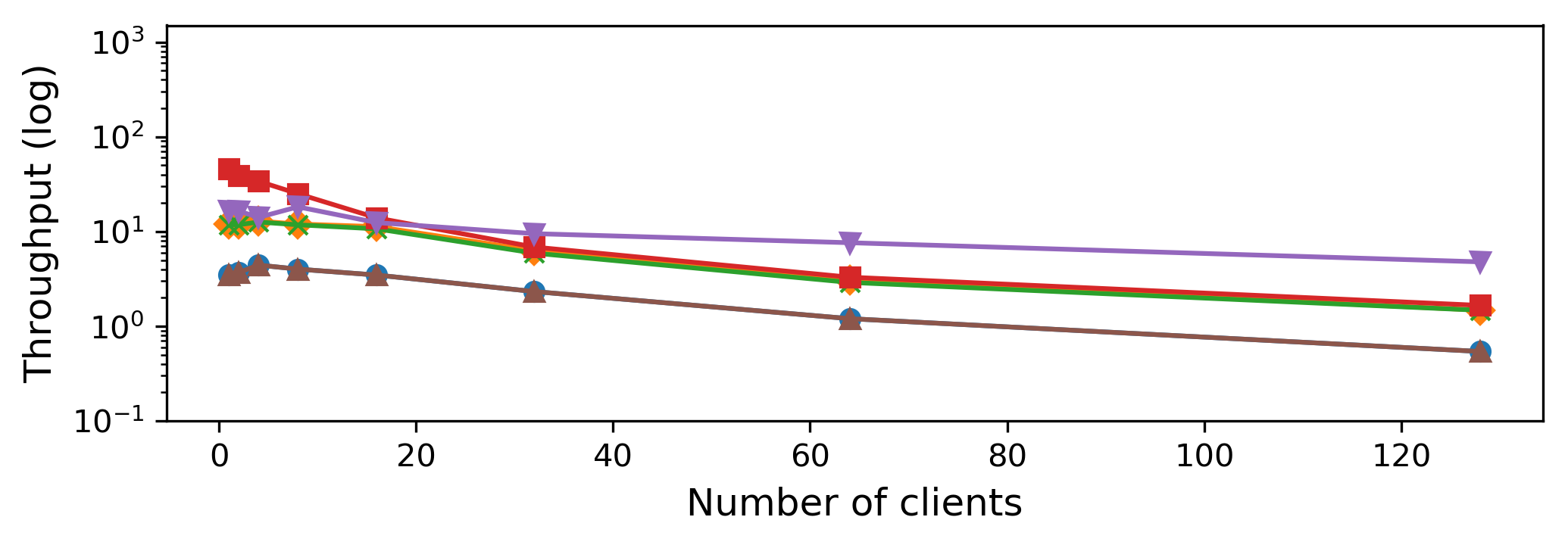}
        \label{fig:tp1_ps_app1}
    }
    \subfloat[Throughput for \texttt{watdiv-paths} over \texttt{watdiv100M} (\textit{log})]{
        \includegraphics[width=.49\textwidth]{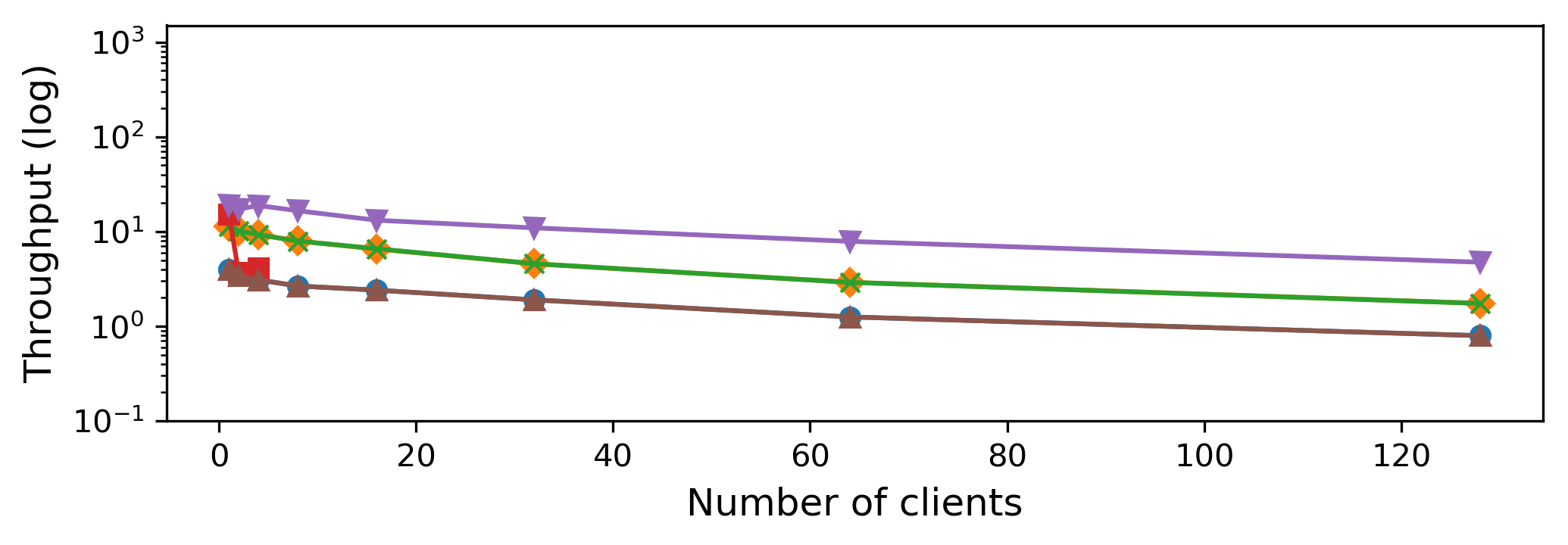}
        \label{fig:tp2_ps_app1}
    }\\\vspace{-1ex}
    \subfloat[Throughput for \texttt{watdiv-paths} over \texttt{watdiv1B} (\textit{log})]{
        \includegraphics[width=.49\textwidth]{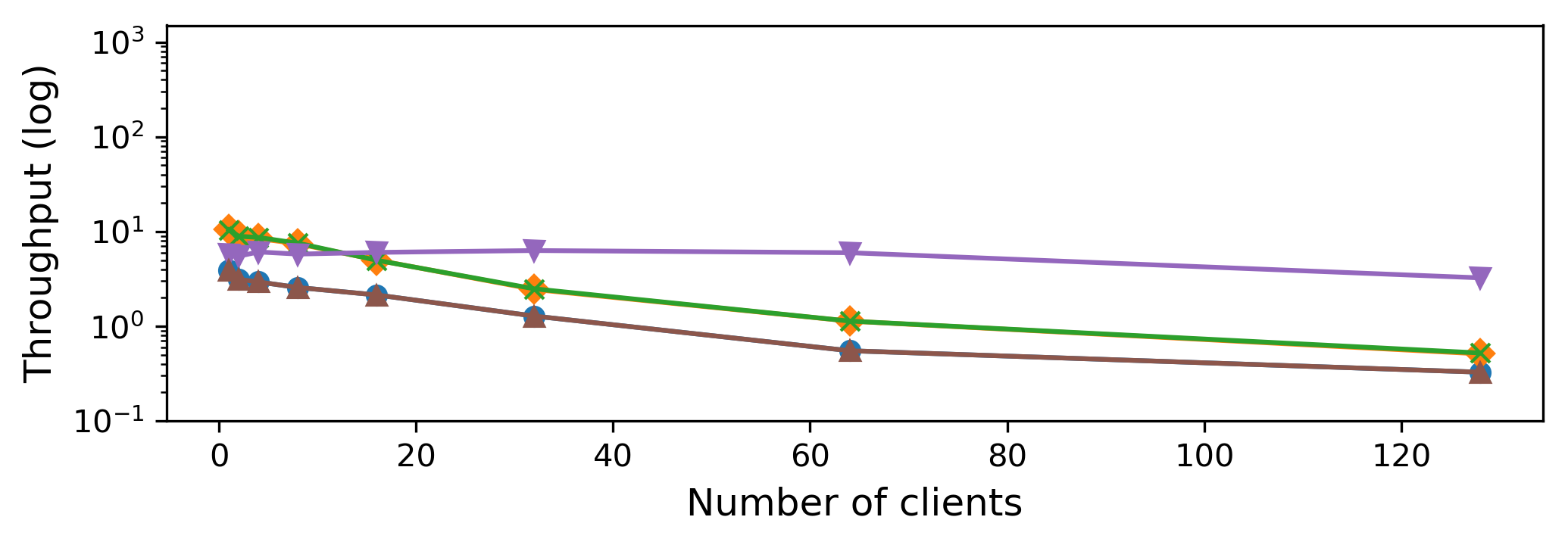}
        \label{fig:tp3_ps_app1}
    }
    \subfloat[Throughput for \texttt{watdiv-paths} over \texttt{watdiv10B} (\textit{log})]{
        \includegraphics[width=.49\textwidth]{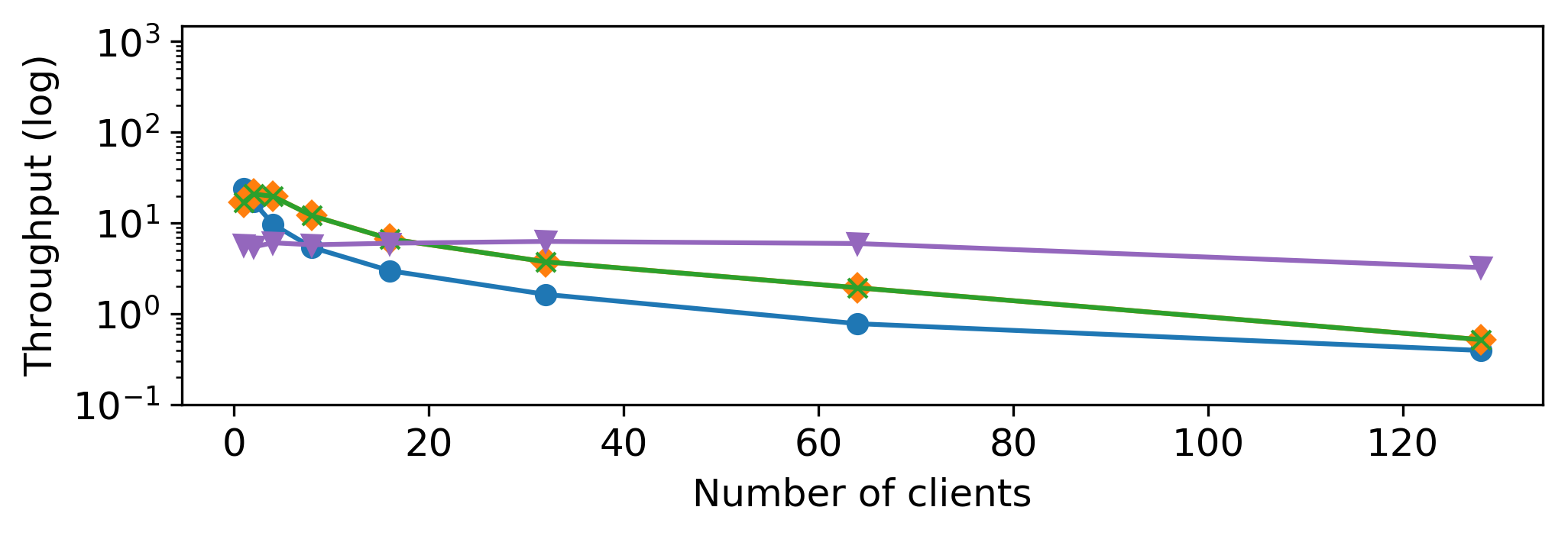}
        \label{fig:tp4_ps_app1}
    }\\\vspace{-1ex}
    \subfloat[Timeouts for \texttt{watdiv-paths} over \texttt{watdiv10M}]{
        \includegraphics[width=.49\textwidth]{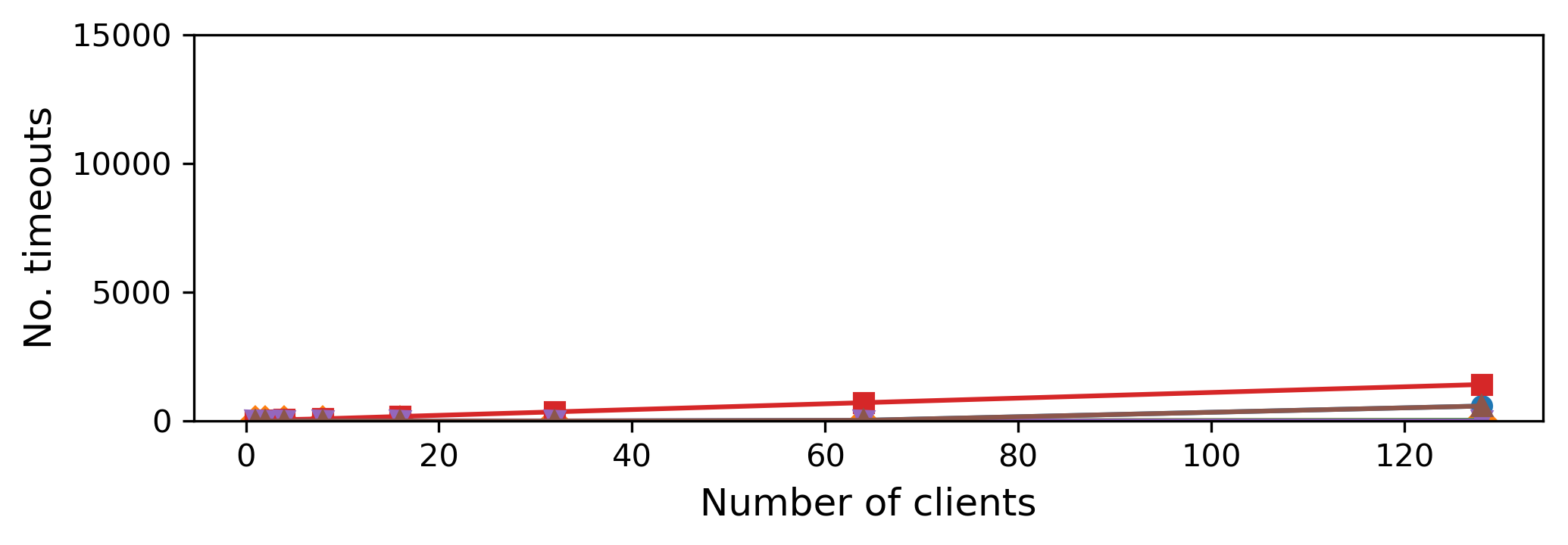}
        \label{fig:to1_ps_app1}
    }
    \subfloat[Timeouts for \texttt{watdiv-paths} over \texttt{watdiv100M}]{
        \includegraphics[width=.49\textwidth]{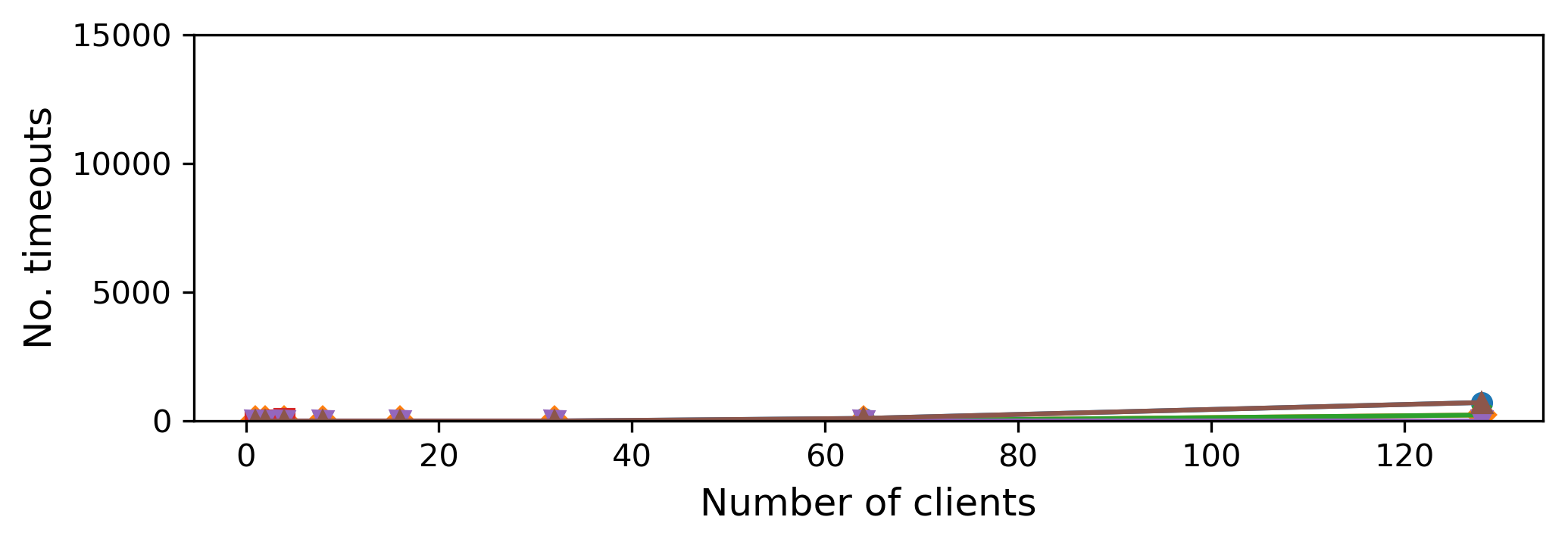}
        \label{fig:to2_ps_app1}
    }\\\vspace{-1ex}
    \subfloat[Timeouts for \texttt{watdiv-paths} over \texttt{watdiv1B}]{
        \includegraphics[width=.49\textwidth]{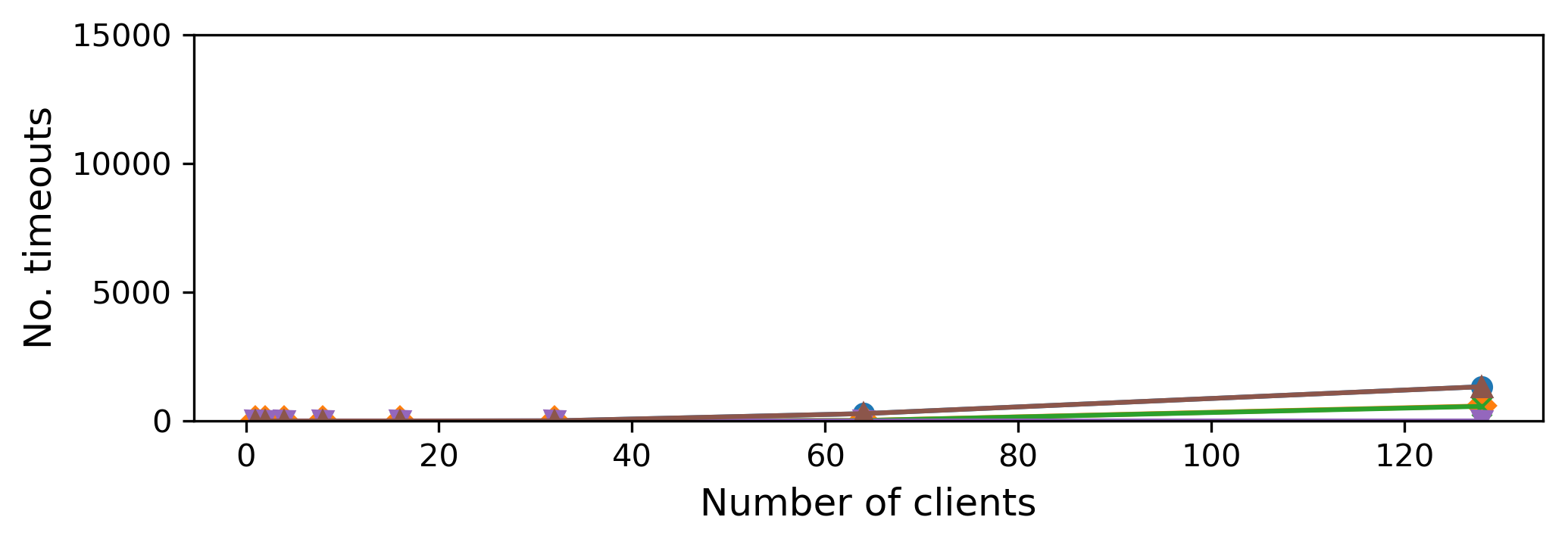}
        \label{fig:to3_ps_app1}
    }
    \subfloat[Timeouts for \texttt{watdiv-paths} over \texttt{watdiv10B}]{
        \includegraphics[width=.49\textwidth]{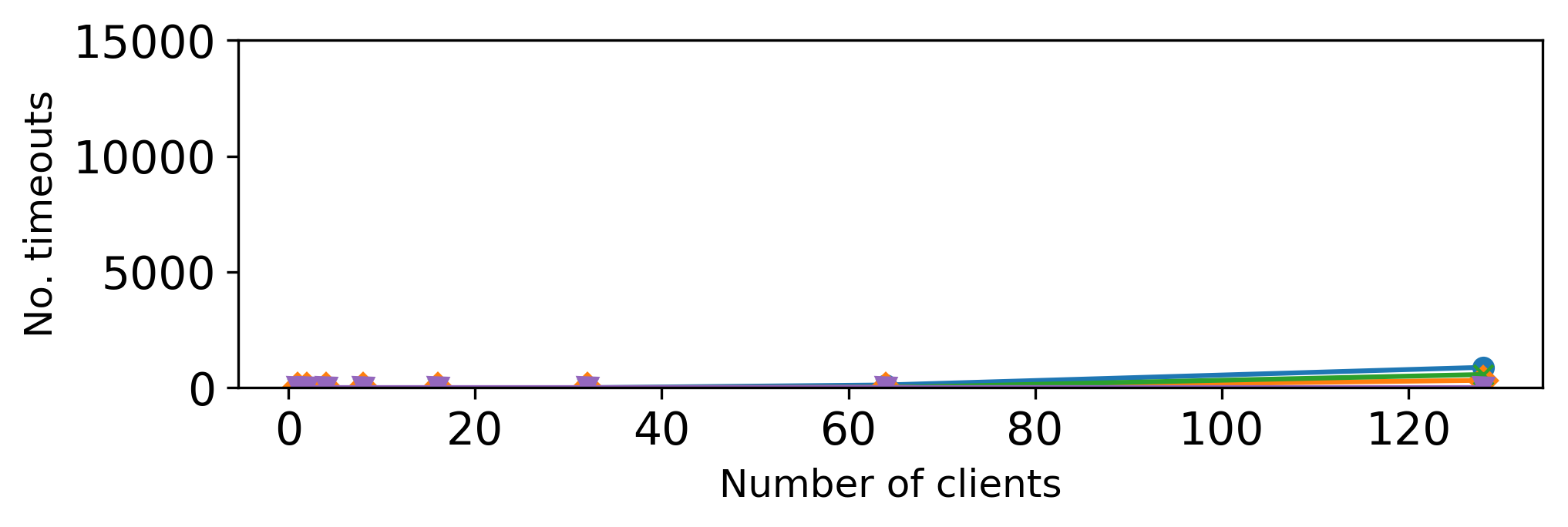}
        \label{fig:to4_ps_app1}
    }
    \begin{minipage}{.6\textwidth}
    \vspace{-214ex}
    \hspace{-100ex}
	\includegraphics[width=\textwidth]{results/graphs/legend1.png}
	\end{minipage}
	\vspace{-1ex}
    \caption{Throughput (\# queries/m) and timeouts for \texttt{watdiv-paths} over the different WatDiv datasets. Includes queries that timed out.}\label{fig:throughput4_app1}
\end{figure*}

\begin{figure*}[p]
    \centering	
    \subfloat[Throughput for \texttt{watdiv-btt} over \texttt{watdiv10M} (\textit{log})]{
        \includegraphics[width=.49\textwidth]{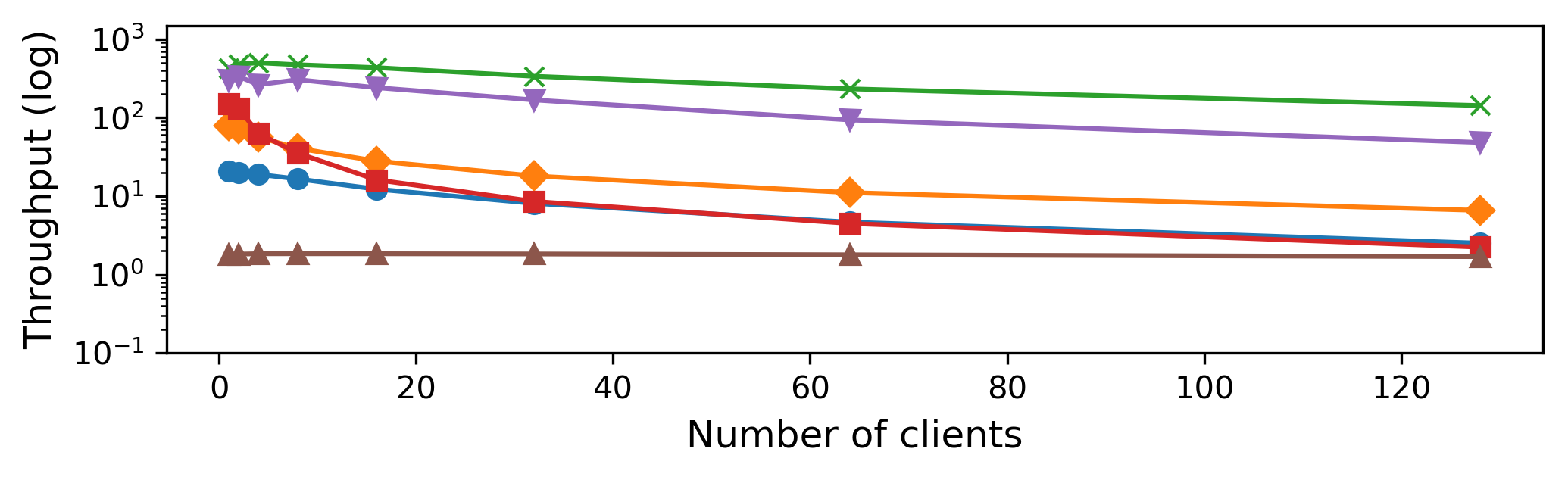}
        \label{fig:tp1_app2}
    }
    \subfloat[Throughput for \texttt{watdiv-btt} over \texttt{watdiv100M} (\textit{log})]{
        \includegraphics[width=.49\textwidth]{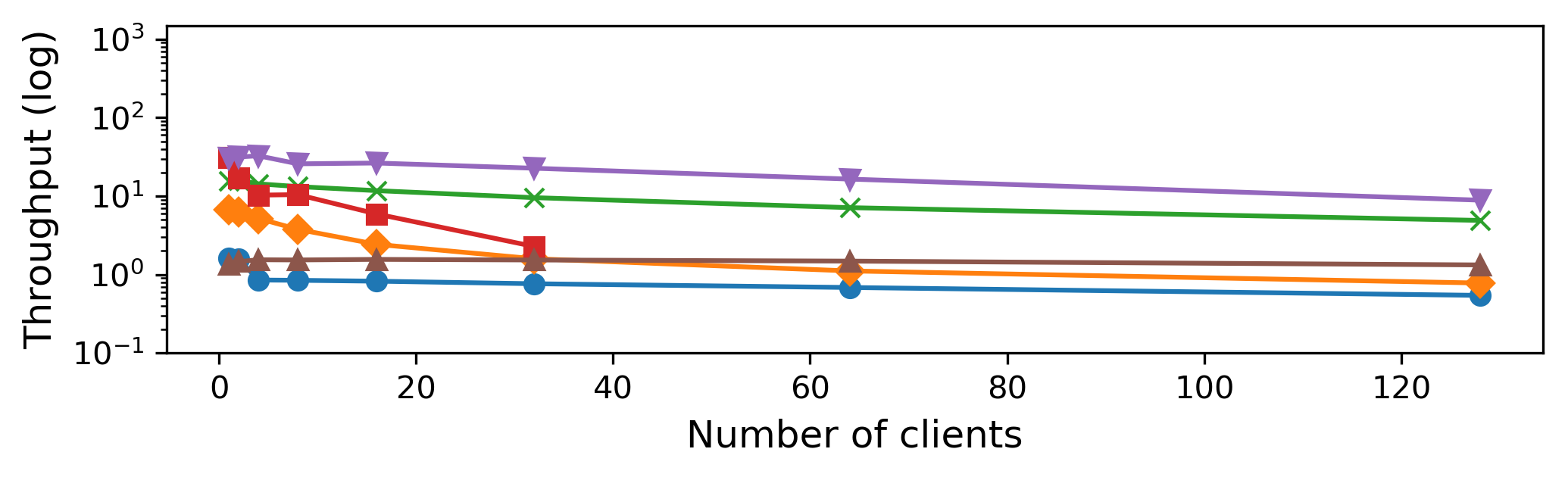}
        \label{fig:tp2_app2}
    }\\\vspace{-1ex}
    \subfloat[Throughput for \texttt{watdiv-btt} over \texttt{watdiv1B} (\textit{log})]{
        \includegraphics[width=.49\textwidth]{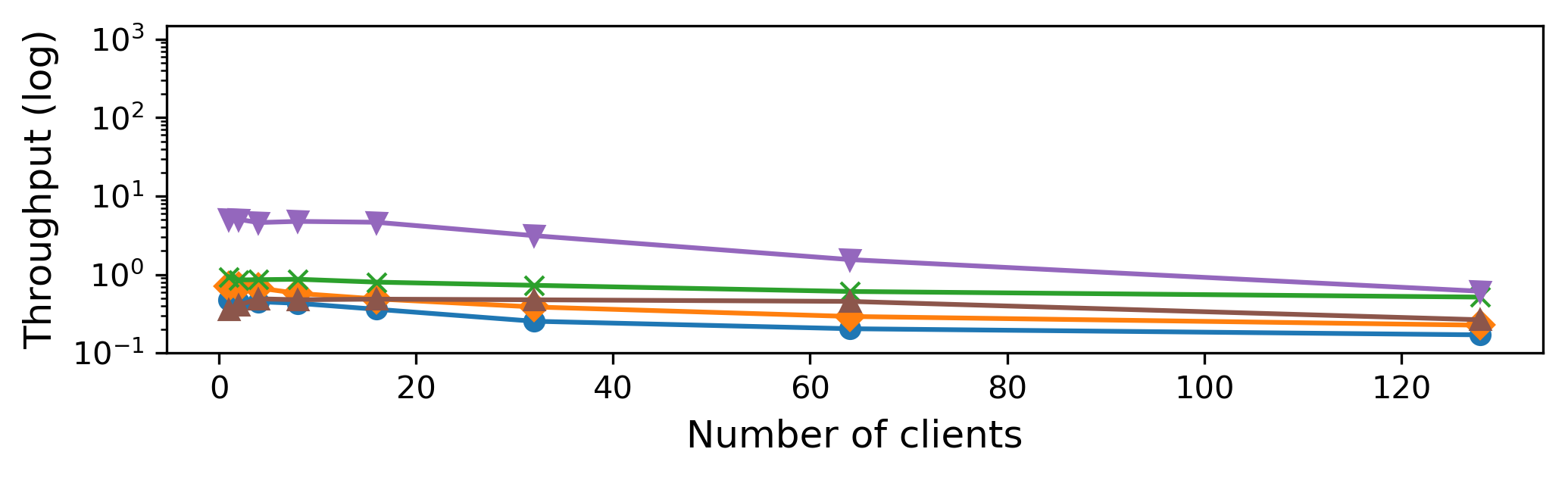}
        \label{fig:tp3_app2}
    }
    \subfloat[Throughput for \texttt{watdiv-btt} over \texttt{watdiv10B} (\textit{log})]{
        \includegraphics[width=.49\textwidth]{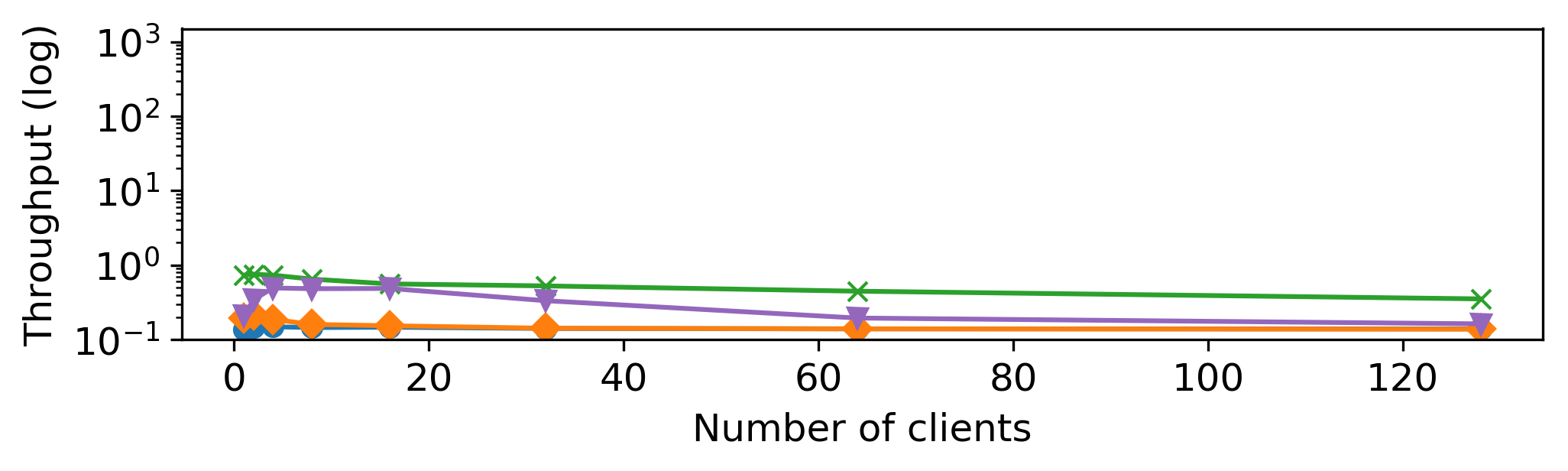}
        \label{fig:tp4_app2}
    }\\\vspace{-1ex}
    \subfloat[Timeouts for \texttt{watdiv-btt} over \texttt{watdiv10M}]{
        \includegraphics[width=.49\textwidth]{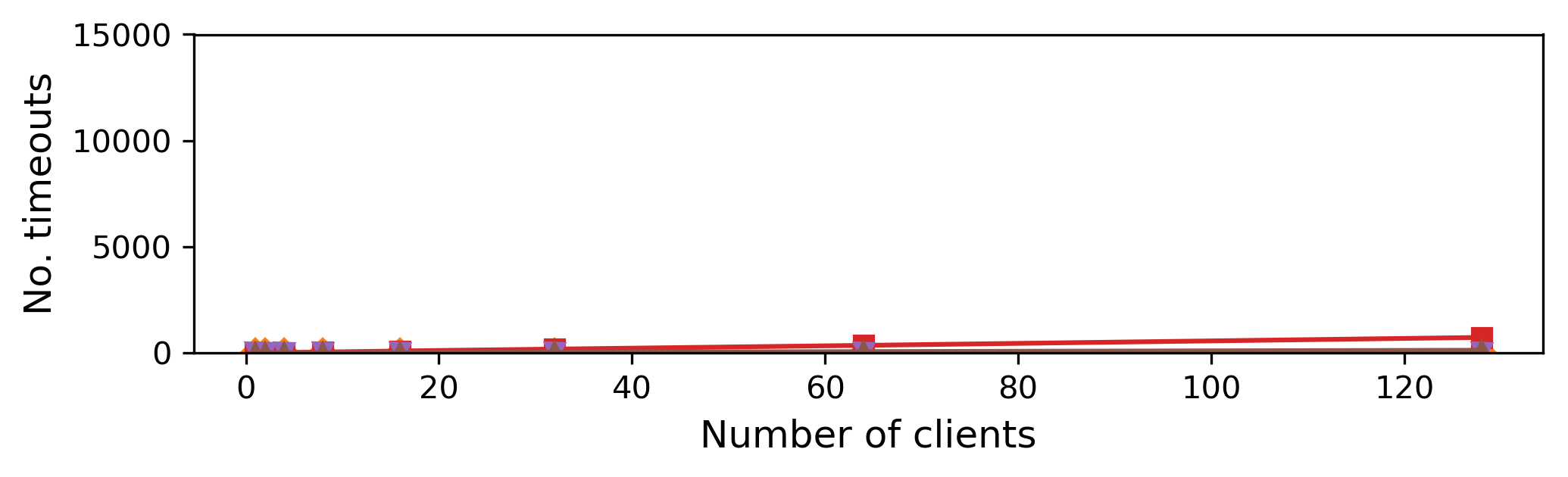}
        \label{fig:to1_app2}
    }
    \subfloat[Timeouts for \texttt{watdiv-btt} over \texttt{watdiv100M}]{
        \includegraphics[width=.49\textwidth]{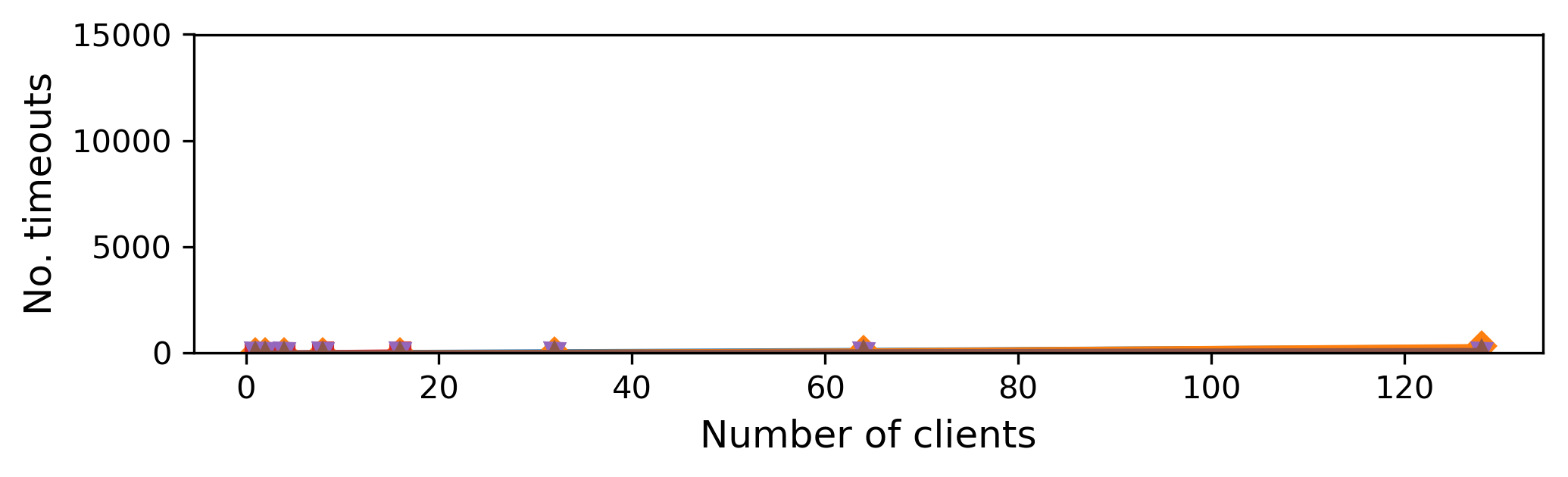}
        \label{fig:to2_app2}
    }\\\vspace{-1ex}
    \subfloat[Timeouts for \texttt{watdiv-btt} over \texttt{watdiv1B}]{
        \includegraphics[width=.49\textwidth]{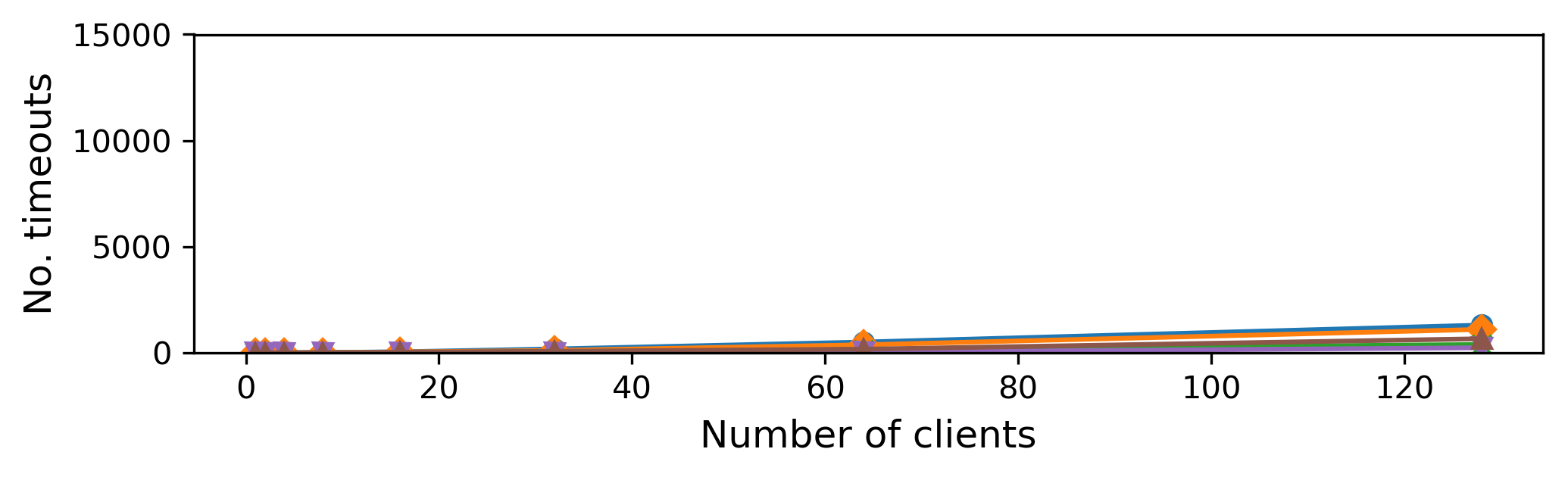}
        \label{fig:to3_app2}
    }
    \subfloat[Timeouts for \texttt{watdiv-btt} over \texttt{watdiv10B}]{
        \includegraphics[width=.49\textwidth]{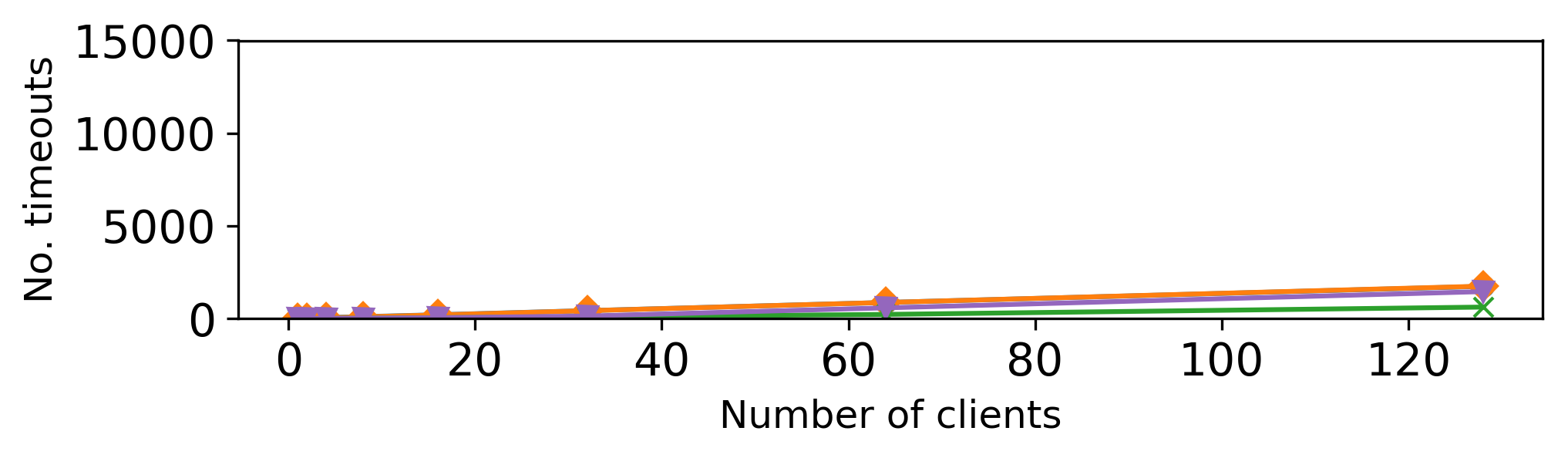}
        \label{fig:to4_app2}
    }\\\vspace{-1ex}
    \subfloat[CPU load for \texttt{watdiv-btt} over \texttt{watdiv10M}]{
        \includegraphics[width=.49\textwidth]{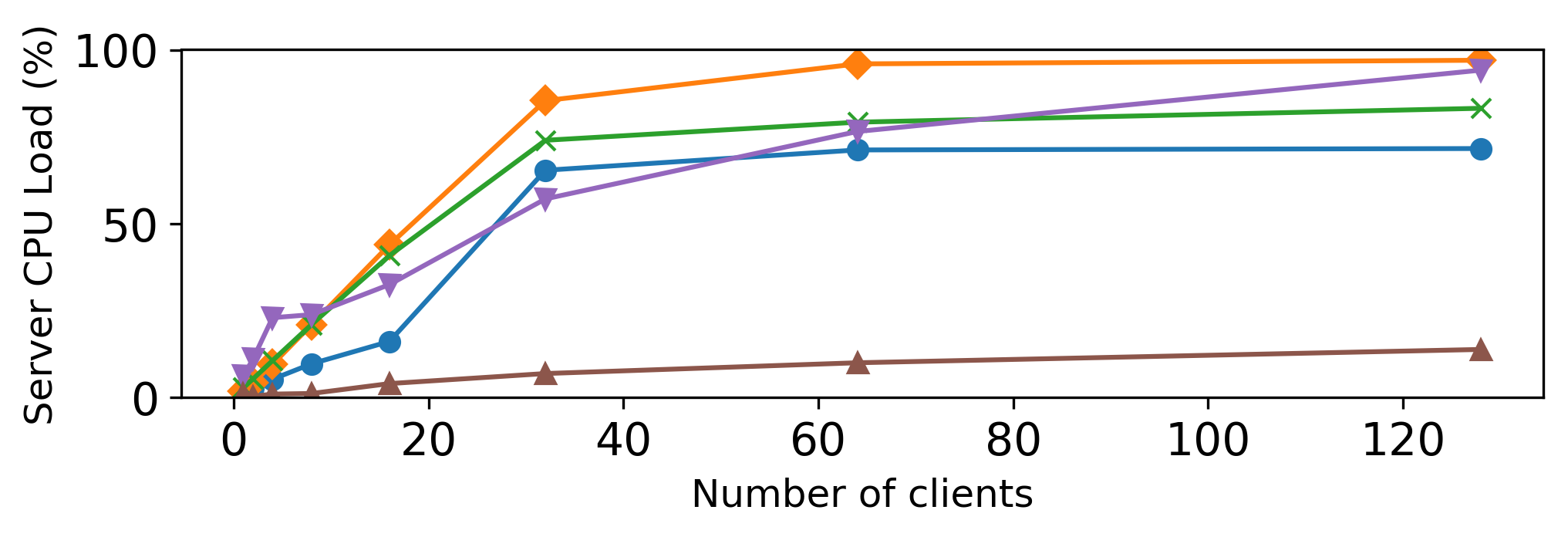}
        \label{fig:cpu1_app}
    }
    \subfloat[CPU load for \texttt{watdiv-btt} over \texttt{watdiv100M}]{
        \includegraphics[width=.49\textwidth]{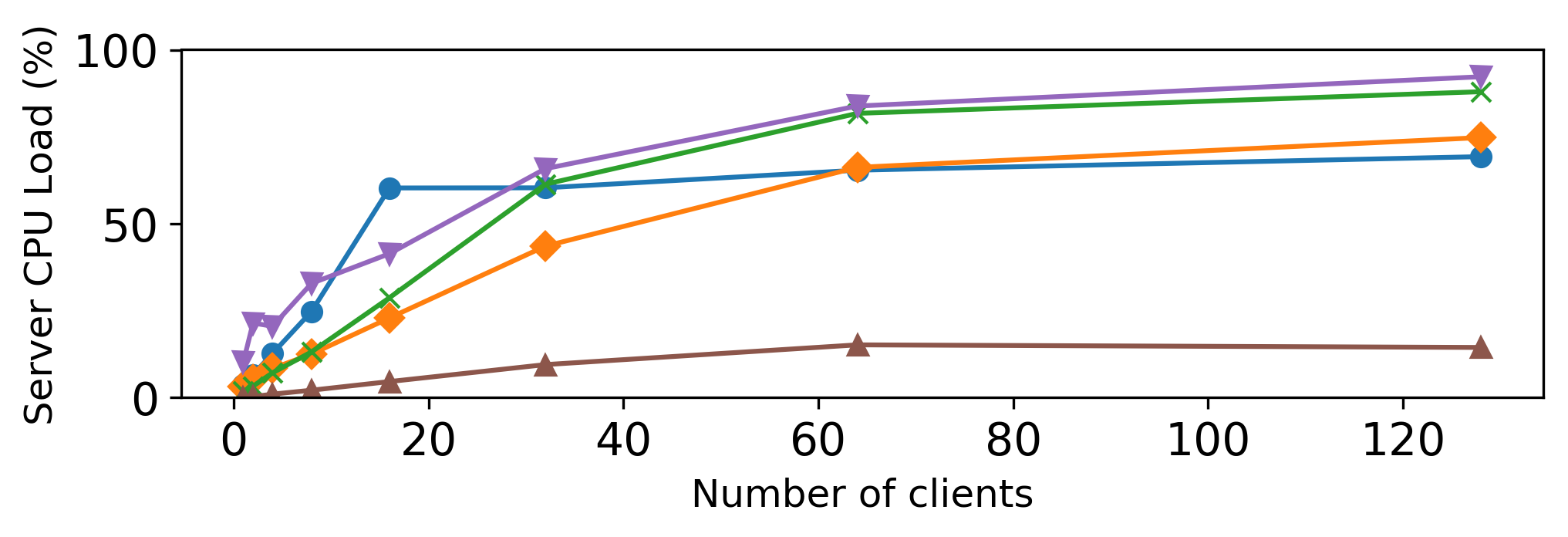}
        \label{fig:cpu2_app}
    }\\\vspace{-1ex}
    \subfloat[CPU load for \texttt{watdiv-btt} over \texttt{watdiv1B}]{
        \includegraphics[width=.49\textwidth]{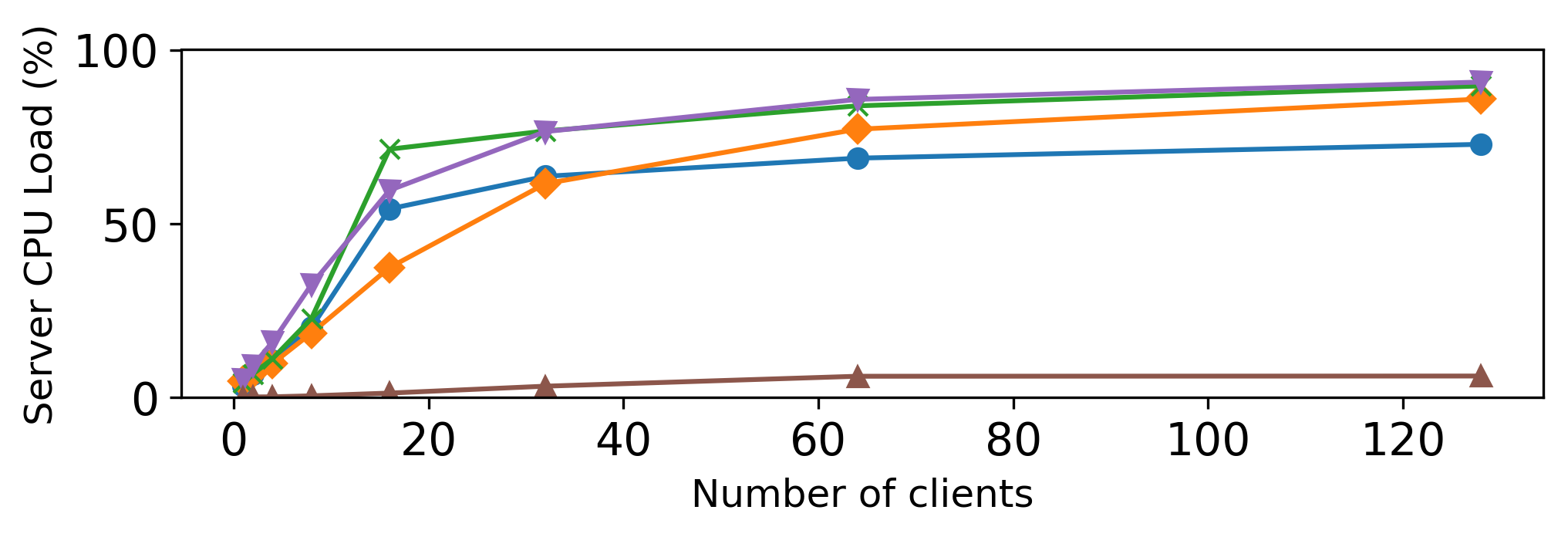}
        \label{fig:cpu3_app}
    }
    \subfloat[CPU load for \texttt{watdiv-btt} over \texttt{watdiv10B}]{
        \includegraphics[width=.49\textwidth]{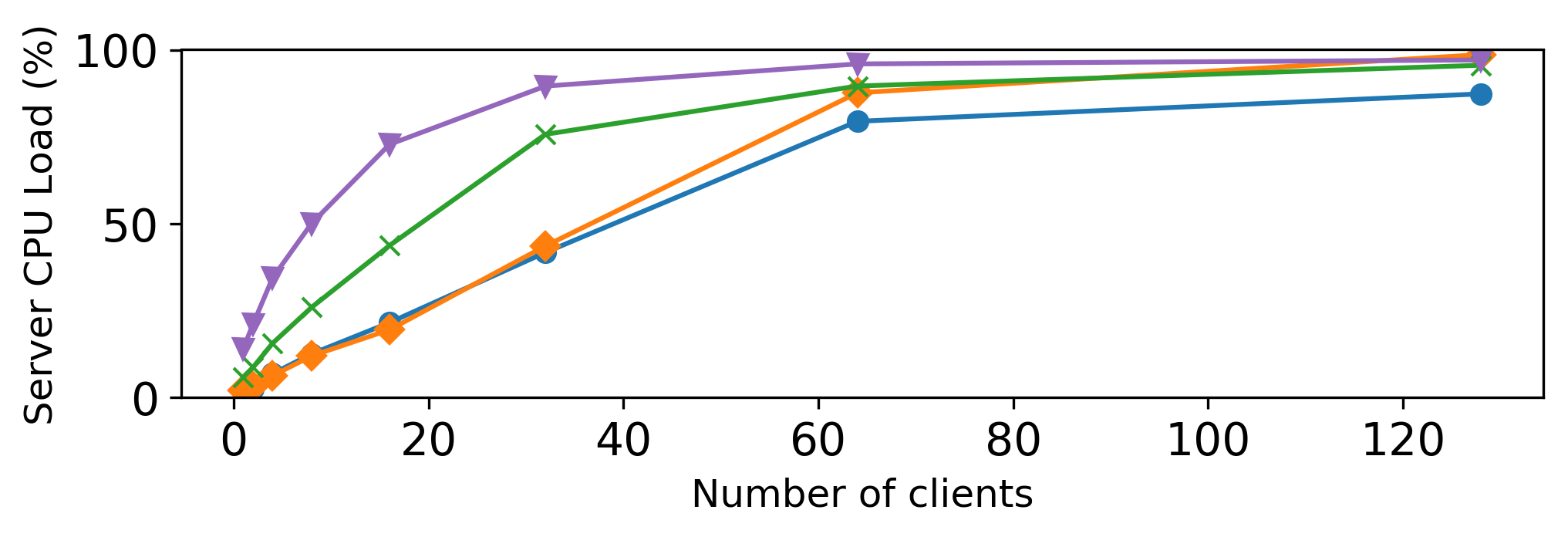}
        \label{fig:cpu4_app}
    }\\\vspace{-1ex}
    \begin{minipage}{.6\textwidth}
    \vspace{-302ex}
    \hspace{3ex}
	\includegraphics[width=\textwidth]{results/graphs/legend1.png}
	\end{minipage}
	\vspace{-1ex}
    \caption{Throughput, number of timeouts and CPU load for \texttt{watdiv-btt} over the different WatDiv datasets.}\label{fig:cpu_app}
\end{figure*}

\begin{figure*}[p]
    \centering
    \subfloat[Timeouts for \texttt{dbpedia-lsq} over \texttt{dbpedia}]{
        \includegraphics[width=.49\textwidth]{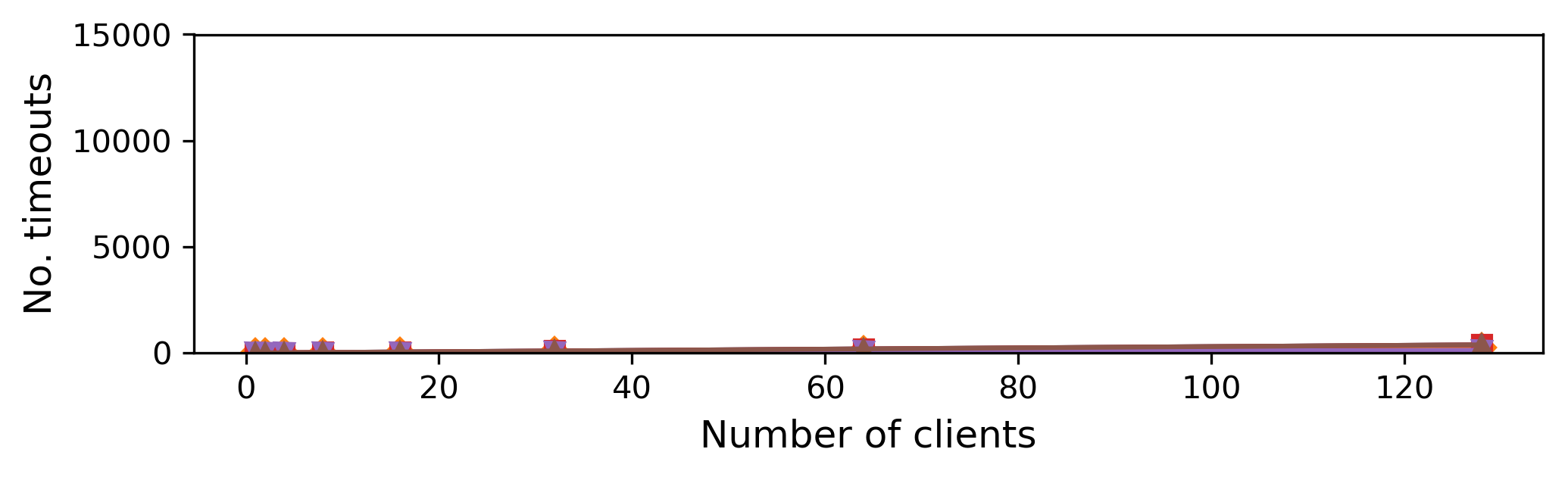}
        \label{fig:to4_app2}
    }
    \subfloat[CPU load for \texttt{dbpedia-lsq} over \texttt{dbpedia}]{
        \includegraphics[width=.45\textwidth]{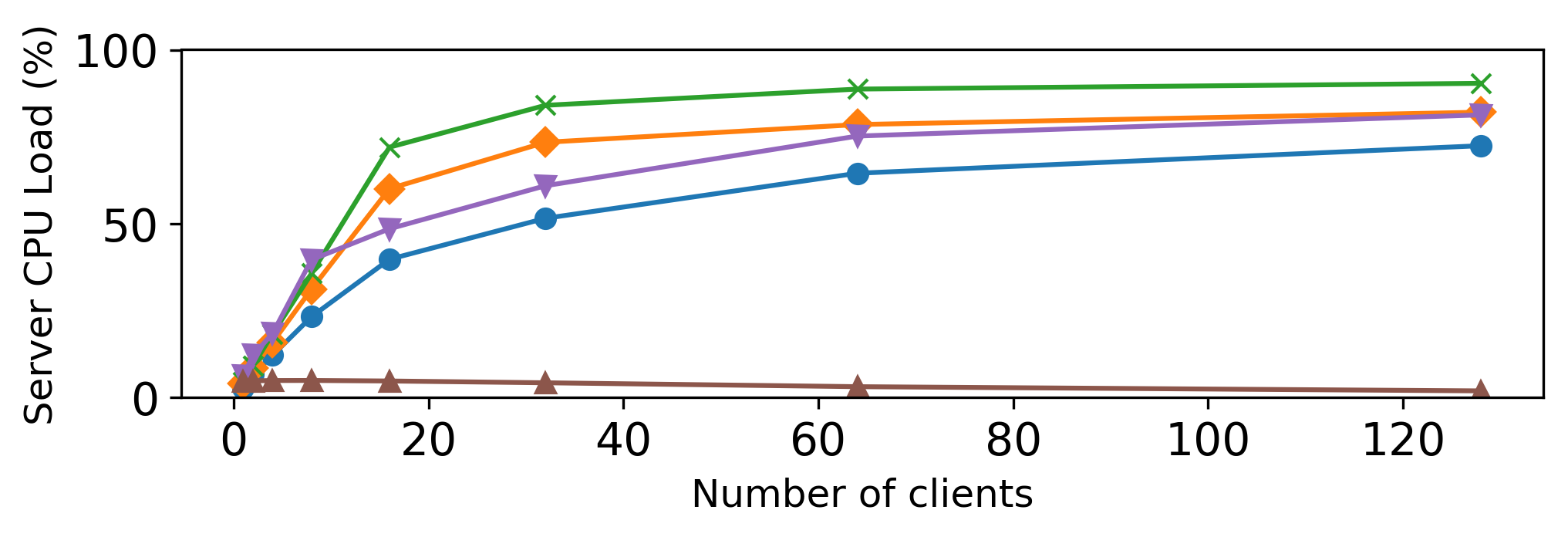}
        \label{fig:cpu4_app}
    }\\\vspace{-1ex}
    \begin{minipage}{.6\textwidth}
    \vspace{-48ex}
    \hspace{3ex}
	\includegraphics[width=\textwidth]{results/graphs/legend1.png}
	\end{minipage}
	\vspace{-1ex}
    \caption{Number of timeouts and CPU load for \texttt{dbpedia-lsq} over \texttt{dbpedia}.}\label{fig:cpu_dbp_app}
\end{figure*}

\begin{figure*}[h]
    \centering
    \subfloat[NRS \texttt{watdiv10M} (\textit{y-axis in log scale}).]{
        \includegraphics[width=.49\textwidth]{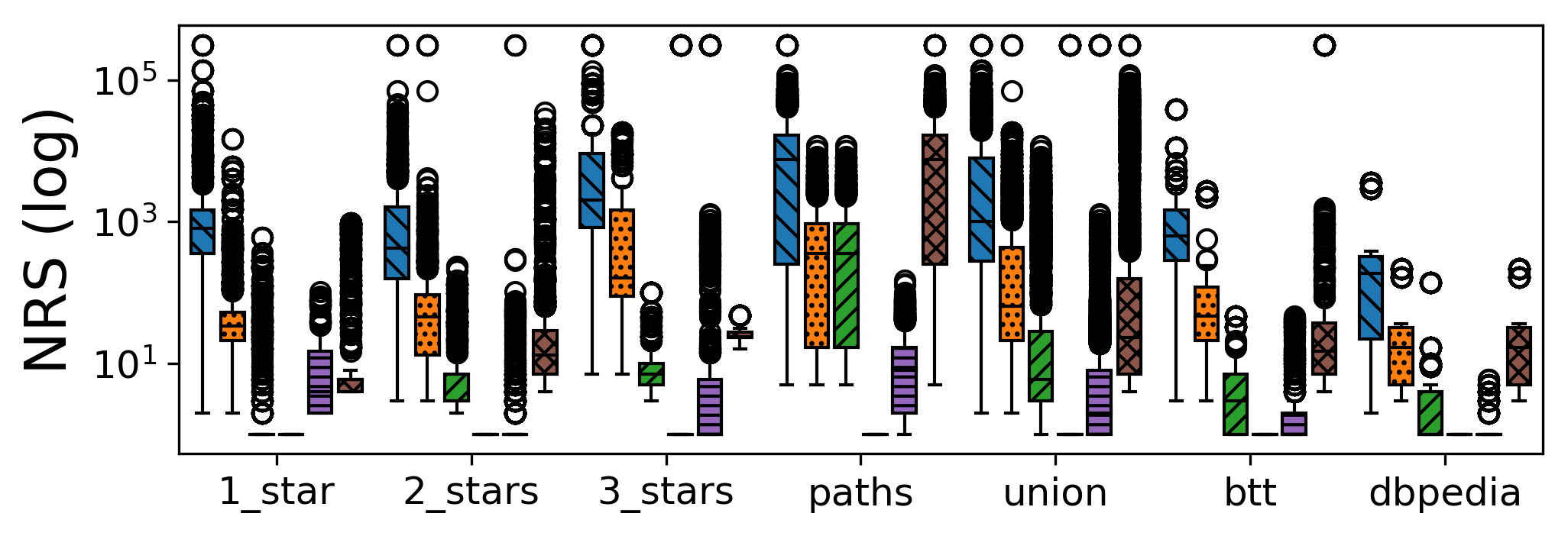}
        \label{fig:req1_app}
    }
    \subfloat[NRS \texttt{watdiv100M} (\textit{y-axis in log scale}).]{
        \includegraphics[width=.49\textwidth]{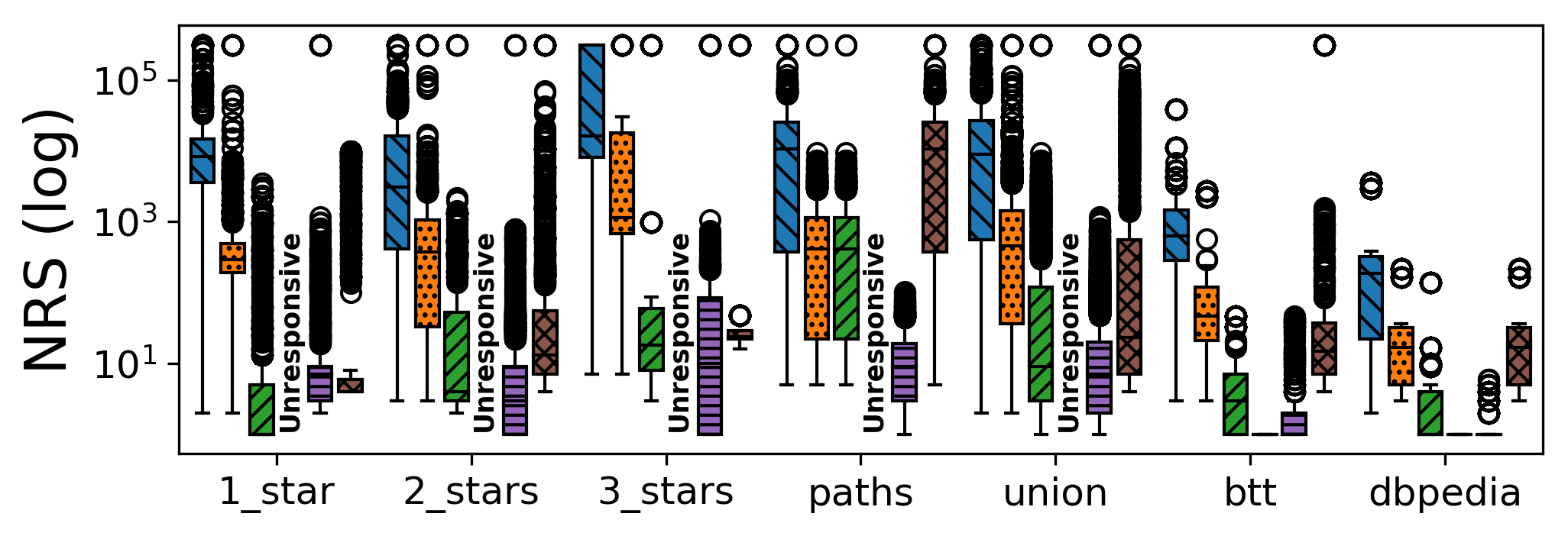}
        \label{fig:req2_app}
    }\\\vspace{-1ex}
    \subfloat[NRS \texttt{watdiv1B} (\textit{y-axis in log scale}).]{
        \includegraphics[width=.49\textwidth]{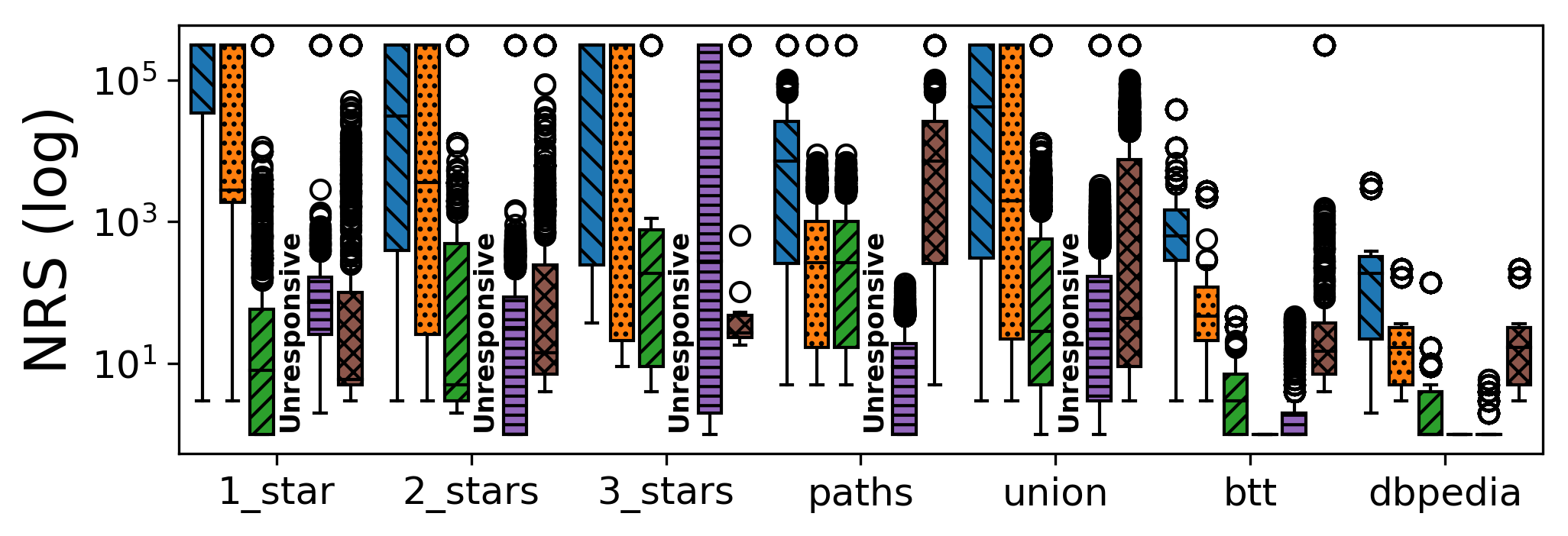}
        \label{fig:req3_app}
    }
    \subfloat[NRS \texttt{watdiv10B} (\textit{y-axis in log scale}).]{
        \includegraphics[width=.49\textwidth]{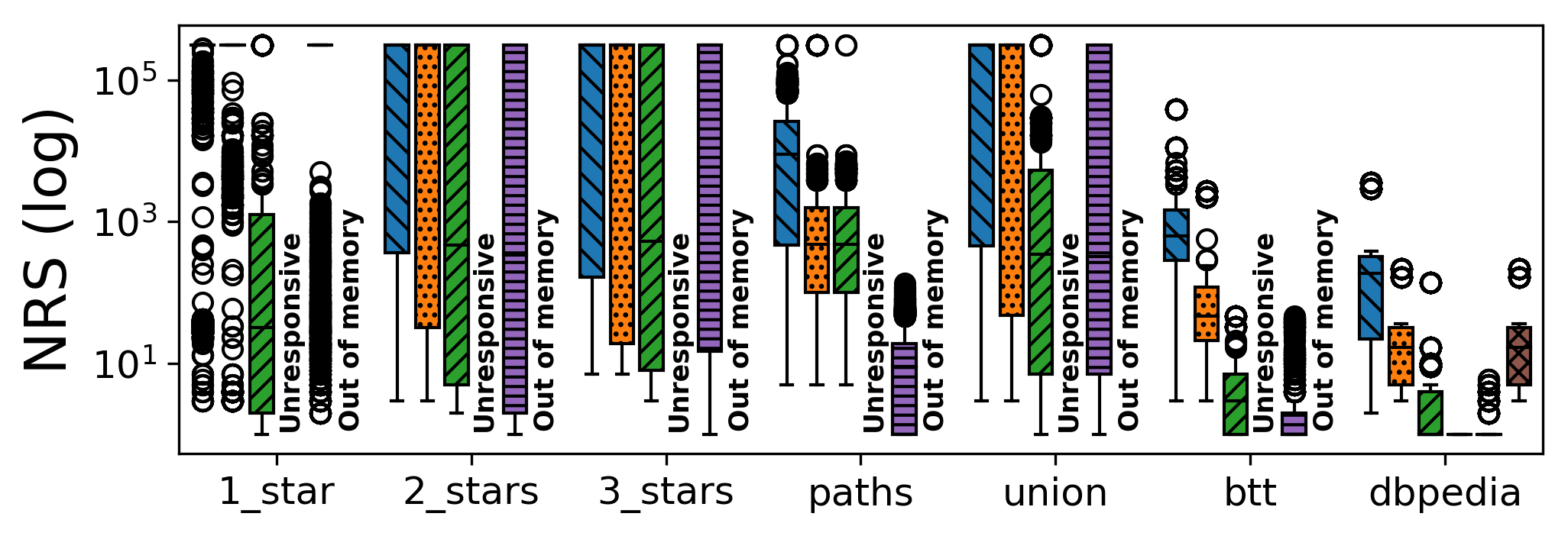}
        \label{fig:req4_app}
    }\\\vspace{-1ex}
    \subfloat[NTB \texttt{watdiv10M} (\textit{y-axis in log scale}).]{
        \includegraphics[width=.49\textwidth]{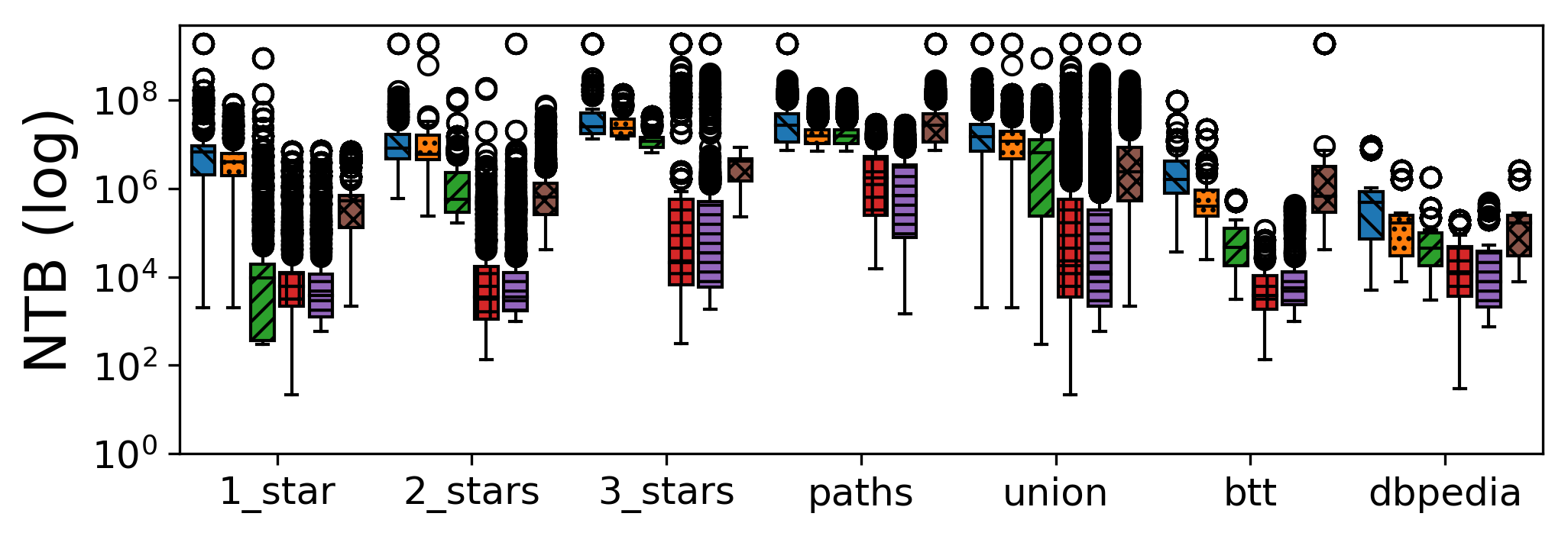}
        \label{fig:ntb1_app}
    }
    \subfloat[NTB \texttt{watdiv100M} (\textit{y-axis in log scale}).]{
        \includegraphics[width=.49\textwidth]{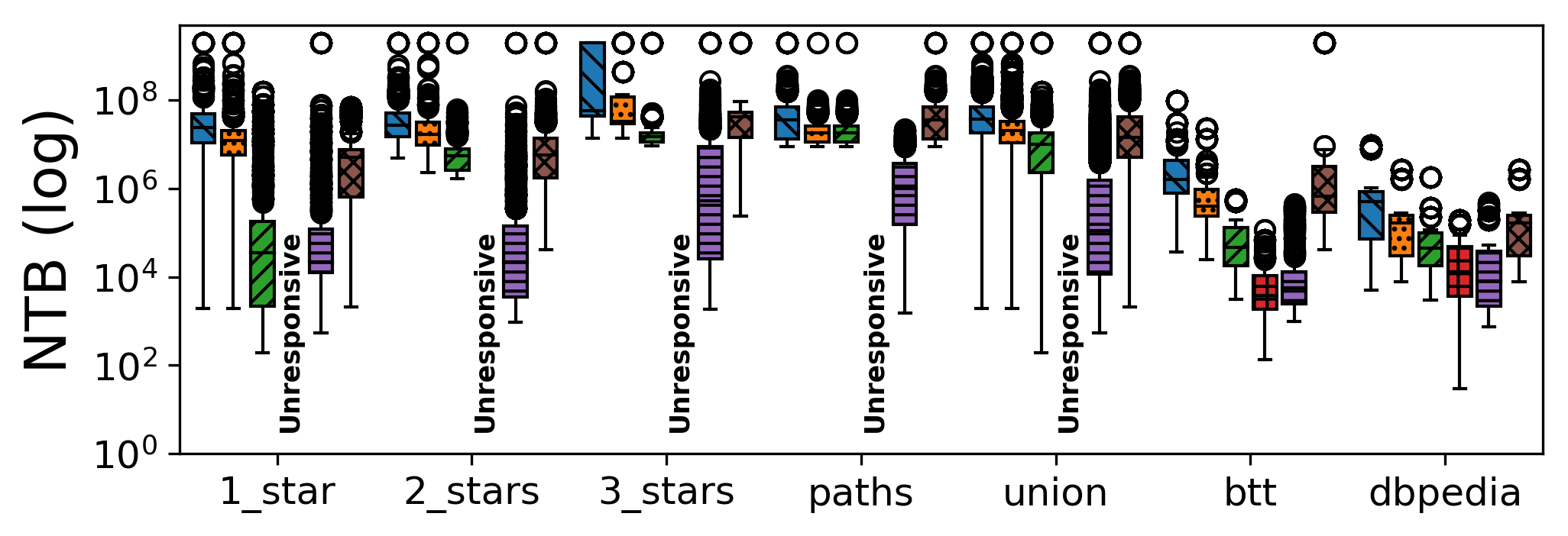}
        \label{fig:ntb2_app}
    }\\\vspace{-1ex}
    \subfloat[NTB \texttt{watdiv1B} (\textit{y-axis in log scale}).]{
        \includegraphics[width=.49\textwidth]{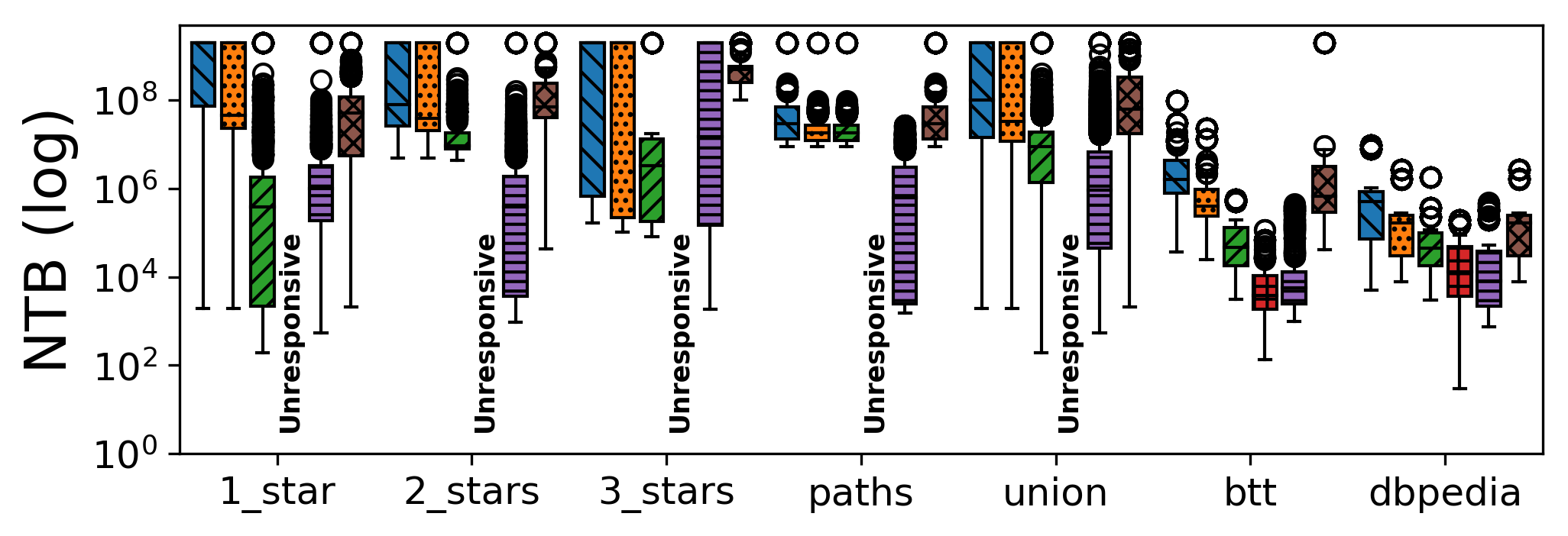}
        \label{fig:ntb3_app}
    }
    \subfloat[NTB \texttt{watdiv10B} (\textit{y-axis in log scale}).]{
        \includegraphics[width=.49\textwidth]{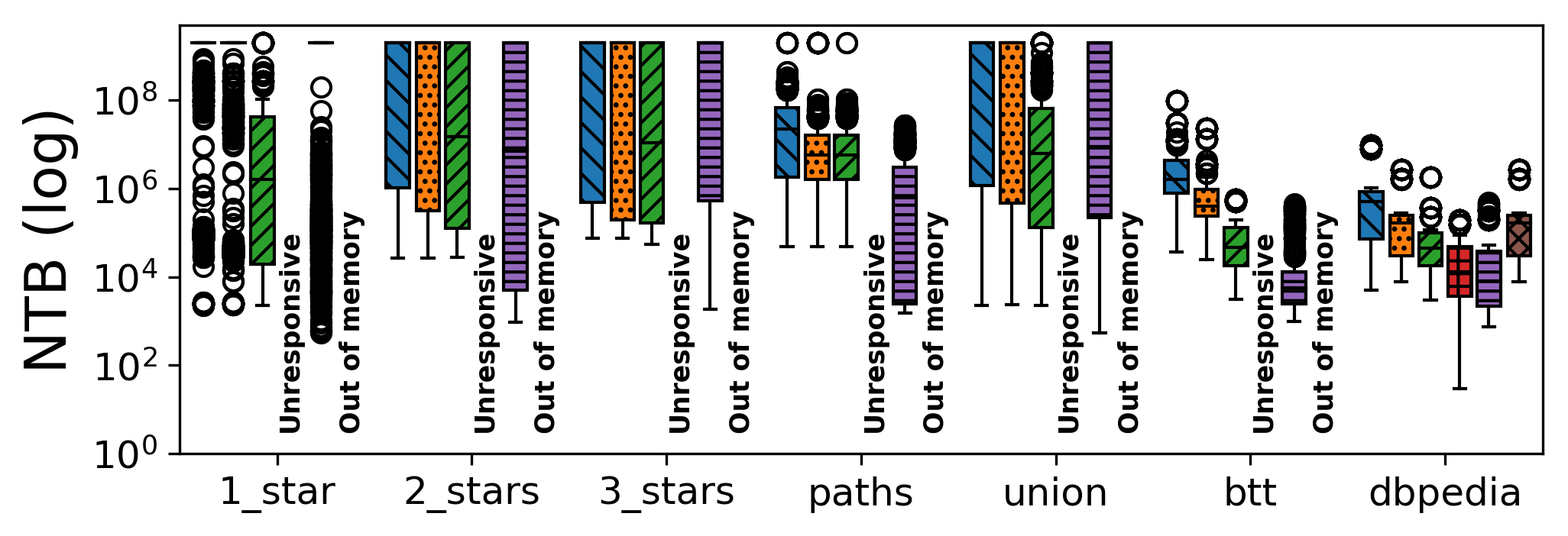}
        \label{fig:ntb4_app}
    }
    \begin{minipage}{.6\textwidth}
    \vspace{-206ex}
    \hspace{-100ex}
	\includegraphics[width=\textwidth]{results/graphs/legend2.png}
	\end{minipage}
	\vspace{-2ex}
    \caption{NRS and NTB with 64 clients including queries that timed out for any approach.}\label{fig:networkusage_app}
\end{figure*}

This appendix contains additional experimental results for query loads left out in Section~\ref{sec:evaluation}.
Figures~\ref{fig:throughput1_app1}-\ref{fig:throughput4_app1} show the throughput and timeouts for \texttt{watdiv-1\_star}, \texttt{watdiv-2\_stars}, \texttt{watdiv-3\_stars}, and \texttt{watdiv-paths} over all WatDiv datasets for all configurations respectively.
Figure~\ref{fig:cpu_app} shows throughput, CPU load and number of timeouts for \texttt{watdiv-btt} over all WatDiv datasets and all configurations, and Figure~\ref{fig:cpu_dbp_app} shows the CPU load and number of timeouts for \texttt{dbpedia-lsq}.
Last, Figure~\ref{fig:networkusage_app} shows the network usage including queries that timed out.

\end{appendix}

\end{document}